\documentclass[aps,pra,reprint,superscriptaddress, usenames,dvipsnames]{revtex4-1}
\usepackage{amsmath}
\usepackage{DejaVuSans}
\usepackage[EULERGREEK]{sansmath}
\usepackage{resizegather}

\usepackage{amssymb}
\usepackage{gensymb}
\usepackage{textcomp}

\usepackage{braket}
\usepackage[T1]{fontenc}
\usepackage{dsfont}
\usepackage{physics}
\newcommand{\rulesep}{\unskip\ \vrule\ }
\usepackage{caption}
\usepackage{mathtools}
\usepackage{siunitx}

\usepackage[ruled]{algorithm2e}

\makeatletter
\def\BState{\State\hskip-\ALG@thistlm}
\makeatother
\usepackage[normalem]{ulem}

\usepackage{tikz} 
\usetikzlibrary{matrix,shapes,arrows,positioning,chains}
\usetikzlibrary{shapes.misc}

\tikzset{cross/.style={cross out, draw=black, minimum size=2*(#1-\pgflinewidth), inner sep=0pt, outer sep=0pt},
cross/.default={1pt}}
\usetikzlibrary{decorations.pathmorphing,patterns}
\usetikzlibrary{backgrounds}
\usetikzlibrary{shapes,arrows}
\usetikzlibrary{decorations.markings}
\usepackage{epstopdf}
\tikzset{
    set arrow inside/.code={\pgfqkeys{/tikz/arrow inside}{#1}},
    set arrow inside={end/.initial=>, opt/.initial=},
    /pgf/decoration/Mark/.style={
        mark/.expanded=at position #1 with
        {
            \noexpand\arrow[\pgfkeysvalueof{/tikz/arrow inside/opt}]{\pgfkeysvalueof{/tikz/arrow inside/end}}
        }
    },
    arrow inside/.style 2 args={
        set arrow inside={#1},
        postaction={
            decorate,decoration={
                markings,Mark/.list={#2}
            }
        }
    },
}
\tikzstyle{line} = [draw, -latex']
\usetikzlibrary{shapes}
\definecolor{myred}{HTML}{FF0000} 

\definecolor{QRc}{HTML}{E64F40} 
\definecolor{nonQRc}{HTML}{3C49E3} 

\usepackage{xcolor}
\usepackage{mdframed}
\usepackage{multienum}

\DeclareMathOperator*{\argmin}{\arg\!\min}

\usepackage{epstopdf}
\usepackage{float}
\usepackage{dblfloatfix}
\usepackage{subcaption}

\makeatletter
\newcommand*{\centerfloat}{%
  \parindent \z@
  \leftskip \z@ \@plus 1fil \@minus \textwidth
  \rightskip\leftskip
  \parfillskip \z@skip}
\makeatother

\usepackage{todonotes}
\usepackage{hyperref}

\usepackage{placeins}
\usepackage{tikz} 
\usetikzlibrary{backgrounds}
\usetikzlibrary{decorations.markings}
\usetikzlibrary{decorations.pathreplacing}
\usetikzlibrary{shapes}
\usetikzlibrary{positioning, shapes.geometric, arrows.meta}

\definecolor{myred}{HTML}{FF0000} 
\definecolor{QRc}{HTML}{E64F40} 
\definecolor{nonQRc}{HTML}{3C49E3} 
\definecolor{mygreen}{RGB}{202, 252, 161}

\usetikzlibrary{shapes.geometric}
\usetikzlibrary{patterns}
\usetikzlibrary{positioning}
\usepackage[T1]{fontenc}
\usepackage{lmodern}
\usepackage{notoccite}
\usepackage{amssymb}

\tikzset{
    buffer/.style={
        draw,
        shape border rotate=180,
        regular polygon,
        regular polygon sides=3,
        fill=gray, fill opacity = 0.2,
        node distance=2cm,
        minimum height=4em
    }
}

\usepackage[noend]{algpseudocode}

\makeatletter
\algrenewcommand\ALG@beginalgorithmic{\footnotesize}
\makeatother

\makeatletter
\def\BState{\State\hskip-\ALG@thistlm}
\makeatother

\algnewcommand{\IfThenElse}[3]{
  \State \algorithmicif\ #1\ \algorithmicthen\ #2\ \algorithmicelse\ #3}

\begin{document}
\title{Optimising repeater schemes for the quantum internet}
\author{Kenneth Goodenough}
\affiliation{QuTech, Delft University of Technology, Lorentzweg 1, 2628 CJ Delft, The Netherlands}
\author{David Elkouss}
\affiliation{QuTech, Delft University of Technology, Lorentzweg 1, 2628 CJ Delft, The Netherlands}
\author{Stephanie Wehner}
\affiliation{QuTech, Delft University of Technology, Lorentzweg 1, 2628 CJ Delft, The Netherlands}
\affiliation{Kavli Institute of Nanoscience, Delft University of Technology}
\begin{abstract}
The rate at which quantum communication tasks can be performed using direct transmission is fundamentally hindered by the channel loss. Quantum repeaters allow, in principle, to overcome these limitations, but their introduction necessarily adds an additional layer of complexity to the distribution of entanglement. This additional complexity - along with the stochastic nature of processes such as entanglement generation, Bell swaps, and entanglement distillation - makes finding good quantum repeater schemes non-trivial. We develop an algorithm that can efficiently perform a heuristic optimisation over a subset of quantum repeater schemes for general repeater platforms. We find a strong improvement in the generation rate in comparison to an optimisation over a simpler class of repeater schemes based on BDCZ repeater schemes. We use the algorithm to study three different experimental quantum repeater implementations on their ability to distribute entanglement, which we dub \emph{information processing} implementations, \emph{multiplexed} implementations, and combinations of the two. We perform this heuristic optimisation of repeater schemes for each of these implementations for a wide range of parameters and different experimental settings. This allows us to make estimates on what are the most critical parameters to improve for entanglement generation, how many repeaters to use, and which implementations perform best in their ability to generate entanglement.
\end{abstract}    
    \maketitle

\section{Introduction}\label{sec:introduction}
The distribution of bipartite entanglement is critical for quantum communication tasks. Examples of such tasks include conference key agreement~\cite{chen2005conference, ribeiro2018fully}, clock synchronisation~\cite{jozsa2000quantum, ben2011optimized, giovannetti2001quantum}, and secure multi-party quantum computation~\cite{crepeau2002secure}. Photonic transfer of quantum states through optical fibre is one of the main candidates for long-distance entanglement generation. This is due to the potential of fast transmission speeds and the potential to be integrated with the hardware of classical networks. However, unlike classical bits, quantum states cannot be copied~\cite{wootters1982single, dieks1982communication}, which prevents us from amplifying the signal at intermediate points. In fact, the rate of entanglement generation over a fibre with transmissivity $\eta \ll 1$ necessarily scales linearly in $\eta = \exp(-\frac{L}{L_0})$~\cite{pirandola2017fundamental, takeoka2014squashed, pirandola2015general, Wilde:2016aa}, where $L$ is the total distance, and $L_0$ is the attenuation length. Thus, for large enough distances, the losses are a limiting factor on the rate of entanglement generation.

Quantum repeaters aim to counteract the effects of loss~\cite{briegel1998quantum,Duer1998,Childress2005, Childress2006}. Quantum repeater schemes are built on the concept of breaking the total length between two parties - Alice and Bob - up into several shorter (elementary) \emph{links}. At the two end points of these elementary links there is a \emph{repeater node}, which is a collection of quantum information processing devices. Depending on the scheme, the nodes have different requirements ranging from storage of quantum states to full-fledged quantum computation. 
By generating and storing entanglement over the elementary links and performing  Bell state measurements on the locally held states, the distance over which entanglement is present can be increased, until the two parties at the end are entangled~\cite{briegel1998quantum,Duer1998,Childress2005, Childress2006}.

However, the imperfect operations during this process lower the quality of the entanglement, potentially ruining the benefits of utilising quantum repeater nodes. The effects of noise can be counteracted by using \emph{entanglement distillation}, which can (in general probabilistically) turn multiple entangled pairs of lower fidelity into a smaller amount of pairs with higher fidelity~\cite{bennett1996concentrating, bennett1996purification, kalb2017entanglement}.

An entanglement generation scheme between two spatially separated parties Alice and Bob consists of the generation of entanglement over elementary links, entanglement swaps and distillation. Our goal is to find schemes that minimise the generation time of the entanglement between Alice and Bob for a given fidelity to the maximally entangled state in a suitable experimental model.  
However, finding optimal schemes is non-trivial for two reasons. First, the amount of schemes that can be performed grows super-exponentially in the number of elementary links/nodes, making a full systematic optimisation infeasible~(see \cite{jiang2007optimal} and Appendix \ref{sec:complexity}). Second, entanglement generation, Bell state measurements, and distillation are all processes that are in general probabilistic. Finding the corresponding probability distributions is believed to be computationally intensive~\cite{brand2019efficient, shchukin2017waiting, santra2019quantum, vinay2019statistical}.

For the reasons mentioned above, it seems necessary to either approximate or simplify the problem. Notably, in~\cite{jiang2007optimal}, an algorithm based on dynamical programming was proposed capable of efficiently optimising repeater schemes over the full parameter space. Under the heuristic approximation that all processes finish at the average time and there is no decoherence over time in the quantum memories, the algorithm constructs the scheme for a large chain combining the optimal solutions over smaller links. 

We take a different route. Instead of approximating the behavior of the schemes by the mean, we simplify the problem by considering a relevant subset of schemes. In particular, we consider schemes that succeed at all levels near-deterministically. Such schemes have the benefit of having a small variance of the fidelity and generation time. 
We note that the requirement of being near-deterministic does not imply that our algorithm cannot handle non-deterministic processes. High success probabilities can be enforced even when certain processes are not deterministic - in that case, the probability of a single success can be increased by repeating the process a number of times, ensuring that the whole process can be made near-deterministic, see Section~\ref{sec:algorithm} for further details. Furthermore, this allows us to calculate the success probability of a scheme exactly, even when more complicated protocols such as distillation and probabilistic swapping are performed. Finally, this approach also allows us to calculate the average noise experienced during storage, in contrast to~\cite{jiang2007optimal}, see Appendix~\ref{sec:avg_noise}.\\

In this paper, we detail an algorithm (publicly available as a Python script at~\cite{repo}) that performs a heuristic optimisation over the set of near-deterministic schemes when there are $n$ elementary links in $\mathcal{O}\left(n^2\log(n)\right)$ time, and $\mathcal{O}\left(n\log(n)\right)$ time if all the nodes have the same parameters and are equidistant. Concretely, the input to our algorithm is given by the experimental parameters of the nodes and connecting fibres, the distances between adjacent nodes, the possible protocols for elementary pair generation, swapping and distillation, and a set of algorithm-specific parameters, see Section~\ref{sec:algdescription}. The algorithm returns a collection of optimised schemes for generating entanglement between Alice and Bob.

We exploit the fact that our algorithm is not specific to any particular experimental setup, which allows for the optimisation over repeater schemes for several types of platforms.

The experimental platforms that we consider can be split up into three types:

\begin{itemize}
	\item \emph{Information processing platforms} - Information processing (IP) implementations have the ability to store quantum states and perform operations on them, such that it is possible to perform distillation. However, the number of quantum states that can be processed at the same time is presently limited to a small number. Examples of information processing implementations include systems such as trapped ions~\cite{monroe2013scaling, inlek2017multispecies, bock2018high}, nitrogen-vacancy centres in diamond~\cite{childress2013diamond, tsurumoto2019quantum}, neutral atoms~\cite{welte2018photon, siverns2019neutral, razavi2006long}, and quantum dots~\cite{hanson2007spins, zak2009quantum}.
	\item \emph{Multiplexed platforms} - Multiplexed (MP) implementations lack the ability to properly perform operations on the stored states, prohibiting distillation. However, a large number ($10^4-10^7$) of states can potentially be generated, transmitted and stored simultaneously with such implementations, effectively increasing the success probability for the elementary pair generation. Examples of such implementations include the different types of atomic ensembles~\cite{chou2005measurement, matsukevich2005entanglement, krovi2016practical}.
	\item \emph{A combination of information processing and multiplexed platforms} - Multiplexed platforms can overcome the effects of losses over the elementary links more easily than information processing platforms, but suffer from the lack of control and long coherence times available to information processing platforms. This motivates a combination of the two. That is, the elementary pair generation is performed with a multiplexed implementation, after which the quantum state is transferred into an information processing system. Such a combined setup benefits from the high success probability of the generation of the elementary pairs, together with the ability to perform entanglement distillation and longer coherence times.
\end{itemize}

We find that the optimisation returns schemes that outperform a simplified optimisation over more structured schemes, similar to those in~\cite{briegel1998quantum,Duer1998,Childress2005, Childress2006}. This highlights the complexity of repeater protocols for realistic repeater chains and the non-trivial nature of the optimisation problem. With such optimised schemes in hand we use our algorithm to study a range of questions, such as which setups hold promise for near-term quantum networks, how many nodes should be implemented, and which experimental parameters are the most important to improve upon. 

In Section \ref{sec:algorithm} we detail the basics of our algorithm, which takes as input an arbitrary repeater chain configuration, and returns a collection of heuristically optimised schemes which generate entanglement between two specified nodes, i.e.~the schemes have an optimal trade-off between the fidelity and generation time (over the set of considered schemes). This section also contains the heuristics we use to reduce the search space/complexity of the algorithm in Section~\ref{sec:heuristics} (with further details in Appendices \ref{sec:complexity} and \ref{sec:heuristics_results} regarding the complexity/runtime) and closes with the pseudocode of our algorithm in Section~\ref{sec:algdescription}. Section \ref{sec:applying} contains an overview of how we model the three experimental platforms considered in this paper, namely information processing (Section \ref{sec:ip}) implementations, multiplexed (Section \ref{sec:nonip}) implementations, and a combination of the two (Section \ref{sec:IPNP}). We then use the algorithm to heuristically optimise over repeater schemes for each of the implementations for several different scenarios in Section \ref{sec:results}. We close with a discussion of the results and the algorithm in Section \ref{sec:conclusions}.

\section{Algorithm description}
\label{sec:algorithm}
In this section we first explain the general structure of quantum repeater schemes (Section \ref{sec:basicconcept}). We then focus on the construction of so-called near-deterministic schemes (Section \ref{sec:ndschemes}). Afterwards, we first detail a non-scalable brute-force algorithm for optimising over such near-deterministic schemes (Section \ref{sec:brute-force}), after which we provide a feasible algorithm by implementing certain heuristics into the brute-force algorithm (Section \ref{sec:heuristics}). Appendices \ref{sec:complexity} and \ref{sec:heuristics_results} contain a more explicit discussion regarding the complexity/runtime with and without the heuristics implemented.

\subsection{Structure of quantum repeater schemes}
\label{sec:basicconcept}
The goal of a quantum repeater scheme is to distribute an entangled state between two remote parties Alice and Bob. Quantum repeater schemes are built up from smaller schemes. Schemes are constructed by performing ]\emph{connection} and \emph{distillation} protocols on pairs of smaller schemes.

\begin{figure}
	\includegraphics[width=0.38\textwidth]{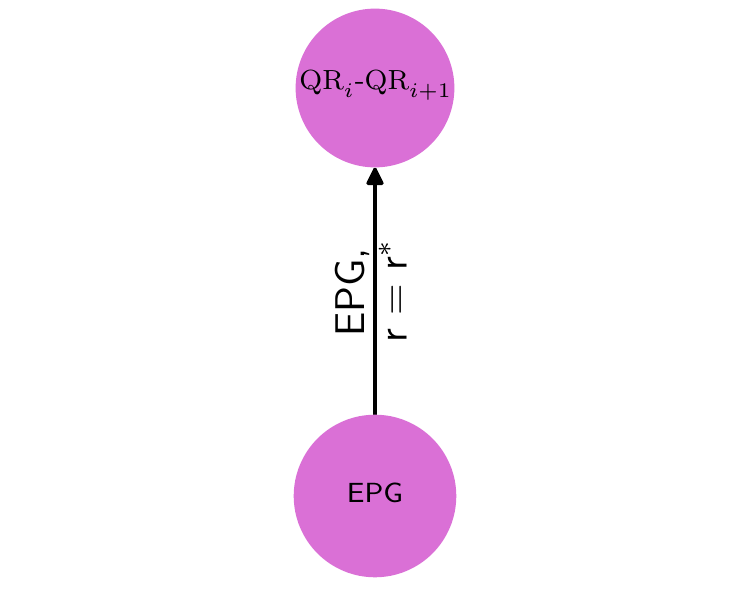}
	\caption{Elementary pair generation (EPG) between adjacent nodes QR$_i$ and QR$_{i+1}$. The schemes take a number of rounds $r=r^*$, even if entanglement is generated at an earlier round. See main text for further details.}
	\label{fig:ELG}
\end{figure}

\begin{figure}
	\includegraphics[width=0.38\textwidth]{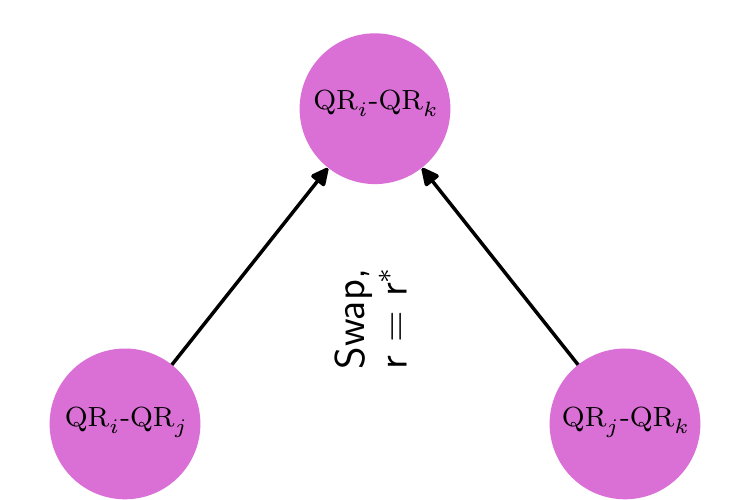}
	\caption{Entanglement swapping between two entangled pairs between links (QR$_{i}$, QR$_{j}$) and (QR$_{j}$, QR$_{k}$). By performing a Bell state measurement on the two local states at QR$_{j}$, the two entangled states turn into one entangled state between (QR$_{i}$, QR$_{k}$). The schemes take a number of rounds $r=r^*$ even if the scheme succeeds at an earlier round, see main text for further details. Note that the distances over which the entanglement has been generated for the links (QR$_{i}$, QR$_{j}$) and (QR$_{j}$, QR$_{k}$) need not be the same.}
	\label{fig:swap}
\end{figure}

Connection protocols extend the range over which entanglement exists. This can be done by elementary pair generation and entanglement swapping. Elementary pair generation (EPG) creates entanglement over elementary links, see Fig.~\ref{fig:ELG}. Entanglement swapping transforms two entangled states over two shorter links to an entangled state over a larger link using a Bell state measurement, see Fig.~\ref{fig:swap}.

\begin{figure}
	\includegraphics[width=0.38\textwidth]{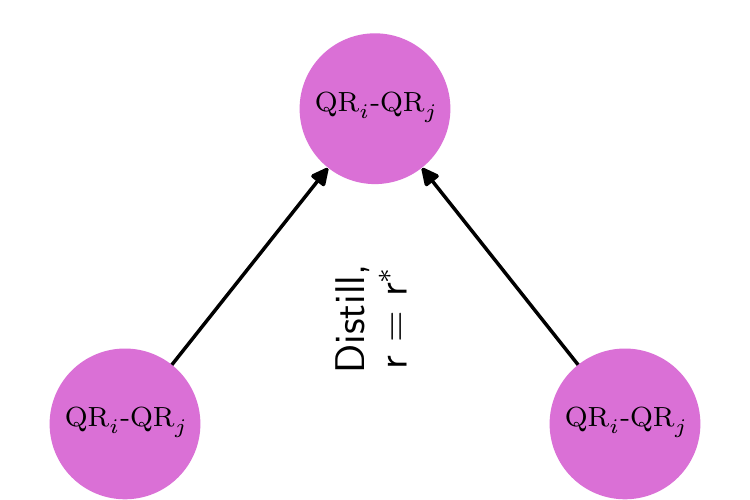}
	\caption{Example of a generic entanglement distillation protocol, transforming (possibly probabilistically) two entangled states to a single, more entangled state between nodes QR$_i$ and QR$_j$, using only local operations and classical communication. The schemes take a number of rounds $r=r^*$ even if distillation succeeds at an earlier round, see main text for further details. Note that QR$_i$ and QR$_j$ do not have to be directly connected by a fibre.}
	\label{fig:distill}
\end{figure}

Distillation protocols allow to (possibly probabilistically) convert two entangled states to a single, more entangled state using only local operations and classical communication~\cite{deutsch1996quantum, bennett1996purification}. There exist more complicated protocols, where an arbitrary number of entangled states are converted to a smaller number of entangled states~\cite{dehaene2003local, horodecki2009quantum}. Here, we only consider distillation protocols taking two states to a single one~\footnote{As discussed later, we actually do consider distillation protocols taking three or more states to a single one, but these are composed of several distillation protocols taking two states to a single one.}. See Fig.~\ref{fig:distill} for an illustration of a distillation protocol.

\subsection{Near-deterministic schemes}
\label{sec:ndschemes}
Entanglement generation schemes should preferably minimise the average generation time for a given fidelity $F$. However, the generation and distribution of entanglement is a \emph{stochastic} process, greatly complicating the optimisation over such schemes. Here, we simplify the problem by demanding that every step of the entanglement generation scheme is \emph{near-deterministic}. This requirement can be enforced even when some of the processes are not deterministic, such as elementary pair generation or Bell swaps. The probability of having at least a single success can be increased by repeating the whole scheme up until that point for multiple attempts~\footnote{Or, in the case of multiplexed platforms, the probability of having a single success for elementary link generation can be increased by having more modes.}. Near-deterministic schemes deliver a state with high probability at a specific time $T$, and it is this generation time $T$ that we use as our metric in this work~\footnote{Since the success of a scheme follows a geometric distribution, the average generation time can be computed from the success probability and the generation time of one attempt.}.

Let us exemplify this idea through a process for elementary pair generation (EPG).
This process might have a very small probability $p$ to succeed in a single attempt, which takes a time $T_{\textrm{attempt}}$ to perform. The probability of having at least a single success after $r$ attempts, is 

\begin{equation}
	p_{\textrm{single success}} = 1-(1-p)^r\label{eq:psuc0}\ .
\end{equation}
Thus, the probability of having at least one success can be increased to no less than $p_\textrm{min}$ by trying for $r= \left \lceil  \frac{\log(1-p_{\textrm{min}})}{\log(1-p)}\right\rceil$ attempts. We now consider protocols where the state is stored until a total time $r\cdot T_{\textrm{attempt}}$ has passed, even if a success occurs before $r$ attempts have passed. This ensures that a state can be delivered near-deterministically (i.e.~with probability at least $p_\textrm{min}$) at a pre-specified time $T = r\cdot T_{\textrm{attempt}}$. However, it comes at the cost of increased decoherence, since the state might have to be stored for a longer time (see~\cite{santra2019quantum} for a related concept).

Consider now the success probability of distillation protocols and (optical) Bell state measurements. Both protocols require the two states to be present, which holds with probability equal to the product of the probabilities of the two individual schemes having succeeded. Furthermore, distilling and swapping typically have a non-zero failure probability, potentially decreasing the success probability even further. However, we can use the same strategy used previously to increase the total success probability. That is, by repeating the whole scheme up to that point, it is possible to increase the success probability to at least the threshold $p_{\textrm{min}}$. Let us consider this concept for the example of a swap operation between two elementary pairs. The total success probability can now be increased by repeating the whole process of generating both elementary pairs and performing the swap operation.

\begin{figure}
	\includegraphics[width=0.38\textwidth]{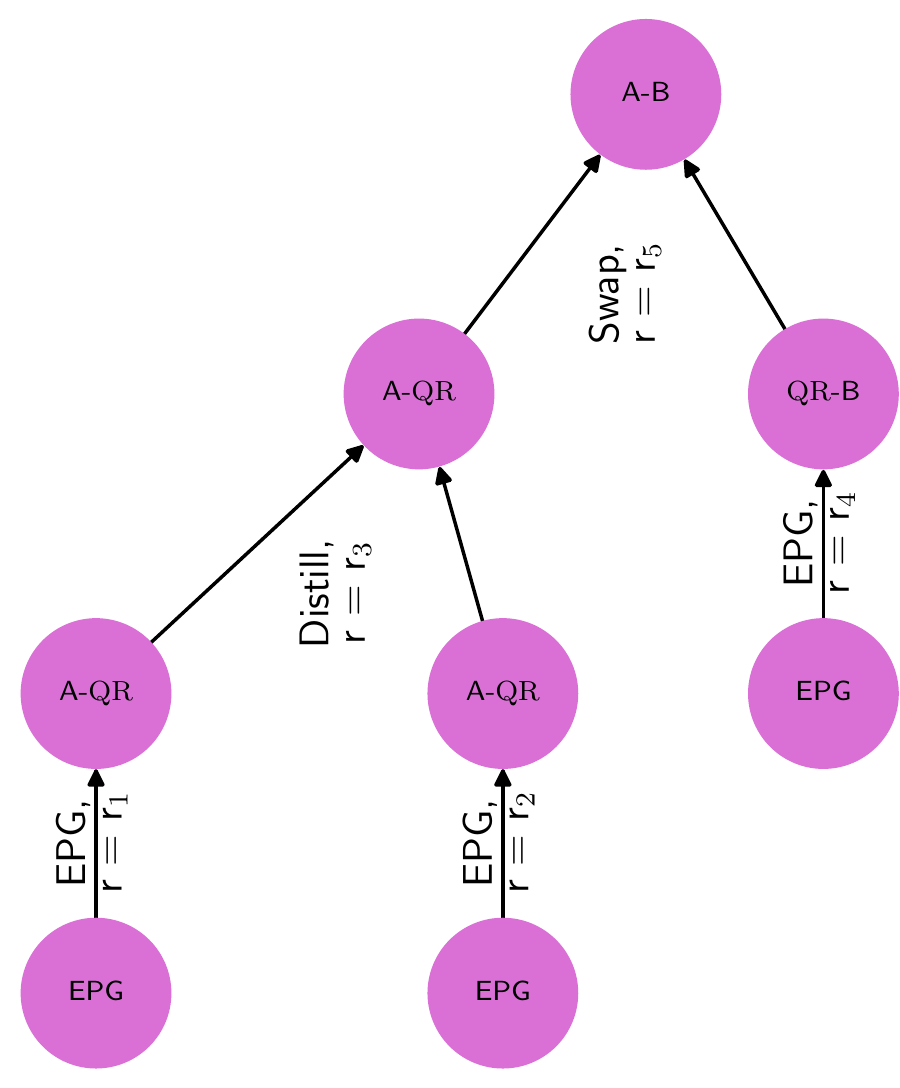}
	\caption{Schematic description of how near-deterministic schemes are constructed from the protocols shown in Figs.~\ref{fig:ELG}, \ref{fig:swap} and \ref{fig:distill}. Here entanglement is generated between the nodes A and B, using an intermediate node labelled by QR. The overall structure is that of a binary tree (modulo the leaves indicating elementary pair generation, indicated by EPG), since swapping and distillation is always performed between exactly two schemes. Each sub-tree is required to succeed with probability at least $p_\textrm{min}$, which can be enforced by repeating the whole sub-tree for a number of attempts $r$. Here, the specific number of attempts is indicated by $r_b$, $b \in \lbrace1, 2, 3, 4, 5 \rbrace$.}
	\label{fig:algorithm}
\end{figure}

This concept can be extended to more complex repeater schemes, ensuring that each step in the repeater scheme succeeds with high probability. A repeater scheme can thus be constructed by combining protocols from the ground up, where the average state, generation time $T$, and success probability $p$ of each scheme are only a function of the number of attempted rounds $r$, the protocol used, the parameters of the repeater chain, and the used schemes. We show an example of how such schemes can be constructed in Fig.~\ref{fig:algorithm}.

We note here that such near-deterministic schemes require us to keep states stored for some time, even if the underlying process has already succeeded, similar to the approaches in~\cite{humphreys2018deterministic, santra2019quantum}. This evidently comes at the cost of increased storage times, and thus a greater amount of average decoherence. Near-deterministic schemes also have benefits, however. 
Firstly, with near-deterministic schemes it is possible to make the variance of the resultant probability distributions arbitrary small by increasing $p_\textrm{min}$. Thus, near-deterministic protocols are able to deliver entanglement at a pre-specified time with high probability, which may be important for quantum information protocols consisting of multiple steps~\cite{humphreys2018deterministic}, such as entanglement routing~\cite{pant2017routing, schoute2016shortcuts}. Secondly, it is possible to calculate exactly the generation times and fidelities of near-deterministic schemes with relative ease, allowing for the optimisation over such schemes.

Let us compare near-deterministic schemes with the more general class of schemes considered in~\cite{jiang2007optimal}. Both frameworks take as building blocks a similar set of probabilistic protocols. In~\cite{jiang2007optimal}, the protocols are freely combined which makes challenging to estimate the average time they take to generate entanglement. This problem is sidestepped in \cite{jiang2007optimal} by heuristically assuming that all protocols take average time. 
In contrast, in  our framework, we combine protocols in blocks that have high success probability and take a fixed amount of time. This reduces the class of schemes but allows us to estimate exactly the generation time and the fidelity of the state generated.

\subsection{Brute-force algorithm}
\label{sec:brute-force}
We now introduce a brute-force algorithm to optimise entanglement distribution over the set of near-deterministic schemes between two distant nodes Alice and Bob. The algorithm takes as input the experimental parameters of the nodes and connecting fibres, the distances between adjacent nodes, a set of protocols for elementary pair generation, swapping and distillation, a minimum success probability and a limit on the maximum number of attempts and the maximum number of distillation rounds. 
The output consists of a data structure containing the schemes that minimise generation time parametrised by success probability and fidelity.

The brute-force algorithm generates and stores every possible scheme that can be created from the input conditions. Then for each achieved fidelity, it walks over the stored schemes to find the scheme minimising the generation time achieving at least that fidelity. In the following we sketch only the first part, as this is enough to argue that such an approach is non-scalable.

First, the algorithm takes the set $\mathcal{E}$ of protocols for elementary pair generation, together with the different number of attempts considered (of which there are at most $r_\textrm{discr}$), and explores all possible combinations of elementary pair generation protocols and number of attempts for each elementary link. Each of these combinations is stored if the success probability is larger than a specified $p_{\min}$. 

Next, the algorithm takes the set of distillation protocols $\mathcal{D}$ and a maximum number of distillation rounds $m$. For each elementary link, the algorithm loops over the number of distillation rounds: $1,\ldots, m$. For each number of rounds, the algorithm explores all combinations of pairs of schemes, number of attempts and distillation protocols and stores the resulting scheme if the success probability is larger than $p_{\min}$. 

The algorithm then proceeds iteratively over links of length $i\in\{2, 3, \ldots, n\}$, where $n$ is the total number of elementary links between the target nodes. Each iteration is divided into a swapping and a distillation step.

In the swapping step the algorithm considers all adjacent links of lengths $i_1,i_2$ such that $i_1+i_2=i$. For each valid pair of adjacent links and for each pair of schemes stored over the adjacent links, the algorithm explores all combinations of number of attempts and protocols in the set of swapping protocols $\mathcal{S}$. 
It stores a resulting scheme if the success probability is larger than $p_{\min}$. 

In the distillation step, the algorithm proceeds analogously to the description above for distillation over elementary links. 

While the approach just described might work for a very small chain, the number of schemes grows too quickly. In particular, the number of schemes to consider in the brute-force approach is lower bounded by

\begin{align}
	\mathcal{O}\left(\left(\left(r_\textrm{discr}\right)^2\cdot \left|\mathcal{E}\right|\cdot \left|\mathcal{S}\right|\right)^n\right)
\end{align}
when distillation protocols are not considered and by
\begin{align}
	\label{eq:complex1}
	\mathcal{O}\left(\left(r_{\textrm{discr}}\cdot \left|\mathcal{E}\right|\cdot \left|\mathcal{S}\right|\cdot \left|\mathcal{D}\right|\right)^{2^{m\cdot n}}\right)
\end{align}
when distillation is considered. Here $n$ is the number of elementary links, $\left|\mathcal{E}\right|$ is the number of ways elementary pairs can be generated (due to for example varying a parameter over some set of values), $\left|\mathcal{S}\right|$ the number of swapping protocols, $\left|\mathcal{D}\right|$ the number of distillation protocols,
$r_\textrm{discr}$ the different number of attempts considered, 
and $m$ the number of distillation rounds (see Appendix \ref{sec:complexity}). 

\subsection{A heuristic algorithm}
\label{sec:heuristics}
Now we introduce an efficient heuristic optimisation algorithm. The heuristic algorithm takes as starting point the brute-force algorithm presented before and incorporates a number of modifications that reduce the search space, thus overcoming the fast-growing complexity of the brute-force algorithm. We divide the modifications into heuristics for the pruning of schemes and heuristics for good schemes and detail them in the following. In the following we first discuss the modifications to the brute-force algorithm before presenting the pseudocode of the algorithm and analysing its complexity.

\subsubsection{Heuristics for the pruning of schemes}
The brute-force algorithm explores a grid of parameters at each step and stores all schemes with success probability above $p_{\min}$ independently of their quality. Instead, we can identify schemes that either are unlikely to combine into good schemes at subsequent steps or are very similar to existing schemes and not store them.

A first strategy is to only store schemes that deliver a state with fidelity above the threshold $F_{\textrm{threshold}} \geq \frac{1}{2}$. 

A second strategy is to \emph{coarse-grain} the fidelity and success probabilities. 
For this, the algorithm rounds the fidelity $F$ and success probability $p$ of each scheme to $\tilde{F}$ and $\tilde{p}$, the closest values in the sets $\left[F_\textrm{threshold},~F_\textrm{threshold} +  \varepsilon_F,~F_\textrm{threshold} + 2\varepsilon_F,~\ldots,~1\right]$ and $\left[p_\textrm{min},~p_\textrm{min} +  \varepsilon_p,~p_\textrm{min} + 2\varepsilon_p,~\ldots,~p_\textrm{max}\right]$ (see Appendix~\ref{sec:heuristics_results}).

If no scheme with the same $\tilde{F}$ and $\tilde{p}$ exists, the scheme is stored. Otherwise, we compare the two generation times of the two schemes. If the old scheme has a lower generation time, the new scheme is not stored. Otherwise, the new scheme replaces the old one. We note here that the actual values of $F$ and $p$ are stored, and not the values $\tilde{F}$ and $\tilde{p}$.

The third strategy consists in pruning sub-optimal protocols after having considered all protocols over a given link. 
A scheme is sub-optimal if there exists another scheme over that link with the same $\tilde{p}$ which has a lower generation time but equal or higher fidelity. 
We detail the implementation of the above pruning heuristics in Algorithm~\ref{algo:pruning}.

\subsubsection{Heuristics for good schemes}
Pruning reduces the amount of sub-optimal schemes that are kept stored. This prevents those schemes from being combined with other schemes, reducing the algorithm runtime. However, it would be preferable if those schemes would not even be considered in the first place. For this reason, we use \emph{heuristics} on what schemes to consider. The heuristics that we use are \emph{banded distillation}, \emph{banded swapping}, and the \emph{bisection heuristic}, which we will detail in what follows.

Many distillation protocols acting on two states yield states of fidelity larger than the input states only when the input states have fidelities that are relatively close to each other~\cite{dur2007entanglement}. This motivates restricting distillation to states that have fidelities $F_1$ and $F_2$ separated at most by some threshold $\varepsilon_{\textrm{distill}}$,
\begin{equation}
	\label{eq:banded_dist}
	\left|F_1 - F_2 \right| \leq \varepsilon_{\textrm{distill}}\ .
\end{equation}
This heuristic, first considered in~\cite{van2009system} is called \emph{banded distillation}.

Inspired by banded distillation we introduce a similar heuristic for entanglement swapping that we dub \emph{banded swapping}. 
A naive extension of banded distillation to swapping would be to require that the absolute difference of the fidelities of the two swapped states be small. However, by investigating the heuristically optimised schemes, our numerical exploration (see Appendix~\ref{sec:heuristics_results}) suggests that the number of nodes over which the entanglement is generated also plays a role. 
In particular, we find that it is sufficient to restrict swapping to states that satisfy,

\begin{equation}
	\label{eq:nodisplength}
	\left|i_1-i_2\right| \leq 2\log(i_1 + i_2 -1),
\end{equation}
and 
\begin{equation}
	\label{eq:banded_swap}
	\left|\frac{\log(F_1)}{i_1} - \frac{\log(F_2)}{i_2} \right| \leq \varepsilon_{\textrm{swap}}\ 
\end{equation}
where $\varepsilon_{\mathrm{swap}}$ controls the granularity of the heuristic, $F_1,~F_2$ are the fidelities of the two states, and $i_1, i_2$ is the number of links over which the entanglement was generated, e.g.~the number of elementary links between QR$_i$-QR$_j$ and QR$_j$-QR$_k$ in Fig.~\ref{fig:swap}, respectively. We note that the first condition was already present in~\cite{jiang2007optimal}. 

The third heuristic - which we call the \emph{bisection heuristic} - is inspired by the BDCZ scheme~\cite{briegel1998quantum}. Similarly to the BDCZ scheme, it applies to \emph{symmetric repeater chains}. That is, repeaters chains where all nodes have the same parameters and are connected by identical elementary links. However, unlike the BDCZ scheme which is only applicable if the number of elementary links is equal to a power of two, the bisection heuristic is applicable independent of the number of elementary links.

The heuristic works as follows. Factorisation allows us to write the total number of elementary links as $n = 2^j \cdot h$, where $j$ is the number of times $n$ is divisible by $2$, and $h$ is the odd part of $n$. First, an optimisation is performed over a link of length $h$. From then on, similar to the BDCZ scheme, swapping only occurs between entanglement that has been generated over a total number of elementary links equal to a multiple of $h$. This heuristic has the possibility of dramatically reducing the algorithm runtime for certain values of $n$.

\subsubsection{Pseudocode of the heuristic algorithm}\label{sec:algdescription}
We now present the pseudocode of the heuristic algorithm. The general algorithm is described in Algorithm~\ref{algo:algo}, while the subroutines for storing the schemes and for the pruning heuristic are given in Algorithm~\ref{algo:store} and Algorithm~\ref{algo:pruning}.

The algorithm takes as input an additional number of parameters on top of the parameters already discussed for the brute-force algorithm. These parameters regard the heuristics and were described in the previous section. These parameters are $\varepsilon_{F},~\varepsilon_p$ (the discretisation used for the pruning of schemes for the fidelity and success probability, respectively), $~F_\mathrm{threshold}$ and $p_{\textrm{max}}$ (the minimum values required to consider a scheme for the fidelity and success probability, respectively). A software implementation requires also a number of experimental parameters for characterising the hardware and estimating the output of each scheme, however we leave the explicit description of the hardware parameters out of the pseudocode. For details of the actual implementation, please refer to the repository~\cite{repo}.

\subsubsection{Complexity and runtime of the heuristic algorithm}
As we show in Appendix \ref{sec:complexity}, the heuristics allow us to go from a number of considered schemes that grows super-exponentially in the number of links, to a number of schemes that is upper bounded by

\begin{align}
	\label{eq:complex2}
	\mathcal{O}\left( 2\cdot r_\textrm{discr}\left(\frac{\left(1-F_{\textrm{threshold}}\right)\left(1-p_{\textrm{min}}\right)}{\varepsilon_F\varepsilon_{p}}\right)^2n^2\log\left(n\right)\right)\ ,
\end{align}
implying that the number of considered schemes is now only on the order of $n^2\log(n)$, as opposed to super-exponential in $n$. 
Here $r_\textrm{discr}$ is the maximum number of values allowed for the number of attempts $r$, $F_\textrm{threshold}$ the minimum fidelity we allow a scheme to have, $p_\textrm{min}$ the minimum accepted success probability, $\varepsilon_F$, $\varepsilon_p$, are the discretisation used for the coarse-graining and $n$ the number of elementary links.
Furthermore, in the case of a symmetric repeater chain (i.e.~every node has the same parameters and the nodes are equidistant), the optimisation can be further simplified. As we show in Appendix \ref{sec:complexity}, the number of schemes to consider in the symmetric case is upper bounded by

\begin{align}
	\label{eq:complex3}
	\mathcal{O}\left(r_\textrm{discr}\left(\frac{\left(1-F_{\textrm{threshold}}\right)\left(1-p_{\textrm{min}}\right)}{\varepsilon_F\varepsilon_{p}}\right)^2 n\log\left(n\right)\right)\ .
\end{align}

In practice, we find that our algorithm runtime ranges from approximately 100 seconds to approximately 100 minutes, when considering 1 and 35 intermediate nodes for a symmetric repeater chain, respectively. We investigate the effects of the heuristics on the algorithm runtime in more detail in Appendix \ref{sec:heuristics_results}, where we perform an experimental analysis of the algorithm runtime and its `accuracy' when varying $\varepsilon_{F}$, $\varepsilon_{p}$, $\varepsilon_\textrm{swap}$, and $\varepsilon_\textrm{distill}$. We use these results to settle on the values for $\varepsilon_{F}$, $\varepsilon_{p}$, $\varepsilon_\textrm{swap}$, and $\varepsilon_\textrm{distill}$. We only investigate the bisection heuristic when going to a larger number of nodes in Section~\ref{sec:resultsIPMP}.

\begin{algorithm}[h]
	\KwIn{{scheme}, {store}, $p_\mathrm{min}$,  $F_\mathrm{threshold}$, {link}, $\varepsilon_F$, $\varepsilon_p$}
	\KwOut{{store} with scheme possibly added}
	\vspace*{1mm}
	\hrule
	\vspace*{1mm}
	$F \gets$ fidelity stored in scheme \;
	$p \gets$ probability stored in scheme \;
	$n_{\varepsilon_p} \gets \argmin_{n \in \mathbb{N}}$ s.t.~$p<p_\mathrm{min} + n\cdot \varepsilon_p$\;
	$n_{\varepsilon_F} \gets \argmin_{n \in \mathbb{N}}$ s.t.~$F<F_\mathrm{threshold} + n\cdot \varepsilon_F$\;
	\If{$F\geq F_\mathrm{threshold}$}{
		\uIf{$\mathrm{store}[\mathrm{link}][n_{\varepsilon_p}][ n_{\varepsilon_F}]$ already exists}{
			$T' \gets$ generation time of ${\textrm{store}}[\textrm{link}][n_{\varepsilon_p}][ n_{\varepsilon_F}]$\;
			\If{T<T'}
			{$\textrm{store}[\textrm{link}][n_{\varepsilon_p}][ n_{\varepsilon_F}] \gets$ scheme
			}
		}
		\Else{
			$\textrm{store}[\textrm{link}][n_{\varepsilon_p}][ n_{\varepsilon_F}] \gets$ scheme}
	}
	\Return{$\mathrm{store}$}
	\caption{{\sc StoreScheme}, subroutine for storage of the schemes.}
	\label{algo:store}
\end{algorithm}

\begin{algorithm}[h]
	\KwIn{{store}, $p_\mathrm{min}$, \textrm{link}, $\varepsilon_p$}
	\KwOut{{store} with sub-optimal schemes over {link} pruned}
	\vspace*{1mm}
	\hrule
	\vspace*{1mm}
	\For{$n\geq 0$ s.t.~$p_\mathrm{min} + n\cdot \varepsilon_p \leq 1$} {
		{orderedSchemes} $\gets$ {$\textrm{store}[\textrm{link}][n], $ ordered by fidelity from high to low}\;
		$N \gets $ size of orderedSchemes\;
		maxTime $\gets$ generation time of {orderedSchemes}[$0$]\;
		\For{$i \gets 1, \ldots,  N$} {
			\uIf{maxTime$\hspace{1mm}\leq$ generation time of orderedSchemes\textrm{[}i\textrm{]}}{
				Remove {orderedSchemes}[$i$] from $\textrm{store}[\textrm{link}][n]$
			}
			\Else{
				maxTime $\gets$ generation time of {orderedSchemes}[$i$]
			}
		}
	}
	\Return{$\mathrm{store}$}
	\caption{{\sc Prune}, prunes the sub-optimal schemes stored for a given link.}
	\label{algo:pruning}
\end{algorithm}

\onecolumngrid

\begin{algorithm}
	\DontPrintSemicolon
	\KwIn{$n:$ number of elementary links $n$ in repeater chain\; 
		\hspace{10.5mm} $\varepsilon_F, \varepsilon_p:$ coarse-graining parameters for the fidelity and probability\;
		\hspace{10.5mm} $\varepsilon_{\mathrm{distill}}, \varepsilon_\mathrm{swap}:$ parameters for the heuristics for distillation and swapping\;
		\hspace{10.5mm} $m:$ maximum number of distillation rounds\;
		\hspace{10.5mm} $p_{\min}, p_{\max}:$ minimum and maximum scheme success probabilities\;
		\hspace{10.5mm} $r_{\textrm{discr}}:$ number of different attempt values\;
		\hspace{10.5mm} $F_\mathrm{threshold}:$ minimum fidelity for schemes to be stored\;
		\hspace{10.5mm} $\mathcal E,\mathcal S, \mathcal D$: sets of protocols for elementary pair generation, swapping and distillation\;
		\hspace{10.5mm} $\mathcal L_i,i\in[1,n]$: set of links of length $i$}
	\KwOut{store: a data structure containing entanglement generation schemes with the minimum generation time parametrised by the link, coarse grained success probability and coarse grained fidelity.}
	\vspace*{1mm}
	\hrule
	\vspace*{1mm}
	Initialise store\;
	\For{$i\gets 1~\mathbf{to}~n$} {
		\For{$\mathrm{link}$ in $\mathcal{L}_i$} {
			\eIf{$i=1$}{
				\tcp{Loop over elementary pair generation protocols}
				\For{$\mathrm{EPGProtocol\ in\ }\mathcal E$}{
					\For{$r$ such that \eqref{eq:psuc0} or \eqref{eq:psuc1} (IP/MP platforms, resp.) is between $p_{\min}$ and $p_{\max}$ in $r_\textrm{discr}$ steps}{
						\textrm{scheme} $\gets$ EPGProtocol$(r, \mathrm{link}, n )$\\
						\textrm{store} $\gets$ {\sc StoreScheme}({scheme}, {store}, $p_\mathrm{min}, F_\mathrm{threshold}$, link, $\varepsilon_F$, $\varepsilon_p$)}}
			}
			{
				\tcp{Loop over all schemes satisfying the swapping heuristic and over all swapping protocols}
				\For{every $\mathrm{link1}$ and $\mathrm{link2}$ such that entanglement can be created over $\mathrm{link}$ by swapping between those links}{
					\For{every pair $(s_1, s_2)$ of stored schemes in $\mathrm{store[link1]}$ and $\mathrm{store[link2]}$ satisfying \eqref{eq:nodisplength} and \eqref{eq:banded_swap}}{
						\For{$\mathrm{swapProtocol\ in\ }\mathcal{S}$}{
							\For{$r$ such that \eqref{eq:psuc0} or \eqref{eq:psuc1} (IP/MP platforms, resp.) is between $p_{\min}$ and $p_{\max}$ in $r_\textrm{discr}$ steps}{
								{scheme} $\gets \mathrm{swapProtocol}(s_1, s_2, r, \mathrm{link}, n )$\\
								{store} $\gets$ {\sc StoreScheme}({scheme}, {store}, $p_\mathrm{min}$, $F_\mathrm{threshold}$, {link}, $\varepsilon_F$, $\varepsilon_p$)}}
					}
				}
			}
			\For{$j\gets 1~\mathbf{to}~m$} {
				\tcp{Loop over all schemes satisfying the distillation heuristic and over all distillation protocols}
				\For{every pair $(s_1, s_2)$ of stored schemes in $\mathrm{store[link]}$ and  satisfying \eqref{eq:banded_dist}}{
					\For{$\mathrm{distillationProtocol\ in\ }\mathcal D$}{
						\For{$r$ such that \eqref{eq:psuc0} or \eqref{eq:psuc1} (IP/MP platforms, resp.) is between $p_{\min}$ and $p_{\max}$ in $r_\textrm{discr}$ steps}{
							{scheme} $\gets$ distillationProtocol$(s_1, s_2, r, \mathrm{link}, n )$\;
							{store} $\gets$ {\sc StoreScheme}({scheme}, {store}, $p_{\min}$, $F_\mathrm{threshold}$, {link}, $\varepsilon_F$, $\varepsilon_p$)}
					}
				}
			}
			{store} $\gets$ {\sc Prune}({store}, $p_{\min}$, $F_\mathrm{threshold}$, {link}, $\varepsilon_p$)
		}
	}
	\Return{\textrm{store}}
	\caption{Heuristic optimisation over near-deterministic schemes for a repeater chain of $n$ elementary links.}
	\label{algo:algo}
\end{algorithm}
\vfill

\twocolumngrid

\section{Platform models}
\label{sec:applying}
The algorithm discussed is independent of the underlying physical implementation, and can thus be applied to several experimental platforms. We use our algorithm to study three different types of platforms encapsulating a large range of technologies. The three platforms share the capability to store quantum information but differ in their quantum information processing capabilities. We call these platforms: information processing platforms, multiplexed platforms, and combined platforms. Information processing platforms have the ability to perform operations on the stored qubits, but are currently limited to a small number of qubits. Multiplexed platforms, on the other hand, lack the ability to perform operations on stored states, but can generate and store a potentially very large number of different states simultaneously. Obviously, these platforms differ greatly, but both approaches have complementary qualities for long-distance entanglement generation. This motivates us to also compare a combination of the two. That is, a setup where the elementary pairs are generated with a multiplexed platform, but swapping and distillation are performed by an information processing platform.

In the rest of the section, we discuss the basics of each of the implementations and the modelling of the underlying processes.

\subsection{Quantum repeaters based on information processing platforms}
\label{sec:ip}
We call information processing (IP) platforms those that have the capability to perform gates on the stored states, thus enabling entanglement distillation. The number of quantum states that can be stored and processed is presently limited. Experimental information processing platforms that have demonstrated excellent control over storage qubits include NV centres in diamond~\cite{humphreys2018deterministic, hensen2015loophole, hensen2016loophole, abobeih2018one, kalb2017entanglement, cramer2016repeated, tsurumoto2019quantum}, neutral atoms~\cite{welte2018photon, siverns2019neutral}, color centers in diamond~\cite{nguyen2019quantum, nguyen2019integrated}, quantum dots~\cite{burkard2000spintronics, chen2018highly, huber2018semiconductor}, and trapped ions~\cite{monroe2013scaling, inlek2017multispecies, bock2018high}. 

\begin{figure}
	\centerfloat
	\includegraphics[clip, trim = 0mm 36mm 0mm 14mm, width = 0.5\textwidth]{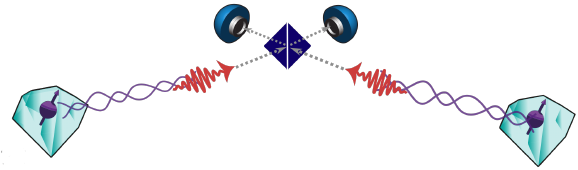}
	\caption{An example of an elementary link implemented with an information processing platform. The two nodes are connected by a fibre with a beamsplitter in the middle and two detectors. For the case considered in this figure, the two nodes are nitrogen-vacancy centres in diamond. For both protocols, the two nodes both send one-half of an entangled state to the middle, which after interference and successful detection leads to a shared state between the two nodes. Figure taken with permission from~\cite{rozpkedek2018near}.}
	\label{fig:IParch}	
\end{figure}

In this work we consider two protocols for the generation of elementary pairs for information processing platforms. These protocols are the single-click-~\cite{cabrillo1999creation, lucamarini2018overcoming, rozpkedek2018near} and double-click protocol~\cite{barrett2005efficient}. We give an example based on nitrogen-vacancy centers in diamond in Fig.~\ref{fig:IParch}. We stress that this is just one example of an information processing platform, and that our algorithm can be applied to other platforms.

The setup for both the single-click and the double-click protocols consists of two nodes with at least one memory qubit. The two nodes are connected via an optical channel to an intermediate beamsplitter station with a detector at each of the output ports (see Fig.~\ref{fig:IParch}).

Let us now detail first the single-click protocol. The qubits at the nodes are prepared in a superposition of the ground state $(\ket{\downarrow})$ and the first excited state $(\ket{\uparrow})$: $\sin(\theta)\ket{\downarrow}+\cos(\theta)\ket{\uparrow}$. 
Upon receiving an appropriate excitation signal, the memory emits a photon ($\ket{1}$) if it is in the excited state, and no photon ($\ket{0}$) otherwise. Since the memory qubit is in a superposition, this results in a memory-photon entangled state $\sin(\theta)\ket{\downarrow}\ket{0}+\cos(\theta)\ket{\uparrow} \ket{1}$. The two photons are then directed to and interfered on the intermediate beamsplitter. One experimental complication here is that the phase picked up by the photons as they travel through the fibre is unknown unless the fibres are stabilised. However, if this is the case, upon the detection of a single photon (single-click) at the beamsplitter station, the creation of an entangled pair can be heralded to the two nodes.

The double-click protocol on the other hand does not rely on phase-stabilisation. For the double-click protocol, each node prepares a qubit in a uniform superposition of the ground and first excited state~\cite{barrett2005efficient}. By applying specific pulses to the qubits, a photon will be coherently emitted in the early or late time-bin, depending on the state of the qubit at the node. The photons are then interfered at the beamsplitter station.
The entanglement between the two qubits is heralded to the two nodes upon the detection of two consecutive clicks at the beamsplitter station.  While the double-click protocol does not require phase-stabilisation, it has a lower success rate in comparison to the single-click protocol.

The parameter $\theta$ is tuneable, which allows for a trade-off between the success probability and the fidelity of the heralded state for the single-click protocol~\cite{campbell2008measurement, kalb2017entanglement, rozpkedek2018near}. For the double-click protocol there is no such trade-off however.

For the single-click protocol we use the error model from~\cite{rozpkedek2018near}. For the double-click protocol we use the error model from~\cite{barrett2005efficient}.

Entanglement distillation across two separated matter qubits has been achieved with an NV-centre setup~\cite{kalb2017entanglement}, where a specific entanglement distillation protocol~\cite{campbell2008measurement} was implemented. This distillation protocol is optimal when the involved states are correlated in a particular manner~\cite{rozpkedek2018optimizing}. In general however, the states that we consider are not of this form. For this reason, we will consider here only the DEJMPS protocol~\cite{deutsch1996quantum}, which was originally designed to work well for maximally entangled states with depolarising noise. In this protocol, we first apply a local rotation on each of the qubits, then two local CNOT operations, and measure the targets of the CNOT operations in the computational basis. We deem the distillation to be a success when the measurement outcomes are equal. 

We now sketch the underlying abstract error models and the various experimental parameters. 

State preparation for the generation of elementary pairs takes some time $t_\textrm{prep}$, performing the gates for distillation takes time $t_\textrm{distill}$, and performing a Bell state measurement takes time $t_\textrm{swap}$. State preparation is also imperfect, which we model as dephasing with parameter $F_\textrm{prep}$. States stored in the memories for a time $t$ are subject to decoherence. We model this decoherence as joint depolarising and dephasing noise, see Appendix \ref{sec:avg_noise} for details on the decoherence model.

The fibre has a refractive index of $n_{\textrm{ri}}$ and an attenuation length $L_0$. The attenuation length is defined such that $\eta = e^{-L/L_0}$, where $\eta$ is the transmissivity and $L$ the length of the fibre. There are three other sources of photon loss that we model~\cite{rozpkedek2018parameter, rozpkedek2018near} - the probability of successfully emitting a photon $p_\textrm{em}$, the probability of emitting a photon with the correct frequency and it not being filtered out (conditioned on having emitted the photon) $p_{\textrm{pps}}$ and the probability of the detector successfully clicking when a photon is incident $p_{\textrm{det}}$.

Applying gates induces noise on the states. Performing a Bell state measurement induces depolarising and dephasing with parameters $\lambda_{\textrm{BSM, depol}}$ and $\lambda_{\textrm{BSM, deph}}$, respectively. Performing the CNOT operations for distillation also leads to depolarising and dephasing with parameters $\lambda_{\textrm{CNOT, depol}}$ and $\lambda_{\textrm{CNOT, deph}}$, respectively.
Furthermore, we model measurement errors by applying depolarising noise with parameter $\lambda_{\textrm{meas. depol.}}$ before measuring a state. Finally, the uncertainty in the phase stabilisation $\Delta \phi$ induces dephasing in the state preparation for the single-click protocol (see~\cite{rozpkedek2018near}). 

\subsection{Quantum repeaters based on multiplexed platforms}
\label{sec:nonip}
Multiplexed (MP) platforms are a promising candidate for quantum repeater implementations~\cite{chou2005measurement, matsukevich2005entanglement, krovi2016practical, sangouard2011quantum}. While multiplexed platforms lack the ability to perform gates on the states stored in the memories, they have the potential to process a large number of states simultaneously, which can dramatically increase the probability at which elementary pairs can be generated. Here we discuss the basics of a model for the quantum repeater scheme proposed in~\cite{krovi2016practical} (see Appendix~\ref{sec:ellinkgenMP}). This repeater scheme uses photon-number and spectrally resolving detectors, frequency-multiplexed multimode memories, and parametric down conversion (PDC) sources. 

\begin{figure}
	\centerfloat
	\includegraphics[clip, trim = 0mm 0mm 0mm 0mm, width = 0.5\textwidth]{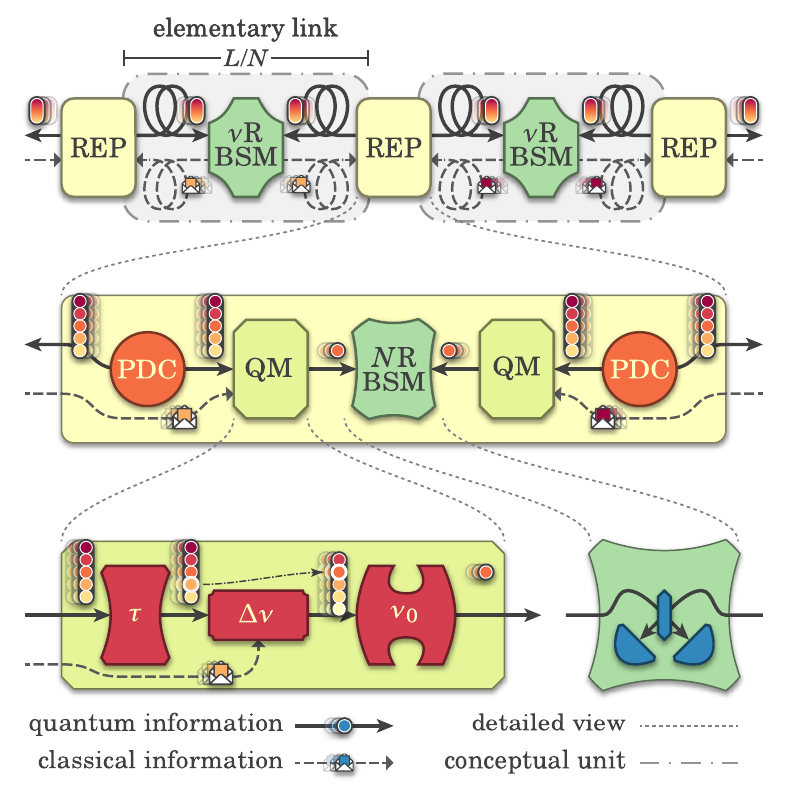}
	\caption{Schematic description of a multiplexed repeater implementation. Top: The total distance $L$ is split into $N$ elementary links, each with a spectrally-resolving BSM (indicated by $\nu$RBSM) in the middle, and with two nodes (each indicated by REP) at the end point of the elementary links. Middle: Zoom in of a node. Each node contains two PDC sources of multiplexed bipartite entanglement, two quantum memories (indicated by QM) and a number-resolving Bell state measurement station (indicated by \emph{N}RBSM). Bottom: Detailed view of QM and \emph{N}RBSM. Each quantum memory not only stores (in the unit indicated by $\tau$), but can also perform a frequency-shift (in a unit indicated by $\Delta \nu$) and a frequency filter (indicated by the unit $\nu_0$), while each \emph{N}RBSM contains a beamsplitter and two single-photon detectors, which performs a Bell state measurement on the frequency-shifted photons. Illustration taken with permission from~\cite{krovi2016practical}.}
	\label{fig:MParch}	
\end{figure}

An elementary link consists of two PDC sources, each located at one of the two nodes. The PDC sources emit entangled states for a large set of frequencies. One half of each entangled state is sent towards a jointly collocated quantum memory, which can store a large number of modes simultaneously. The other half is sent to an intermediate station between the two nodes, where it interferes on a spectrally-resolving beamsplitter with the corresponding state sent from an adjacent node. If at least one successful click pattern is detected at the output of the beamsplitter, the information of the corresponding mode is sent to the nodes. The information is used to filter out the other modes, after which frequency conversion is performed to a predetermined frequency at each of the nodes. The frequency conversion to a predetermined frequency ensures that at each node the successful modes from the two adjacent links can interfere at a local beamsplitter station. Photon-number resolving detectors are collocated at the output of the local beamsplitter to identify and discard multiphoton events. A schematic description can be found in Fig.~\ref{fig:MParch}.

Let us now investigate the parameters underlying the scheme we have just described. Consider a PDC source emitting entangled states with time-bin encoding. An ideal source would emit states of the form $\frac{1}{\sqrt{2}}\left(\ket{10, 01} + \ket{01, 10}\right)$, where the notation $\ket{nm, mn}$ indicates $n/m$ photons in the `early/late' bin in one half of the state and $m/n$ photons in the `early/late' bin in the other half. However, realistic PDC sources include additional terms. The resulting state can be approximated~\cite{krovi2016practical} by a state of the form

\begin{align}
	\ket{\psi_{N_s}} &= \sqrt{p_0} \ket{00, 00} + \sqrt{\frac{p_1}{2}}\left(\ket{10, 01} + \ket{01, 10}\right)\nonumber\\
	+ &\sqrt{\frac{p_2}{3}}\left(\ket{20, 02} - \ket{11, 11} + \ket{02, 20}\right)\ ,
\end{align}
with
\begin{align}
	p_0 = \frac{1}{(N_s + 1)^2},\nonumber\\
	p_1 = \frac{2N_s}{(N_s + 1)^3},\nonumber\\
	p_2 = 1-p_0 - p_1\label{eq:photondistribution}\ .
\end{align}
Here $N_s$ is the mean photon number present in the state and is a tuneable parameter.
Increasing the mean photon number $N_s$ increases the probability of detecting two clicks at the middle station (as can be seen from the decrease in the parameter $p_0$), while at the same time lowering the fidelity of the state conditioned on detecting two clicks.

Note that \eqref{eq:photondistribution} is a truncated version of the state derived in~\cite{kok2000postselected}, i.e.~all the higher order terms are included in $p_2$. As described in~\cite{krovi2016practical}, the multiphoton components limit the ability to generate entanglement without the use of photon-number resolving detectors.

The number of modes $N_{\textrm{modes}}$ increases the success probability of elementary pair generation. If the success probability of the creation of a single elementary pair is given by  $p_{\textrm{el}}$, the success probability of generating at least one elementary pair is given by $1-(1-p_{\textrm{el}})^{N_{\textrm{modes}}}$. Thus, $N_{\textrm{modes}}$ should be on the order of $\frac{1}{p_{\textrm{el}}}$, since $\lim_{p_\textrm{el}\rightarrow 0}1-(1-p_{\textrm{el}})^{\frac{\alpha}{p_\textrm{el}}}=1-e^{-\alpha}$. Finally, while a purely deterministic Bell state measurement is impossible using only linear optics~\cite{lutkenhaus1999bell, vaidman1999methods}, there are theoretical workarounds to increase the success probability~\cite{olivo2018investigating, ewert20143, grice2011arbitrarily, lee2015nearly1, lee2015nearly2, lee2013near, zaidi2013beating}. We consider the approach introduced in~\cite{grice2011arbitrarily}, where the success probability of the Bell state measurement can be increased to $1-\frac{1}{2^{N+1}}$ by using $2^{N+1}-2$ ancillary photons.

We assume the states can be retrieved from the memories on-demand. On-demand retrieval is necessary for our algorithm to work, since the storage times are not fixed. This is due to the uncertainty in which attempt entanglement will be generated. On-demand retrieval can be achieved with rare-earth ion ensembles by, for example, switching coherence from electronic levels to spin levels, as done in~\cite{Holzaepfel2019, Timoney2013}. Besides allowing for on-demand recall, this also has the added benefit of increased memory life-time~\cite{saglamyurek2011broadband}.

We consider the same type of noise for operations as we did for information processing platforms. This means that measurements have an associated amount of depolarising and dephasing.
Finally, `decoherence' over time for the memory manifests as an exponential decay in the output efficiency of the memory, not in a reduction of the fidelity of the state~\cite{afzelius2009multimode, saglamyurek2011broadband}. Thus, the longer a state is stored, the smaller the probability it can be retrieved for measuring or further processing.

\subsection{Combining the two setups}
\label{sec:IPNP}
An information processing implementation has the benefit of long coherence times and control over the memory qubits, which allows for distillation. On the other hand, multiplexed platforms do not support distillation, but have the benefit of emitting and storing a large number of modes, increasing the success probability of the elementary pair generation significantly. Optimistically, one could imagine a futuristic setup which combines the strengths of the two setups. That is, elementary pair generation is performed by a multiplexed platform, after which the successfully generated pairs are frequency-converted into a frequency that can be stored in an information processing platform. The state is then stored in a memory, which can be done using, for example, a reflection-based heralded transfer~\cite{kalb2015heralded, nemoto2016photonic}. 
For simplicity, we assume that the transfer and frequency conversion do not introduce any further noise or losses.

\section{Results}
\label{sec:results}
In this section, we study information processing platforms, multiplexed platforms and the combination thereof with the algorithm that we introduced in Section \ref{sec:applying}.
In order to compare different simulation results, we have chosen four sets of parameters for both platforms. With these sets, we first investigate the performance of information processing platforms for short ($\approx 15$-$50$ km), intermediate ($50$-$200$ km) and large (i.e.~$\approx 200$-$800$ km) distances. We then perform a similar investigation for multiplexed platforms, after which we investigate the combination of the two. In order to get an understanding of the necessary parameters to generate remote entanglement with each platform or combination, the four sets of parameters for each platform are strictly ordered, with set 4 having the best parameters. We begin each three of the investigations with a specification of the input to our algorithm, which consists of the used elementary pair generation, swapping and distillation protocols, experimental parameters and the parameters specific to the algorithm discussed previously.

In order to investigate longer repeater chains, we consider only \emph{symmetric repeater chains} (see Section \ref{sec:heuristics}) in this section unless specified otherwise.

\subsection{Scheme optimisation results for IP platforms}
\label{sec:IPresults}
In the following we discuss the heuristic optimisation results for information processing platforms. Let us first briefly discuss the protocols that we include in the optimisation.

We consider two protocols for elementary pair generation: the single- and double-click protocol, see Section~\ref{sec:ip}. The single-click protocol has an additional parameter $\theta$, which modulates the weight of the zero and one photon component~\cite{rozpkedek2018near}. We optimise over all single-click protocols with $\theta$ taking values between $\frac{1}{2}$ and $\pi$, equally spaced in 300 steps. 

Both for swapping and distillation we consider a single protocol. For swapping we perform a deterministic Bell state measurement on matter qubits while for distillation we implement the DEJMPS protocol. For swapping and distillation, we optimise over all pairs of schemes that satisfy the banded swapping and distillation heuristics, see Section~\ref{sec:heuristics}.

For all of the schemes, $r$ ranges from $r_\textrm{min}$ to $r_\textrm{max}$ in (at most) $r_\textrm{discr} = 200$ steps, where $r_\textrm{min}$ and $r_\textrm{max}$ are chosen such that the success probabilities are at least $p_\textrm{min}$ and $p_\textrm{max}$, respectively.

We set $\varepsilon_{\textrm{swap}} = \varepsilon_{\textrm{distill}} = 0.05$, $\varepsilon_F = 0.01$ and $\varepsilon_p = 0.02$. These parameters were settled on by investigating the trade-off between the accuracy of the algorithm and its runtime, see Section \ref{sec:heuristics_results} of the Appendix for a detailed analysis. We only consider $m = 2$ distillation rounds. Finally, we set $p_\textrm{min} = 0.9$.

We now specify four sets of parameters for information processing platforms. We fix the parameters in Table \ref{tab:Parameters} as a baseline common to all sets. We then choose sets of parameters for the efficiency coherence times, efficiencies and gate fidelities, which can be found in Table~\ref{tab:parameter_sets_IP}.

\begin{table}[h]
	\begin{centering}
		\begin{tabular}{l | c}
			\hline
			$t_{\textrm{prep}}$ (entanglement preparation time)  & 6 $\mu$s~\cite{hensen2015loophole}\\
			$F_{\textrm{prep}}$ (dephasing for state preparation)  & $0.99$~\cite{hensen2015loophole}\\
			DcS (dark count rate) & $10$ Hz~\cite{hensen2015loophole}\\
			$L_{0}$ (attenuation length) & $22$ km~\cite{LaserEncyclopedia}\\
			$n_{\textrm{ri}}$ (refractive index of the fibre)  &  $1.44$~\cite{LaserEncyclopedia} \\
			$\Delta\phi$ (optical phase uncertainty) & \ang{14.3}~\cite{humphreys2018deterministic} \\
			$F_{\textrm{gates, deph}}$ (dephasing for all gates) & 1
		\end{tabular}
	\end{centering}
	\caption{Base parameters used for information processing platforms.}
	\label{tab:Parameters}
\end{table}

\begin{table}[h]
	\begin{centering}
		\begin{tabular}{l | c c c c}
			& Set 1 & Set 2 & Set 3 & Set 4\\
			\hline
			$T_{\textrm{deph}}$ (dephasing with time) & $3$ s & $10$ s & $50$ s & $100$ s\\
			$T_{\textrm{depol}}$ (depolarising with time) & $3$ s& $10$ s& $50$ s & $100$ s\\
			$p_{\textrm{em}}$ (probability of emission) & 0.8 & 0.9 & 0.95 & 0.99\\
			$p_{\textrm{ps}}$ (probability of post-selection) & 0.8 & 0.9 & 0.95 & 0.99 \\
			$F_\textrm{gates}$ (depolarisation of all gates) & 0.98 & 0.99 & 0.995 & 0.999\\
		\end{tabular}
		\caption{Four different sets of example parameters considered for information processing platforms.}
		\label{tab:parameter_sets_IP}
	\end{centering}
\end{table}

\begin{figure}
	\centerfloat
	\includegraphics[clip,  width = 0.5\textwidth, trim = 3.9mm 3.9mm 3.9mm 8mm]{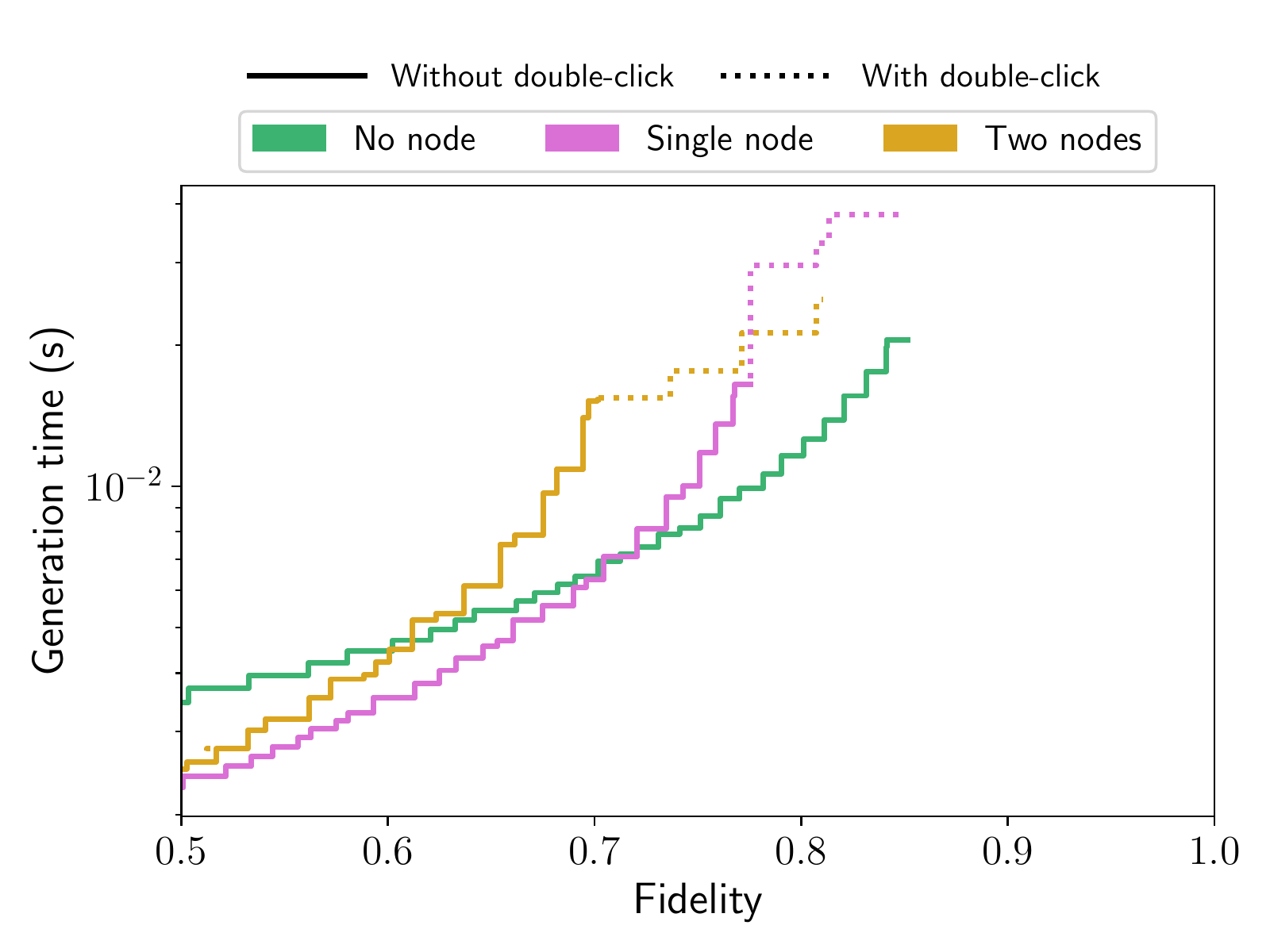}
	\caption{Results of the achieved fidelity and generation time for a total distance of 50 kilometre using parameter set 1 (see Table \ref{tab:parameter_sets_IP}) for information processing nodes, where we consider having 0 (green), 1 (purple), or 2 (yellow) of such intermediate nodes. The solid line corresponds to a heuristic optimisation where we have excluded the double-click protocol, and the dotted line corresponds to a heuristic optimisation with both the single- and double-click protocol. The double-click protocol does not provide a benefit for direct transmission, since the double-click protocol suffers more strongly from losses than the single-click protocol.}
	\label{fig:IPshort_distances}
\end{figure}

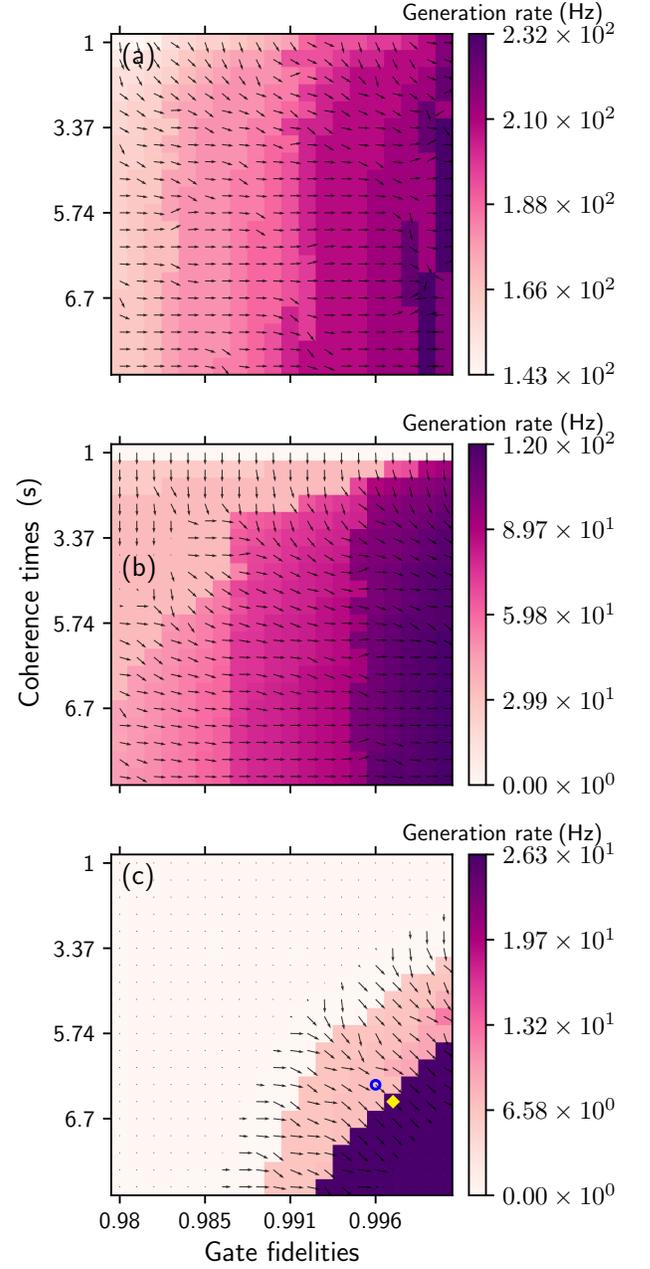
\begin{figure}[h]
	\vspace*{2mm}
	\begin{tikzpicture}[font=\sffamily]
	\centering
	\node[anchor=south west,inner sep=-7mm] (image) at (-1,0) {\includegraphics[clip, trim =  0mm 4mm 0mm 3mm, width=0.5\textwidth]{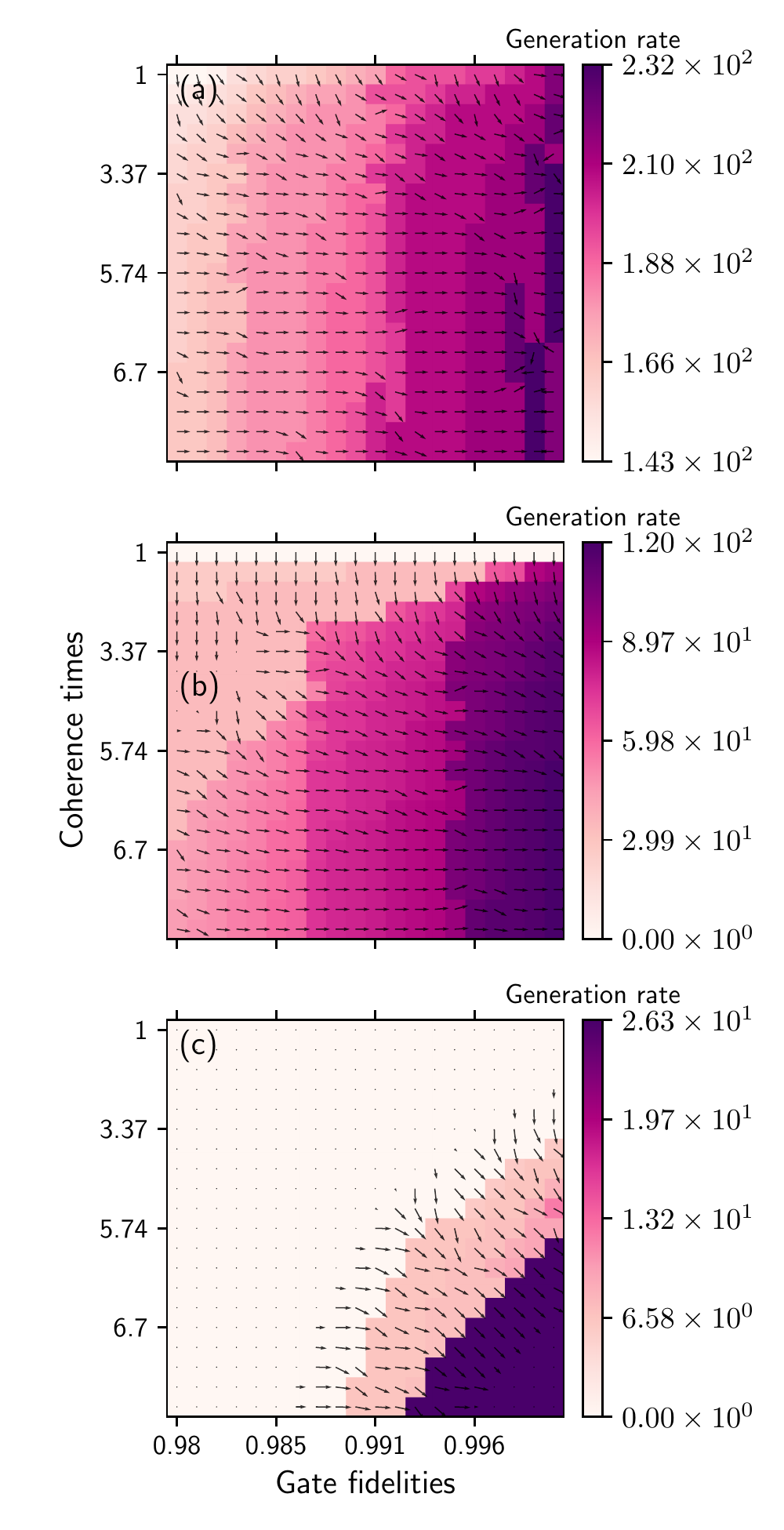}};
	\begin{scope}[x={(image.south east)},y={(image.north west)}]
	\draw (0.5840, 0.1105) node[scale = 1.5, draw=blue, rotate=45, circle, inner sep=0.2ex, line width=0.4mm] {};
	\draw (0.61715, 0.096) node[scale = 0.6,fill, yellow, rotate=45] {};
	\draw (1.031, 0.3255) node[scale=0.95] {(Hz)};
	\draw (1.088, 0.6788) node[scale=0.95] {(Hz)};
	\draw (1.140, 1.0321) node[scale=0.95] {(Hz)};
	\draw (-0.0384, 0.6200) node[scale=1.05, rotate=90] {(s)};
	\end{scope}
	\end{tikzpicture}
	\vspace{4.5mm}
	\caption{Maximum generation rates for several different values of the coherence times (1-10s) and gate fidelities (0.98 to 1) and for several different target fidelities, for a distance of 50 kilometre and a single information processing node. All the other parameters are fixed to those of set 2 (Table \ref{tab:parameter_sets_IP}) or the base parameters (Table \ref{tab:Parameters}). The target fidelities are (a) $F=0.7$, (b) $F=0.8$, (c) $F=0.9$, respectively. We also plot the gradient, indicating the direction and magnitude of steepest ascent. The blue ring and yellow diamond indicate the schemes we investigate in Fig.~\ref{fig:IP_blue_green_scheme_graphs}.}
	\label{fig:IPparameterexploration1}
\end{figure}

\begin{figure}
	\centerfloat
	\begin{subfigure}{0.25\textwidth}
		\begin{tikzpicture}
		\centering
		\node[anchor=south west,inner sep=-7mm] (image) at (-1,0) {\includegraphics[clip,  width = 0.98\textwidth, trim = 0mm 0mm -5mm 0mm]{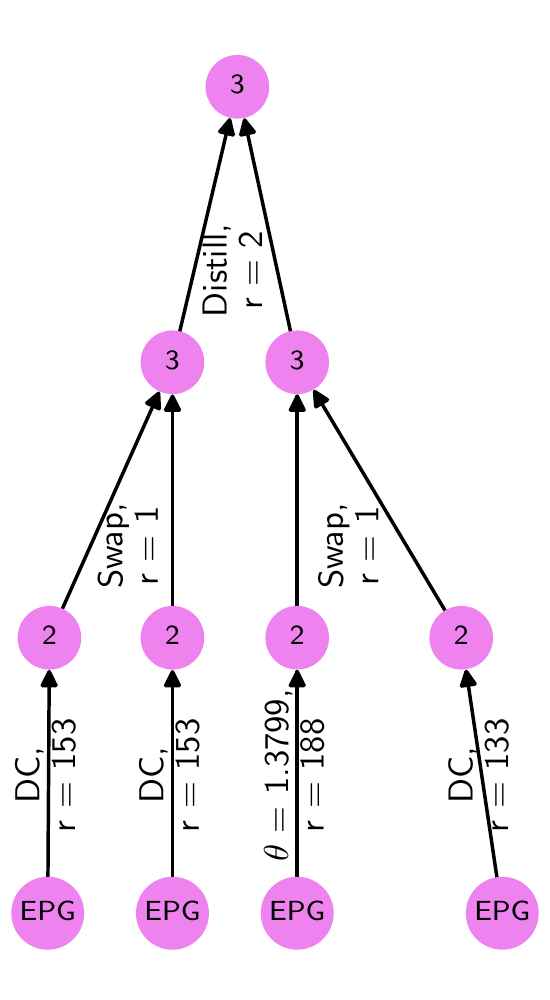}};
		\draw (-1.30735, 4.951) node[scale = 3.21, draw=black, rotate=45, circle, inner sep=0.3ex, line width=1.5mm] {};
		\draw (-1.30735, 4.951) node[scale = 3.2, draw=blue, rotate=45, circle, inner sep=0.3ex, line width=1.5mm] {};
		\end{tikzpicture}
	\end{subfigure}%
	\rulesep
	\begin{subfigure}{0.225\textwidth}
		\begin{tikzpicture}
		\centering
		\node[anchor=south west,inner sep=-7mm] (image) at (-1,0) {\includegraphics[clip,  width = 0.98\textwidth, trim = -1mm 0mm 0mm 3.1mm]{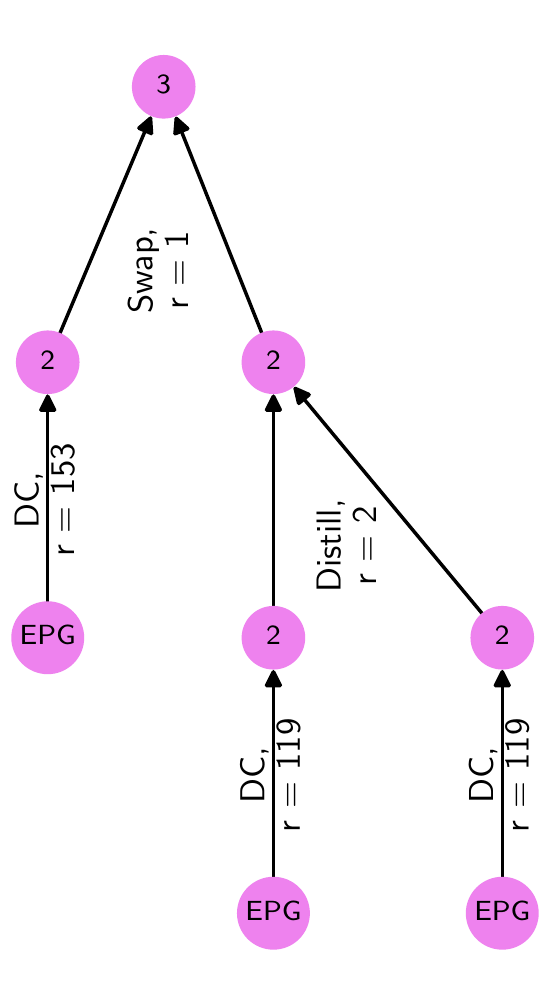}};
		\draw (-1.26735, 5.011) node[scale = 1.51, fill, black, rotate=45] {};
		\draw (-1.26735, 5.011) node[scale = 1.5, fill, yellow, rotate=45] {};
		\end{tikzpicture}
	\end{subfigure}
	\vspace{4mm}
	\caption{Visualisation of the two schemes indicated in the bottom of Fig.~\ref{fig:IPparameterexploration1} by the blue ring (left) and the yellow diamond (right). The numbers in the purple nodes indicate the number of nodes over which entanglement has been established, or elementary pair generation (EPG) has been performed. The `DC' indicates the double-click protocol, and the `$\theta=\theta^*$' indicates a single-click protocol with the $\theta$ parameter set to $\theta^*$. The `$r$' here indicates the number of rounds the corresponding subtree is attempted. Note the necessity of combining disparate schemes - in both cases the EPG protocols used are not the same, and the yellow diamond scheme requires a swap on a distilled and undistilled pair.}
	\label{fig:IP_blue_green_scheme_graphs}
\end{figure}
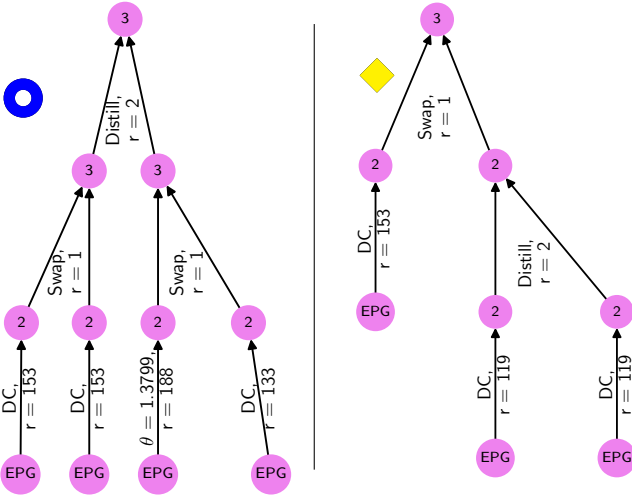

\begin{figure}[h]
	\vspace*{2mm}
	\begin{tikzpicture}[font=\sffamily]
	\centering
	\node[anchor=south west,inner sep=-7mm] (image) at (-1,0) {
		\includegraphics[clip, trim =  7mm 4mm 0mm 3mm, width = 0.5\textwidth]{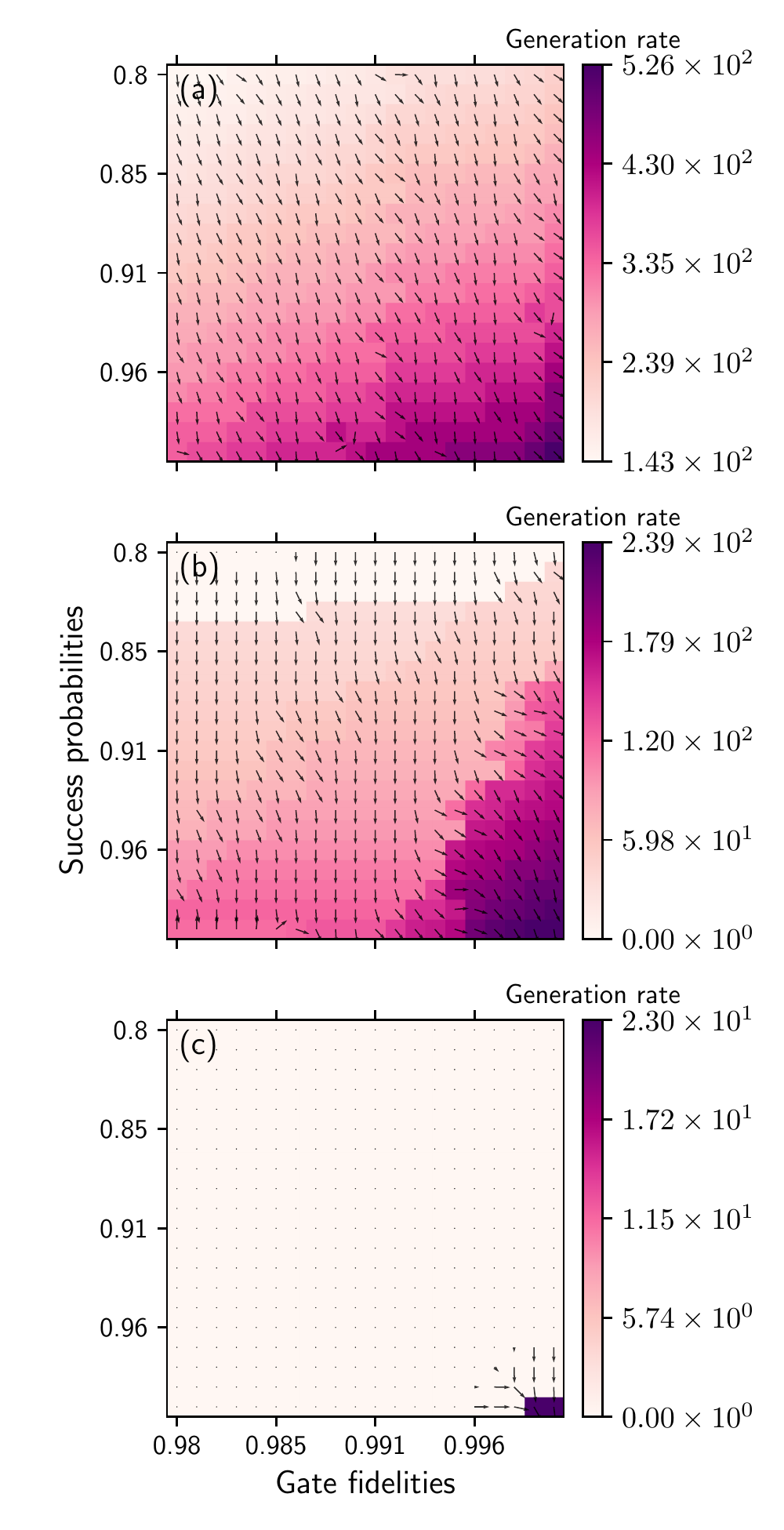}	
	};
	\begin{scope}[x={(image.south east)},y={(image.north west)}]
	\draw (1.0303, 0.3261) node[scale=0.95] {(Hz)};
	\draw (1.084, 0.6788) node[scale=0.95] {(Hz)};
	\draw (1.140, 1.0295) node[scale=0.95] {(Hz)};
	\end{scope}
	\end{tikzpicture}
	\vspace{4.5mm}
	\caption{Maximum achieved generation rates for several different values of the success probabilities (i.e.~we vary $p_\textrm{det}=p_\textrm{em} = p_\textrm{ps}$ simultaneously from 0.8 to 1) and gate fidelities (0.98 to 1), and for several different target fidelities, for a distance of 50 kilometre and a single intermediate node for information processing platforms. All the other parameters are fixed to those of set 2 (Table \ref{tab:parameter_sets_IP}) and the base parameters (Table \ref{tab:Parameters}). The target fidelities are (a) $F=0.7$, (b) $F=0.8$, (c) $F=0.9$, respectively. We also plot the gradient, indicating the direction and magnitude of steepest ascent.}
	\label{fig:IPparameterexploration2}
\end{figure}

\subsubsection{Entanglement generation for short distances with IP platforms}\label{sec:shortIP}
Small-scale experiments relevant for entanglement distribution with information processing platforms have already been performed~\cite{blok2015towards, hensen2015loophole, hensen2016loophole, cramer2016repeated, kalb2017entanglement, reiserer2016robust, humphreys2018deterministic}, demonstrating the potential of such platforms for quantum networks. It is therefore of interest to understand what is within reach for information processing platforms, and what are the relevant parameters to improve. Thus, in this section we investigate how well we can perform entanglement generation with a small number of nodes and near-term parameters over short distances with information processing platforms. In particular, we are interested in when the introduction of a node becomes useful. To this end, we first consider entanglement generation over a distance of 50 kilometres with parameter set 1. We show the results from our heuristic optimisation in Fig.~\ref{fig:IPshort_distances}, where we consider the scenarios with no node, a single node, and two intermediate nodes. Furthermore, we plot the results where we include only the single-click protocol, and both the single- and double-click protocol.

\begin{figure}[h]
	\centering
	\begin{subfigure}[b]{0.5\textwidth}
		\begin{tikzpicture}[font=\sffamily]
		\centering
		\node[anchor=south west,inner sep=-0mm] (image) at (-1,0) {\includegraphics[clip, trim =  3mm 0mm 0mm 1mm, width = 1\textwidth]{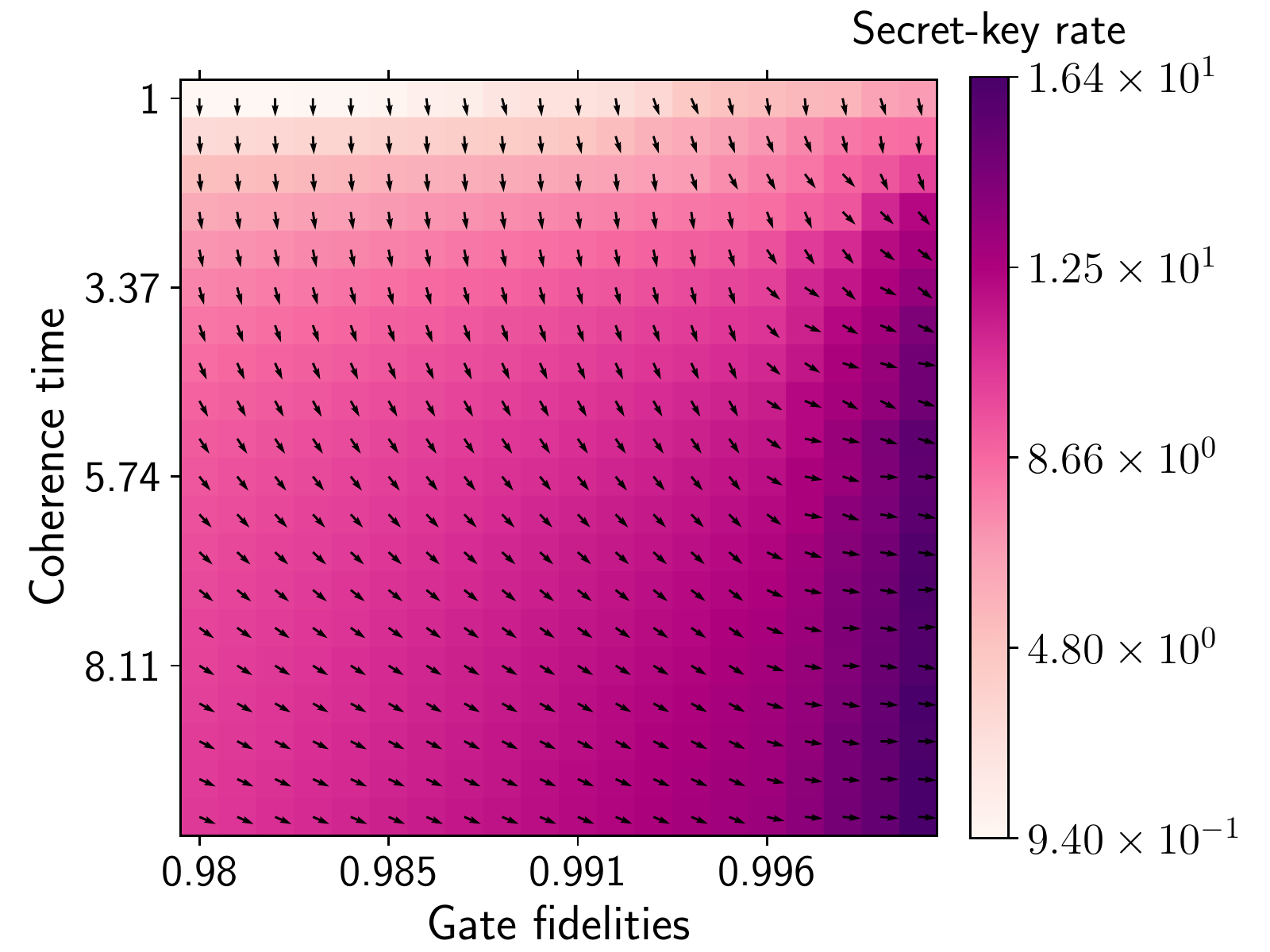}};
		\begin{scope}[x={(image.south east)},y={(image.north west)}]
		\draw (0.1811, 0.875) node[scale=1.2] {{\fontfamily{lmss}\selectfont (a)}};
		\draw (1.0623, 0.9725) node[scale=0.95] {(bits/s)};
		\end{scope}
		\end{tikzpicture}
	\end{subfigure}
	\vspace*{-2.8mm}
	\begin{subfigure}[b]{0.5\textwidth}
		\begin{tikzpicture}[font=\sffamily]
		\centering
		\node[anchor=south west,inner sep=-0mm] (image) at (-1,0) {\includegraphics[clip, trim =  3mm 0mm 0mm 1mm, width = 1\textwidth]{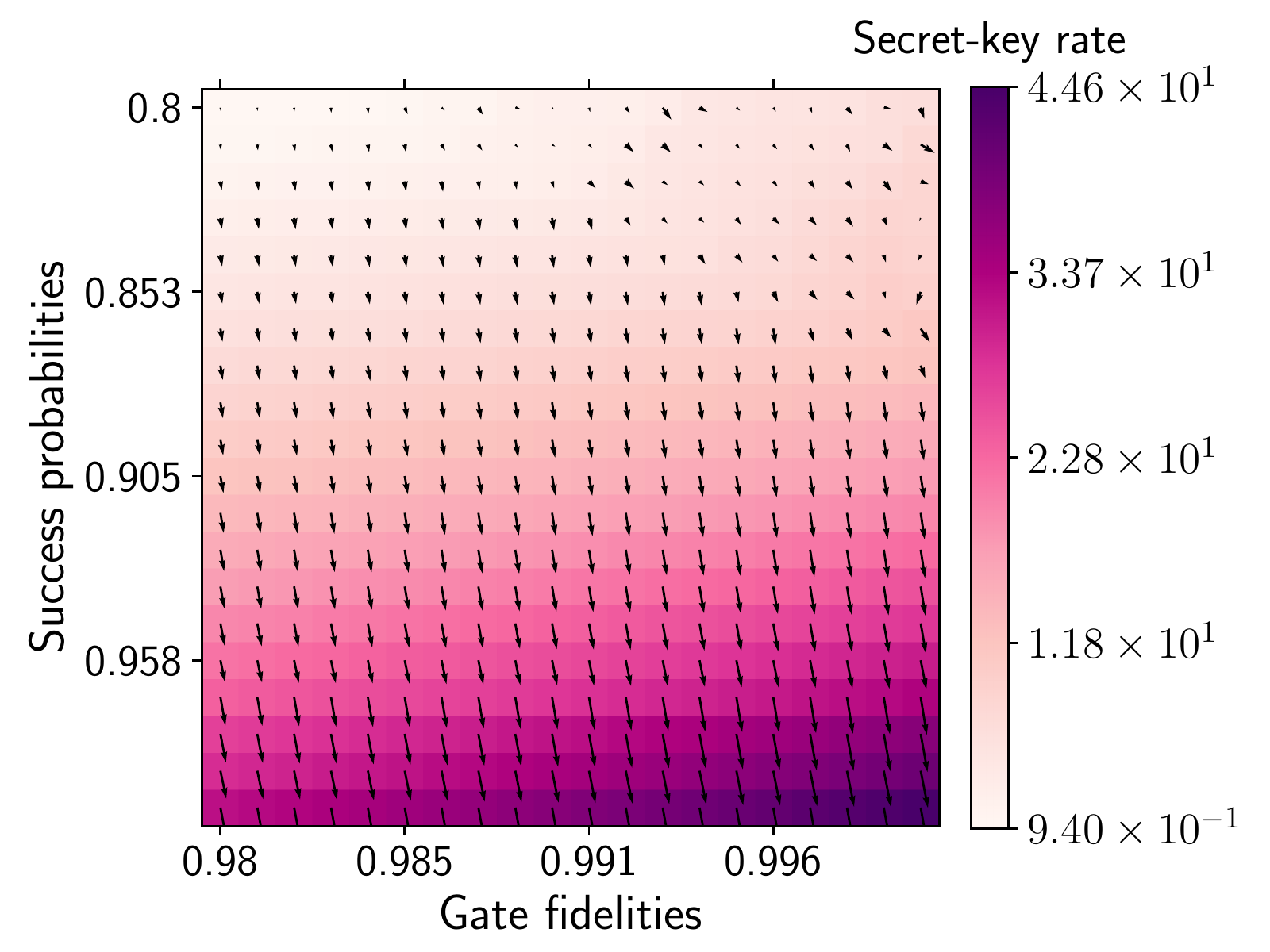}};
		\begin{scope}[x={(image.south east)},y={(image.north west)}]
		\draw (0.1931, 0.855) node[scale=1.2] {{\fontfamily{lmss}\selectfont (b)}};
		\draw (1.0623, 0.9615) node[scale=0.95] {(bits/s)};
		\end{scope}
		\end{tikzpicture}
	\end{subfigure}
	\caption{Secret-key generation using the six-state protocol, for several different values of (a) the coherence times (1-10 s) and gate fidelities (0.98 to 1), and (b) the success probabilities (i.e.~we vary $p_\textrm{det}=p_\textrm{em} = p_\textrm{ps}$ simultaneously from 0.8 to 1) and gate fidelities (0.98 to 1) for a distance of 50 kilometre and a single intermediate node for information processing platforms. All the other parameters are fixed to those of set 2 (Table \ref{tab:parameter_sets_IP}) and the base parameters (Table \ref{tab:Parameters}). We also plot the gradient, indicating the direction and magnitude of steepest ascent.}
	\label{fig:IPparameterexploration3}
\end{figure}

First off, the double-click protocol provides only a benefit for higher fidelities and for the scenarios with one and two intermediate nodes. This can be attributed to the fact that the double-click protocol is inherently less noisy if there are no losses, but is more sensitive to losses than the single-click protocol. However, this does not necessarily imply that all the elementary pairs have been generated with the double-click protocol. As we will see in later results, we will find schemes where elementary pairs are generated using both the single and double-click protocol, indicating the importance of considering such complex schemes in our optimisation.

Secondly, we observe that there is a cross-over point for $F \approx 0.7$ below which adding a node allows for a shorter generation time. Thus, implementing a quantum node over a modest distance of less than 50 kilometres, can in fact increase the generation time by a moderate amount for low fidelities ($\lesssim 0.7$). However, increasing the total distance does not shift this cross-over point, since the maximum achieved fidelity with a single node also drops down if the parameters do not change. 

\begin{figure}
	\centerfloat
	\includegraphics[clip,  width = 0.49\textwidth, trim = 2mm 3mm 4mm 3mm]{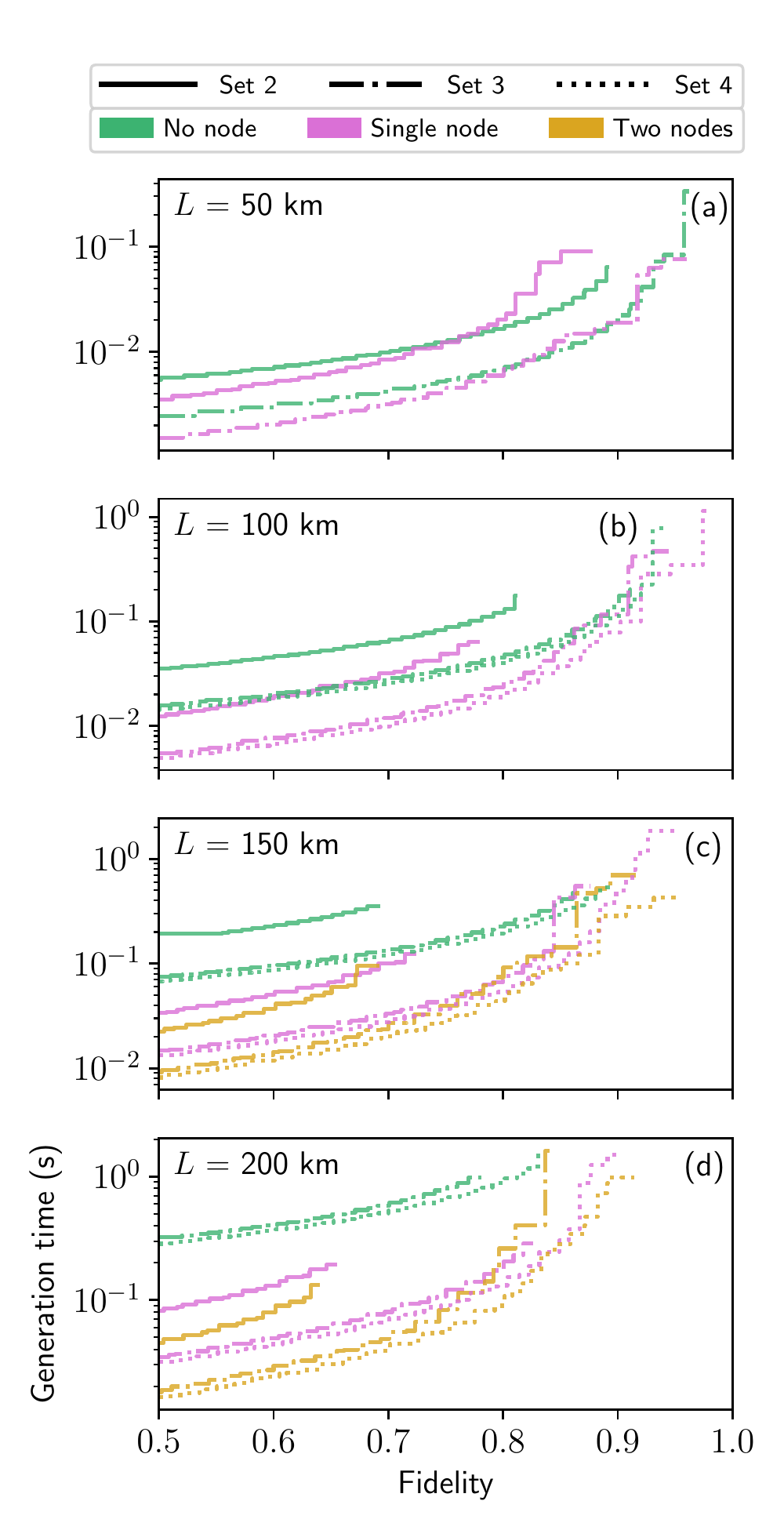}
	\caption{Results of the achieved fidelity and generation time for total distances of 50 (a), 100 (b), 150 (c) and 200 (c) kilometre using parameter sets 2 (solid), 3 (dashed-dotted) and 4 (dashed) (see Table \ref{tab:parameter_sets_IP}) for information processing nodes, where we consider having 0 (green), 1 (purple), or 2 (yellow) of such intermediate nodes.}
	\label{fig:intermediate_distances_IP_nodes}
\end{figure}

Next, we explore the impact of a single parameter in the performance of implementations expected in the longer term. To this end, in Fig.~\ref{fig:IPparameterexploration1} we investigate how the minimum generation time for several fixed target fidelities ($F = 0.7, 0.8, 0.9$) scales, when varying the gate fidelities and coherence times and using parameter set 2. More specifically, we vary the gate fidelities from $0.98$ to $1$ and the coherence times $T_{\textrm{deph}}$ and $T_{\textrm{depol}}$ from 1 to 100 seconds. 
We perform a similar investigation in Fig.~\ref{fig:IPparameterexploration2}, where instead of varying the coherence times, we vary the success probabilities of the detector successfully clicking ($p_\textrm{det}$), successfully emitting a photon from a node ($p_\textrm{em}$), and the probability of emitting a photon of the correct frequency ($p_\textrm{pps}$) simultaneously from 0.8 to 1.

From Fig.~\ref{fig:IPparameterexploration1} we observe that increasing the gate fidelities has a bigger impact on the ability to generate entanglement than increasing the coherence times. In the bottom plot of Fig.~\ref{fig:IPparameterexploration1} we choose two points, indicated by a blue ring and a yellow diamond. The schemes corresponding to those two points are visualised in Fig.~\ref{fig:IP_blue_green_scheme_graphs}.

We make two observations about the algorithm from Fig.~\ref{fig:IP_blue_green_scheme_graphs}. First, the two schemes in Fig.~\ref{fig:IP_blue_green_scheme_graphs} require swaps and distillation on states that have been created in different ways. This shows that already for only a single node entanglement distribution benefits from combining schemes in asymmetric fashion, even if the repeater chain itself is symmetric.
Secondly, the algorithm is sensitive to parameter changes. We see that a small change in the parameters allows the diamond scheme to achieve a generation rate approximately four times as large as the ring scheme. This demonstrates further that the large space of explored schemes can provide a benefit.

The trade-off between the success probability and the gate fidelities in Fig.~\ref{fig:IPparameterexploration2} appears more complex. Not surprisingly, we observe that increasing the success probabilities has the greatest effect on the generation time and the ability to generate entangled states. In contrast to the previous scenario where only varying the gate fidelities leads to jumps in the generation time, we do not observe a similar phenomenon when varying the success probabilities. This is due to the fact that changing the success probabilities changes the generation time primarily by reducing the required number of attempts. Thus, if the minimal number of attempts $r_\textrm{min}$ is well approximated by a continuous function $\hat{r}_\textrm{min}(p)$ in $p$, we expect to see no jumps in the generation time as we vary $p$. More formally, we say that $r_\textrm{min}$ approximates $\hat{r}_\textrm{min}$ well if $\frac{r_\textrm{min}(p)-\hat{r}_\textrm{min}(p)}{r_\textrm{min}(p)} \approx 0$. Since $r_\textrm{min}(p) = \left \lceil \frac{\log(1-p_\textrm{min})}{\log(1-p)}\right \rceil$, an obvious choice for $\hat{r}_\textrm{min}$ is $\frac{\log(1-p_\textrm{min})}{\log(1-p)}$. Note that we then have that $\left|r_\textrm{min}(p)-\hat{r}_\textrm{min}(p)\right| \leq 1$ , and that for $p$ small enough, $\left \lceil \frac{\log(1-p_\textrm{min})}{\log(1-p)}\right \rceil \gg 1$. Since the total success probability of establishing an elementary pair is small, we have indeed that $\frac{r_\textrm{min}(p)-\hat{r}_\textrm{min}(p)}{r_\textrm{min}(p)} \approx 0$, explaining the lack of sudden jumps.
Furthermore, we find from Fig.~\ref{fig:IPparameterexploration2} (c) that, for almost all values of success probabilities and gate fidelities, it is impossible to generate a state with a fidelity of 0.9. 

One of the near-term applications of a quantum repeater chain is the generation of secret-key. This motivates investigating the rate at which secret-key can be generated per unit time for several parameter ranges. Concretely, in Fig.~\ref{fig:IPparameterexploration3} (a) and (b) we investigate the same experimental settings and parameters as in Fig.~\ref{fig:IPparameterexploration1} and Fig.~\ref{fig:IPparameterexploration2}. Each point corresponds to the maximum achieved secret-key per unit time generated using a six-state protocol with advantage distillation~\cite{watanabe2007key} for each of the schemes in the output of our algorithm. 

As in Fig.~\ref{fig:IPparameterexploration1}, we find in Fig.~\ref{fig:IPparameterexploration3}(a) that for both increasing the generation rate or secret-key rate, increasing the coherence times is most beneficial only up to a certain point, after which the gate fidelities become more important. As in Fig.~\ref{fig:IPparameterexploration2}, we observe in Fig.~\ref{fig:IPparameterexploration3}(b) that almost always the success probabilities are more critical than the gate fidelities for increasing the secret-key rate.

\begin{figure}
	\centerfloat
	\begin{subfigure}{0.25\textwidth}
		\hspace*{-16mm}
		\begin{tikzpicture}
		\centering
		\node[anchor=south west,inner sep=-0mm] (image) at (-1,0) {\includegraphics[clip,  scale =0.87, trim = 4.5mm 5.5mm 4.5mm 4mm]{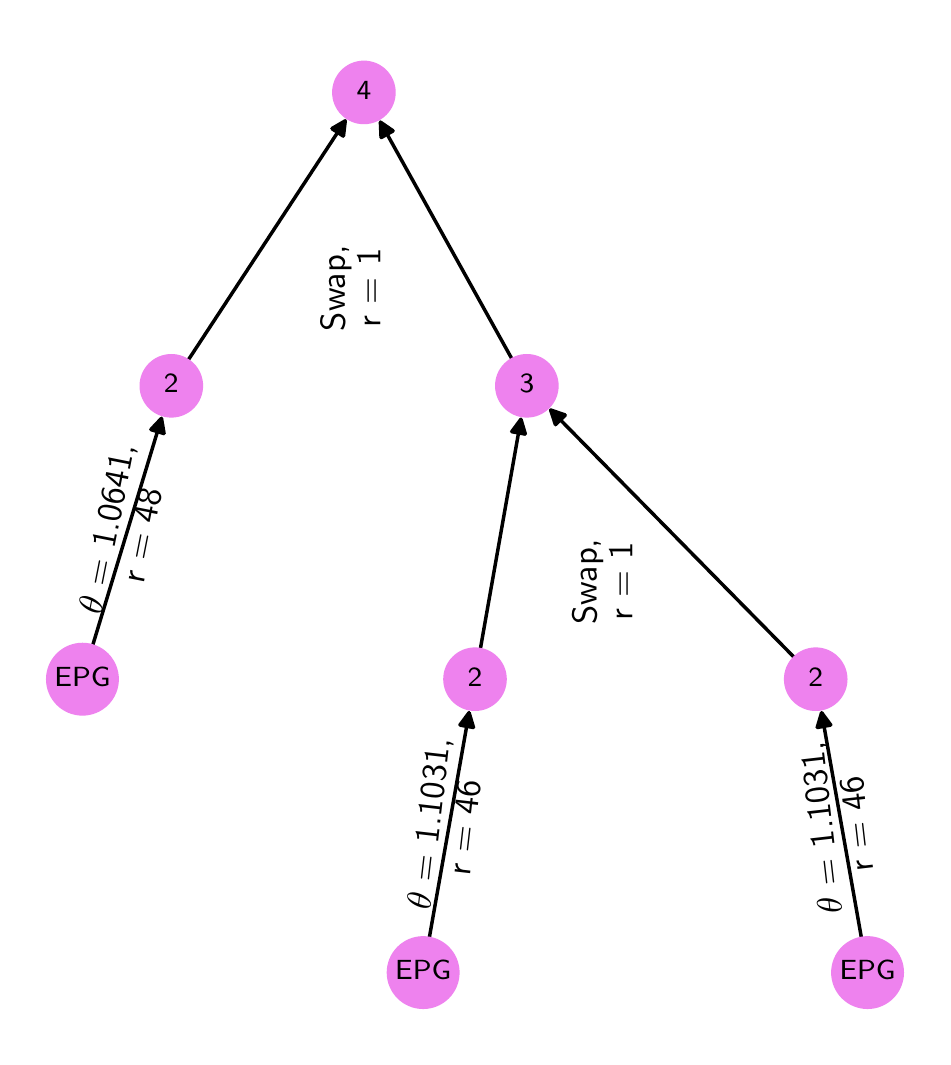}};
		\node[scale = 1.15] at (5.8201, 7.815) {(a)};
		\end{tikzpicture}
	\end{subfigure}
	\vspace{1.1mm}
	\hrule
	\vspace*{1.1mm}
	\hspace*{-11mm}\begin{subfigure}{0.38\textwidth}
		\begin{tikzpicture}
		\centerfloat
		\node[anchor=south west,inner sep=-0mm] (image) at (-1,0) {\includegraphics[clip,  scale = 0.87, trim = 4.6mm 7.0mm 3.5mm 7.7mm]{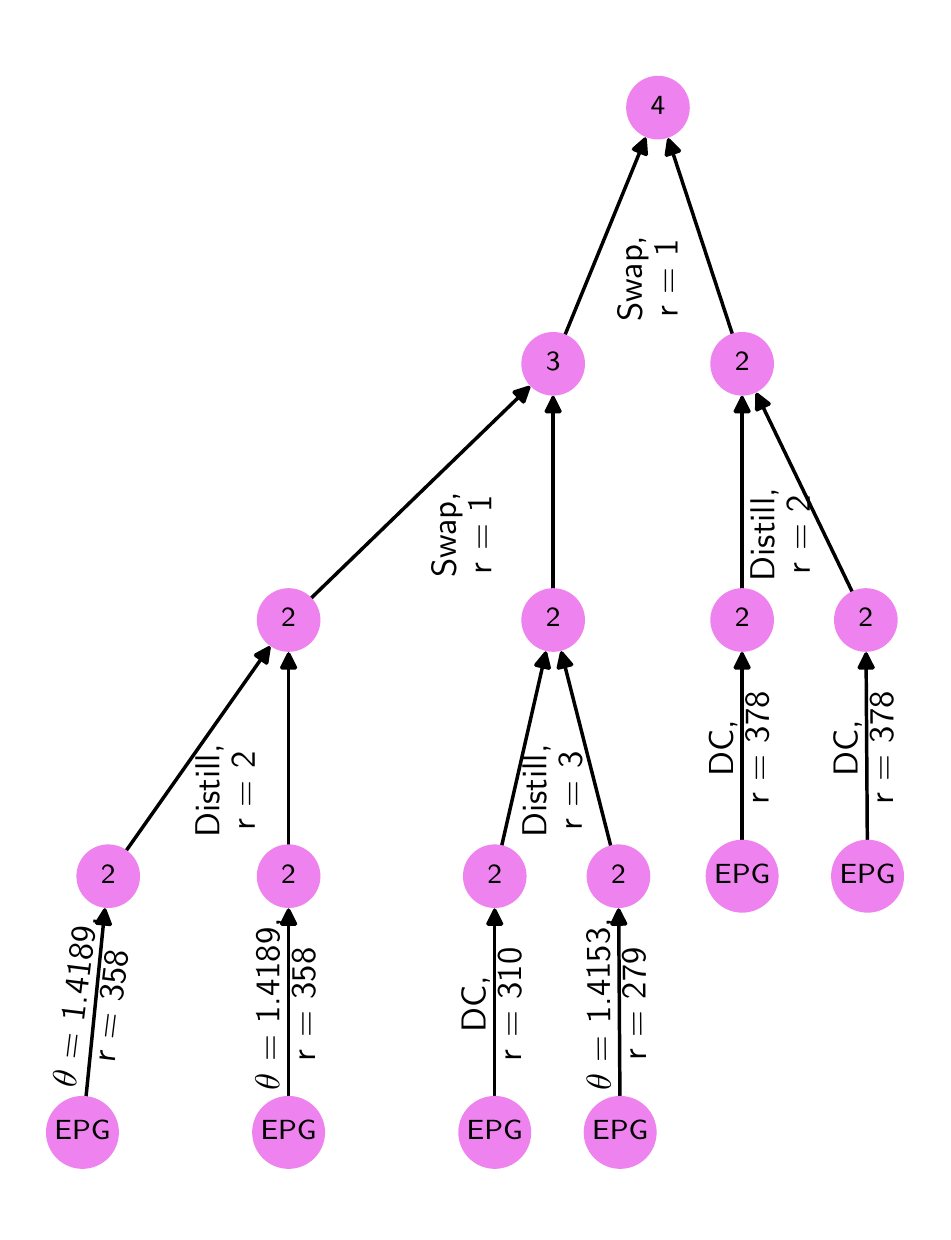}};
		\node[scale = 1.15] at (6.0101, 8.215) {(b)};
		\end{tikzpicture}
	\end{subfigure}
	\vspace*{-3mm}
	\caption{Visual representation of the schemes with the lowest non-trivial fidelity (a) and highest fidelity (b), for a distance of 200 kilometres with information processing platforms using parameter set 4 (see Table \ref{tab:parameter_sets_IP}) and two intermediate nodes. The numbers in the vertices indicate the number of nodes over which entanglement has been established. The `$\theta=\theta^*$' indicates a single-click protocol with the $\theta$ parameter set to $\theta^*$. The `$r$' indicates the number of rounds the corresponding subtree is attempted. We find that the second scheme performs distillation between two elementary pairs generated with a single- and double-click protocol, demonstrating the benefit of including such distillation protocols in our optimisation.}
	\label{fig:IPparameterexploration_schemes}
\end{figure}

\subsubsection{Intermediate-distance entanglement generation using IP platforms}\label{section:intermediateIP}
We expect the addition of nodes to become more beneficial as the distance over which entanglement is generated increases, conditioned on the fact that the experimental parameters are sufficiently high. In this section, we aim to quantify how good the experimental parameters need to be for this to be true.
This motivates us to perform the heuristic optimisation for the entanglement generation for greater distances, and with improved parameter sets. 
More concretely, we investigate the achieved generation times and fidelities for intermediate distances (i.e.~50 to 200 kilometre) for the different experimental parameters proposed in Table \ref{tab:parameter_sets_IP}. We start with Fig.~\ref{fig:intermediate_distances_IP_nodes}(a), where we re-examine the scenario of Fig.~\ref{fig:IPshort_distances} of a total distance of 50 kilometre. We now perform the heuristic optimisation with parameter sets 2 and 3, where we consider implementing either no or a single intermediate node. It is clear from Fig.~\ref{fig:intermediate_distances_IP_nodes}(a) that introducing a node over a distance of 50 kilometre only improves the generation time by a modest amount for low fidelities, even with increased parameters. If we increase the total distance to 100 kilometre, where we now also include parameter set 4, we find in Fig.~\ref{fig:intermediate_distances_IP_nodes}(b) that a single node proves advantageous for almost all fidelities over all three considered parameter sets. In Fig.~\ref{fig:intermediate_distances_IP_nodes}(c) and (d) we consider greater distances of 150 and 200 kilometre, where we also include the heuristic optimisation with two intermediate nodes. We observe that while having no node is clearly inferior to having at least one, introducing two nodes also outperforms a single node for most fidelities and sets of parameters for these distances. This suggests that the values of parameter set 3 (see Table~\ref{tab:parameter_sets_IP}) are a relevant objective to reach for fast near-deterministic entanglement generation with information processing platforms.

We investigate the schemes for the above scenario of 200 kilometres in Fig.~\ref{fig:IPparameterexploration_schemes}, where we depict the schemes that achieve the lowest (non-trivial) fidelity and the highest fidelity. Interestingly, the scheme that achieves the highest fidelity requires that the different elementary pairs are generated both with the double- and single-click protocol. This exemplifies the need for including such asymmetric schemes in our optimisation, which appears to become more important for higher fidelities.

The numerical investigation until this point has been dedicated to symmetric repeater chains. However, realistic quantum networks will be inhomogeneous and nodes will not be equally separated. In Fig.~\ref{fig:bdcz_comp} in Appendix~\ref{sec:additional_results} we show the optimisation results when considering an asymmetric repeater chain over 200 kilometres with three intermediate nodes equally separated. The parameters used are: parameter set 4 for the three intermediate nodes, and parameter set 2 for the nodes corresponding to Alice and Bob (see Table~\ref{tab:parameter_sets_IP}). Such a situation can arise if the end users have access to different technology than the network operator. 
In this setting, we compare the results of a full optimisation with an optimisation over BDCZ schemes, a class of schemes similar to the ones proposed in~\cite{briegel1998quantum,Duer1998}. In particular, we include under the BDCZ class schemes that only combine identical pairs of schemes for connection and distillation. This class is different than the one in \cite{jiang2007optimal} as it allows optimisation over the elementary pair generation protocols but, on the other hand, it does not include distillation schemes based on pumping \cite{jiang2007optimal}. We find that the full optimisation gives an increased generation rate of up to a factor of 10 over BDCZ schemes. 

\subsubsection{Long-distance entanglement generation using IP platforms}
Generating near-deterministic entanglement over larger distances requires excellent experimental control. It is not clear how the number of nodes and the experimental parameters affect our ability to generate entanglement. To this end, we consider here the generation of high fidelity entanglement over distances of 200, 400, 600 and 800 kilometre. To gain an understanding of the relevant parameters, we study the effects of increasing gate fidelities and the memory coherence separately in Fig.~\ref{fig:IPparameterexploration-10-repeaters} in Appendix~\ref{sec:additional_results}.
We observe in Fig.~\ref{fig:IPparameterexploration-10-repeaters} that increasing the coherence times yields a greater benefit than increasing the gate fidelities for these distances and parameters. In particular, increasing the coherence times allows for the generation of entanglement over larger distances, while increasing the gate fidelities effectively extends the ranges of fidelity over which entanglement is generated with the same generation time. We note here that the parameters $p_\textrm{em}$, $p_\textrm{pps}$ and $p_\textrm{det}$ (corresponding to the probability of emitting a photon from the memory, emitting in the correct mode/frequency, and the probability of detecting a photon successfully, respectively) remain fixed, which inhibits the potential benefits of including more nodes.

We have found that information processing platforms, with sufficiently high parameters are a good candidate for near-term entanglement generation. In particular the success probabilities are an important factor for the generation of entanglement. However, even with multiple nodes, the maximum fidelity that can be reached is limited when attempting entanglement generation at large distances.

\FloatBarrier
\subsection{Optimisation results for MP platforms}
\label{sec:MPresults}
Having investigated the performance of information processing platforms with regards to entanglement generation, we now explore entanglement generation with multiplexed platforms. Not only are we interested in how well entanglement can be generated with a repeater chain built using a multiplexed implementation, but also in how the performance differs from information processing platforms. As explained in Section~\ref{sec:introduction}, we expect that multiplexed platforms perform better than information processing platforms for larger distances, provided the experimental parameters are high enough. Our aim for this section is thus to investigate for which parameters and network configurations this becomes true.

First, let us discuss the set of protocols, the algorithm parameters and the hardware parameters we will consider. 

We consider one protocol for elementary pair generation, one for swapping and no protocol for distillation.

The elementary pair generation protocol (see Section~\ref{sec:nonip}) has one free parameter, the mean photon number $N_s$. Similar to information processing platforms, we also optimise over values of the mean-photon number by considering a range of values of $N_s$. In this case, the range is from $2\cdot 10^{-4}$ to $\frac{1}{2}\left(\sqrt{5+\frac{2\sqrt{F_\textrm{threshold}\left(F_\textrm{threshold}+3\right)}}{F_\textrm{threshold}}}-3\right)$, in steps of $10^{-4}$. The lowest value of $2 \cdot 10^{-4}$ was empirically found from the simulations to be a good conservative lower bound, while the upper bound corresponds to achieving a fidelity of the elementary pair with fidelity equal to $F_\textrm{threshold}$ when $\eta \rightarrow0$~\footnote{Obviously, when $\eta \rightarrow 0$ the probability of getting a successful click pattern is zero. However, here we are only interested in the worst-case scenario/upper bound, which corresponds to detecting a successful click pattern as $\eta\rightarrow 0$.}, see Eq.~\ref{eq:nsF}.

The swapping protocol is a photonic Bell state measurement with fixed efficiency depending on the number of ancillary photons, see Table \ref{tab:parameter_sets_MP}.  Similar to the optimisation with information processing platforms, to reduce the parameter space, we implement the banded swapping heuristic, see \ref{sec:heuristics}.

We use the same algorithm parameters as with the information processing platform optimisation. For all of the schemes $r$ ranges from $r_\textrm{min}$ to $r_\textrm{max}$ in (at most) $r_\textrm{discr} = 200$ steps, where $r_\textrm{min}$ and $r_\textrm{max}$ are chosen such that the success probabilities are equal to $p_\textrm{min}$ and $p_\textrm{max}$, respectively. We set $\varepsilon_{\textrm{swap}} = \varepsilon_{\textrm{distill}} = 0.05$, $\varepsilon_F = 0.01$, $\varepsilon_p = 0.02$ and $p_{\textrm{min}} = 0.9$. We consider only symmetric repeater chains, i.e.~all the node have the same parameters and are equidistant.

Regarding the hardware parameters, the base parameters are given in Table~\ref{tab:ParametersMP}, while the four sets of parameters are given in Table~\ref{tab:parameter_sets_MP}.

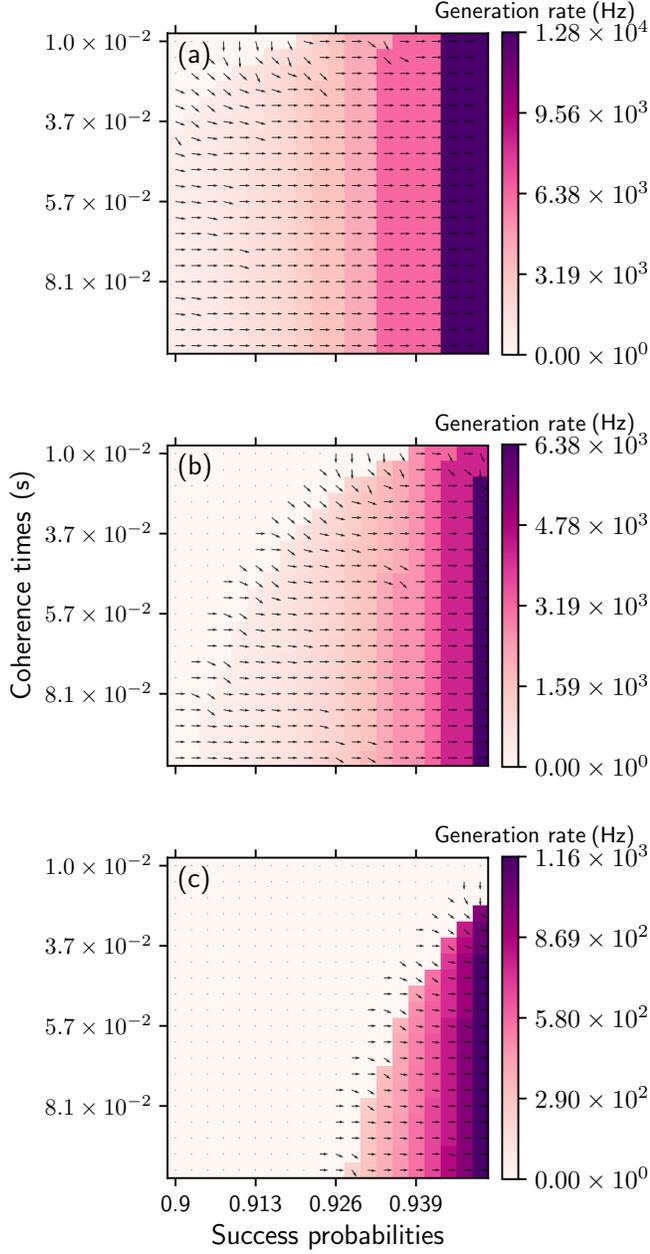
\begin{figure}[h]
	\vspace*{2mm}
	\begin{tikzpicture}[font=\sffamily]
	\centering
	\node[anchor=south west,inner sep=-7mm] (image) at (-1,0) {\includegraphics[clip,  width = 0.48\textwidth, trim = 2mm 3mm 2.5mm 3mm]{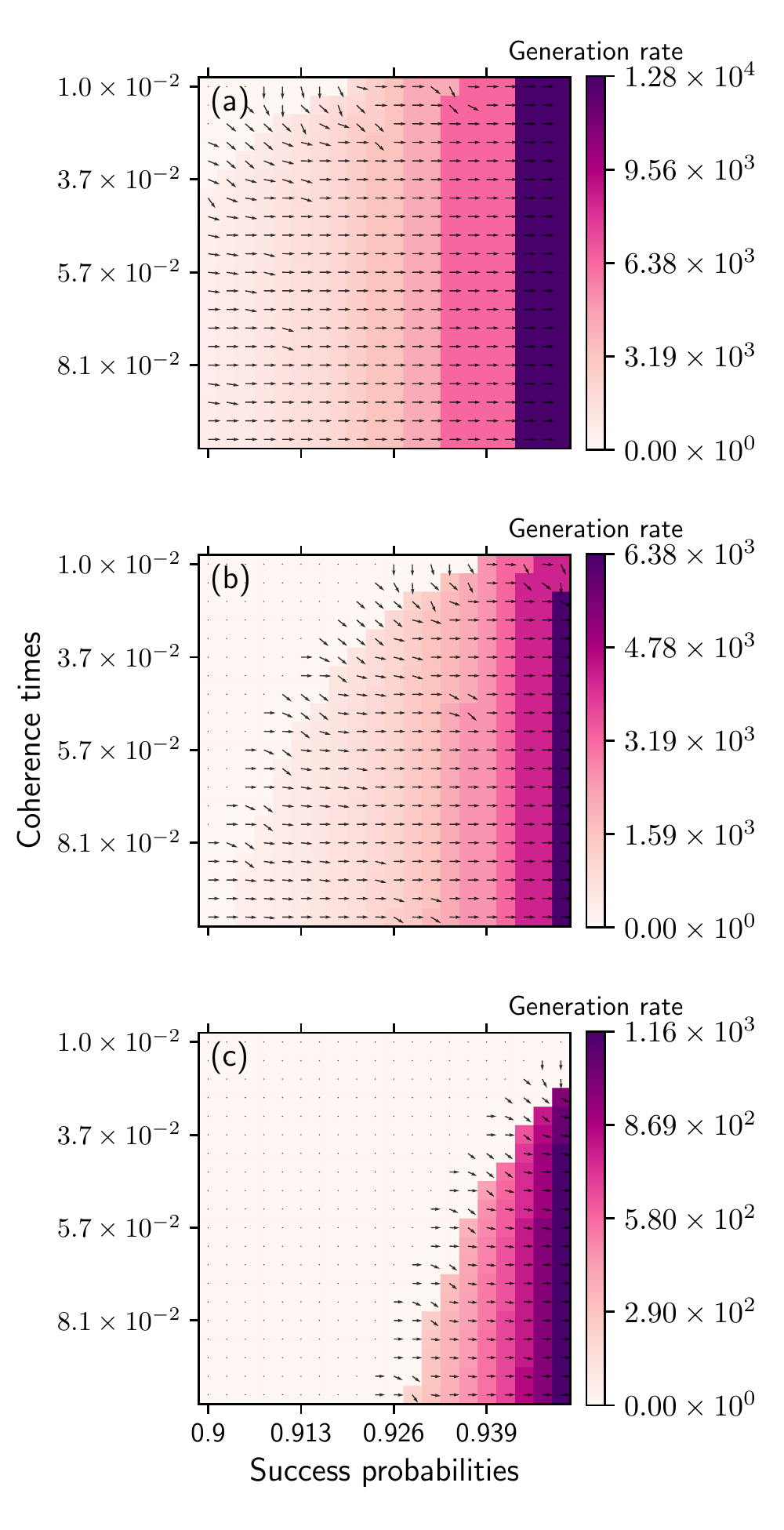}};
	\begin{scope}[x={(image.south east)},y={(image.north west)}]
	\draw (1.071, 0.3208) node[scale=0.935] {(Hz)};
	\draw (1.128, 0.6718) node[scale=0.935] {(Hz)};
	\draw (1.185, 1.0221) node[scale=0.935] {(Hz)};
	\draw (-0.144, 0.6200) node[scale=1.05, rotate=90] {(s)};
	\end{scope}
	\end{tikzpicture}
	\vspace{4.5mm}
	\centerfloat
	\caption{Maximum generation rates for several different values of the success probabilities (i.e.~we vary $p_\textrm{det}=p_\textrm{em} = p_\textrm{ps}$ simultaneously) and efficiency coherence times, and for several different target fidelities, for a distance of 15 kilometre and a single node for multiplexed platforms. All the other parameters are fixed to those of set 2 (Table \ref{tab:parameter_sets_MP}) or the base parameters (Table \ref{tab:ParametersMP}). The target fidelities are (a) $F=0.7$, (b) $F=0.8$, (c) $F=0.9$, respectively. We also plot the gradient, indicating the direction and magnitude of steepest ascent.}
	\label{fig:param15MP1}
\end{figure}

\begin{table}[h]
	\centerfloat
	\begin{tabular}{l | c}
		\hline
		$t_{\textrm{prep}}$ (entanglement preparation time)  & 6 $\mu$s\\
		DcS (dark count rate) & $10$ per second\\
		$L_{0}$ (attenuation length) & $22$ km \\
		$n_{\textrm{ri}}$ (refractive index of the fibre)  &  $1.44$~\cite{LaserEncyclopedia} \
	\end{tabular}
	\caption{Base parameters used for the multiplexed platforms considered in this paper.}
	\label{tab:ParametersMP}
\end{table}

\begin{table}[h]
	\centerfloat
	\begin{tabular}{l | c c c c}
		& Set 1 & Set 2 & Set 3 & Set 4\\
		\hline
		$T_{\textrm{coh}}$ (efficiency coherence times) & $10^{-2}$ s & $10^{-1}$ s & $10^0$ s & $10^1$ s\\
		$N_\textrm{modes}$ (number of modes) & $10^4$ & $10^5$ & $10^6$ & $10^7$\\
		$p$ (success probabilities) & 0.9 & 0.95 & 0.99 & 0.999\\
		$p_{\textrm{BSM}}$ (BSM efficiency) & $\frac{1}{2}$ & $\frac{3}{4}$ & $\frac{7}{8}$ & $\frac{15}{16}$
	\end{tabular}
	\caption{The different sets of parameters considered for multiplexed platforms in this paper.}
	\label{tab:parameter_sets_MP}
\end{table}

\FloatBarrier
\subsubsection{Entanglement generation for short distances with MP platforms}
\begin{figure}
	\centerfloat
	\includegraphics[clip,  width = 0.48\textwidth, trim = 2mm 3mm 2.5mm 3mm]{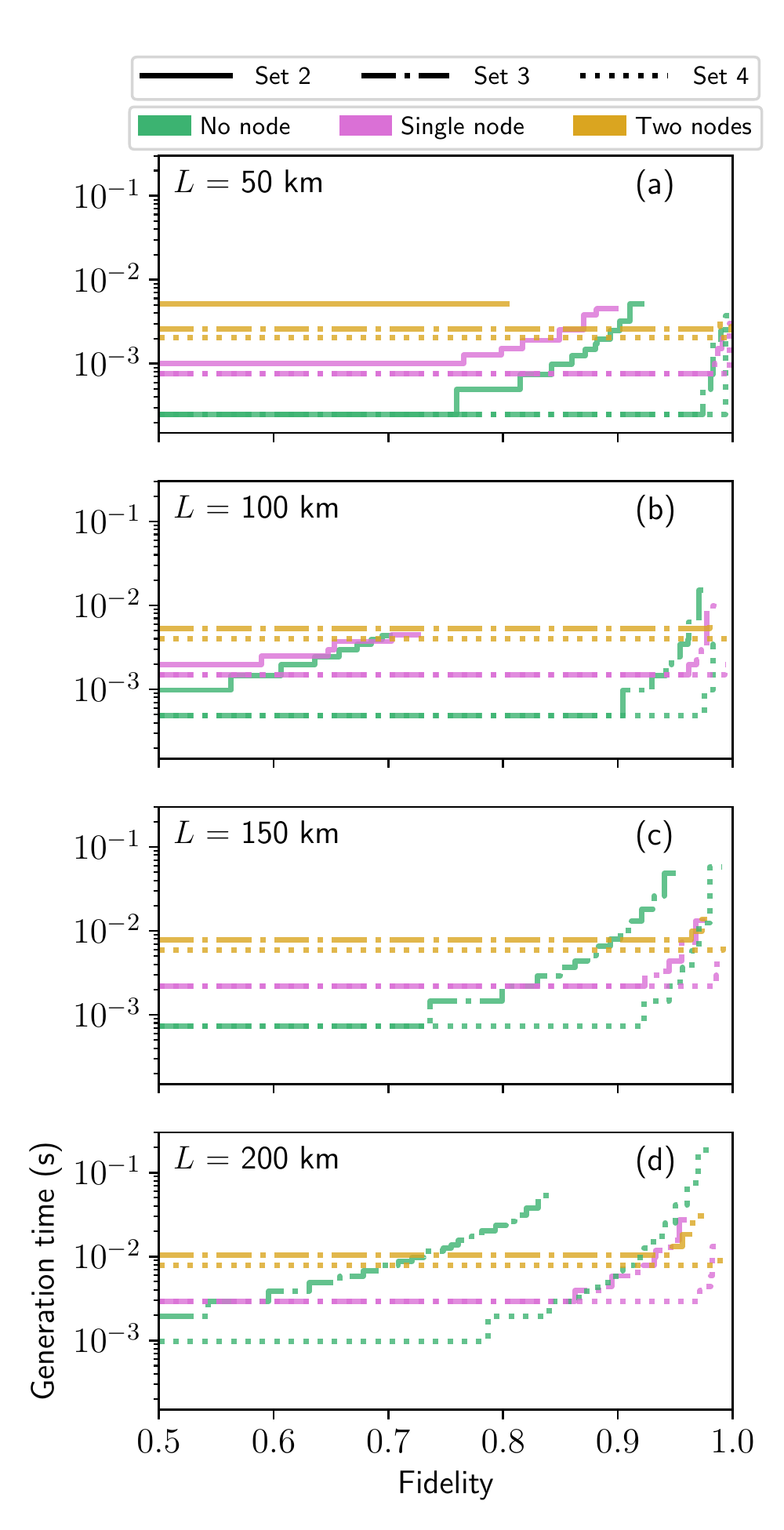}
	\caption{Results of the achieved fidelity and generation time for total distances of 50 (a), 100 (b), 150 (c) and 200 (c) kilometre using parameter sets 2 (solid), 3 (dashed-dotted) and 4 (dashed) (see Table \ref{tab:parameter_sets_MP}) for multiplexed platforms, where we consider having 0 (green), 1 (purple), or 2 (yellow) of such intermediate nodes.}
	\label{fig:intermediate_distances_MP_nodes}
\end{figure}

\begin{figure}[h]
	\vspace*{2mm}
	\begin{tikzpicture}[font=\sffamily]
	\centering
	\node[anchor=south west,inner sep=-7mm] (image) at (-1,0) {\includegraphics[clip,  width = 0.49\textwidth, trim = 2mm 1.8mm 4mm 1mm]{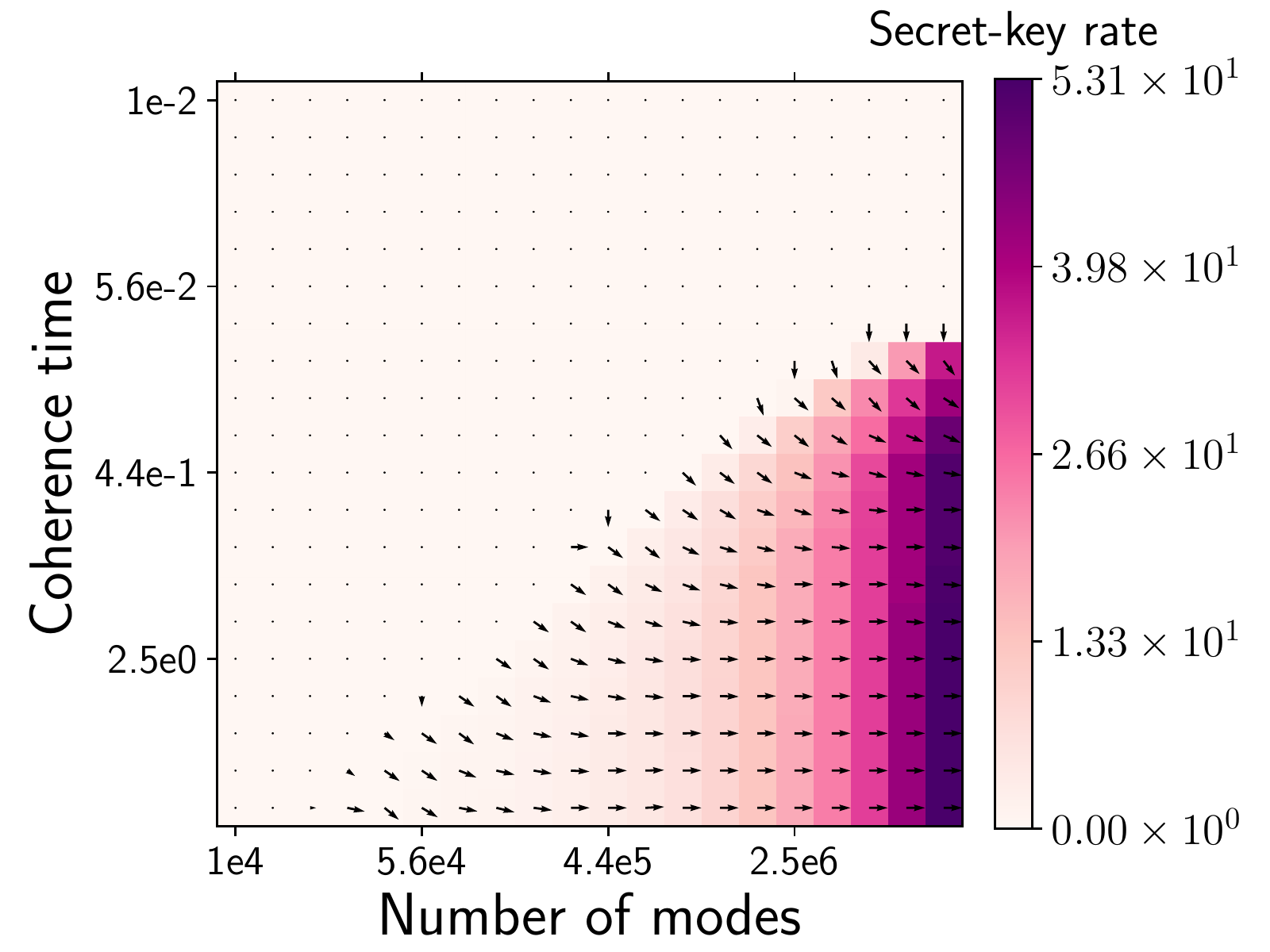}};
	\begin{scope}[x={(image.south east)},y={(image.north west)}]
	\draw (1.240, 1.0931) node[scale=0.935] {(Hz)};
	\draw (-0.094, 0.8190) node[scale=1.05, rotate=90] {(s)};
	\end{scope}
	\end{tikzpicture}
	\vspace{4.5mm}
	\centerfloat
	\caption{Secret-key generation using the six-state protocol, for several different values of the efficiency coherence times ($10^{-2}$-$10^{0}$~s) and number of modes ($10^{4}$-$10^{7}$), for a distance of 200 kilometre and a single node for multiplexed platforms. All the other parameters are fixed to those of set 2 (Table \ref{tab:parameter_sets_MP}) and the base parameters (Table \ref{tab:ParametersMP}). We also plot the gradient, indicating the direction and magnitude of steepest ascent.}
	\label{fig:parameter_explorationMP1}
\end{figure}

We expect that multiplexed platforms provide mostly a benefit over information processing platforms for larger distances. However, it is still of interest to investigate the performance of multiplexed platforms for shorter distances. This is to gain an understanding of what can be done experimentally in the very near-term. Thus, as in Section~\ref{sec:shortIP}, we first explore entanglement generation with multiplexed platforms for short distances. We performed the heuristic optimisation with parameter set 1 for distances of $15$, $25$ and $50$ kilometre, with $0$, $1$ or $2$ intermediate nodes. We found that, except for a distance of 15 kilometres with no nodes, no entanglement could be generated. Even in the scenario of 15 kilometres with no nodes, the maximum fidelity that could be generated was approximately $0.56$. It is thus clear that, at least with the used parameters, information processing platforms are better than multiplexed platforms for entanglement generation over short distances. We now investigate what are the relevant parameters to increase for the multiplexed platforms for entanglement generation over short distances. To this end, we perform a parameter exploration for a distance of 15 kilometres. In particular, we vary the success probabilities and the efficiency coherence times from the values of parameter set 1 to those of set 2 in Table~\ref{tab:parameter_sets_MP}, see Fig.~\ref{fig:param15MP1}.  

We observe that with modest increases in the efficiency coherence times and success probabilities, entanglement generation becomes significantly more efficient. In particular, parameter set 1 (i.e.~top left corner of the parameter plots) is only good enough for the generation of entanglement of very low fidelity ($\sim0.56$), while already a secret-key rate of $\sim500$ bits per second can be achieved for parameter set 2, see Table~\ref{tab:parameter_sets_MP}. We conclude from the plots that, for current and near-term parameters and short distances, increasing the success probabilities is more important than increasing the efficiency coherence times.

\subsubsection{Intermediate-distance entanglement generation using MP platforms}
In the previous section we have found that at short distances multiplexed platforms do not fare as well as information processing platforms. This motivates us to investigate for which parameters and distances this does become the case. We thus investigate here entanglement distribution over distances of 50, 100, 150 and 200 kilometre, where we consider the improved parameters found in sets 2, 3, and 4 in Tables~\ref{tab:ParametersMP} and \ref{tab:parameter_sets_MP} in Fig.~\ref{fig:intermediate_distances_MP_nodes}.

We find that, for most target fidelities in Fig.~\ref{fig:intermediate_distances_MP_nodes}(a), (b) and (c), that the generation time is relatively independent of the desired fidelity. We now explain this behaviour. The fidelity is most strongly controlled by the parameter $N_s$ - lowering $N_s$ allows us to increase the fidelity, but lowers the success probability $p$ of the elementary pair generation. However, the \emph{total} success probability of generating at least one elementary pair $1-(1-p)^{N_{\textrm{modes}}}$ does not decrease significantly, due to the large number of modes $N_{\textrm{modes}}$. In Appendix~\ref{sec:interplay} we investigate how the minimum number of modes changes, as a function of the desired fidelity of the elementary pairs. We find that the required number of modes scales at least as $\frac{\exp(\frac{L}{L_0})}{(1-F)^2}$, where $L$ is the distance between nodes and $L_0$ the attenuation length.

Since multiplexed platforms are expected to have an advantage over information processing platforms for longer distances, we investigate the secret-key rate per unit time for a total distance of 200 kilometre (instead of 50 kilometres for information processing platforms, see Fig.~\ref{fig:IPparameterexploration1} and~\ref{fig:IPparameterexploration3}), where we vary the number of modes and the efficiency coherence time. In Fig.~\ref{fig:parameter_explorationMP1} we find that for most parameters the secret-key rate per unit time is zero. As in the previous parameter explorations performed, we observe that increasing the efficiency coherence times is only (strongly) beneficial up to a certain point (which depends on the number of modes in this case), after which increasing the efficiency coherence times further does not help. Interestingly, increasing the number of modes has the greatest effect on the secret-key per unit time. Increasing the number of modes allows us to push the mean photon number to smaller numbers, effectively increasing the fidelity that can be generated within the same time-window. 

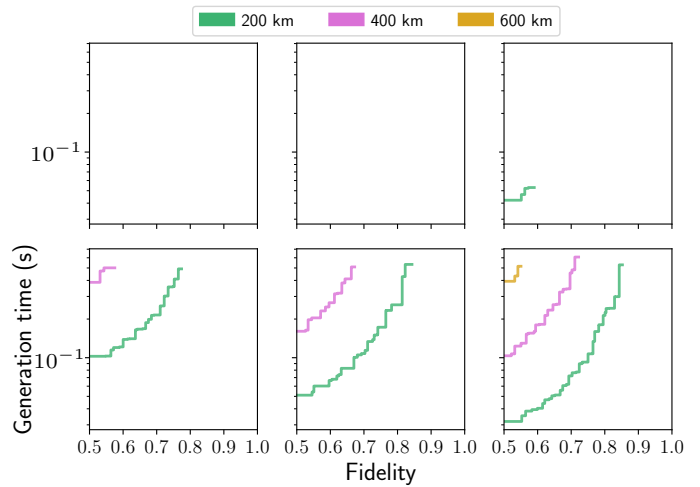
\begin{figure}
	\centerfloat
	\begin{tikzpicture}
	\centering
	\node[anchor=south west,inner sep=-5mm] (image) at (-1,0) {\includegraphics[clip, trim =  3mm 3mm 4mm 1.5mm, width = 0.5\textwidth]{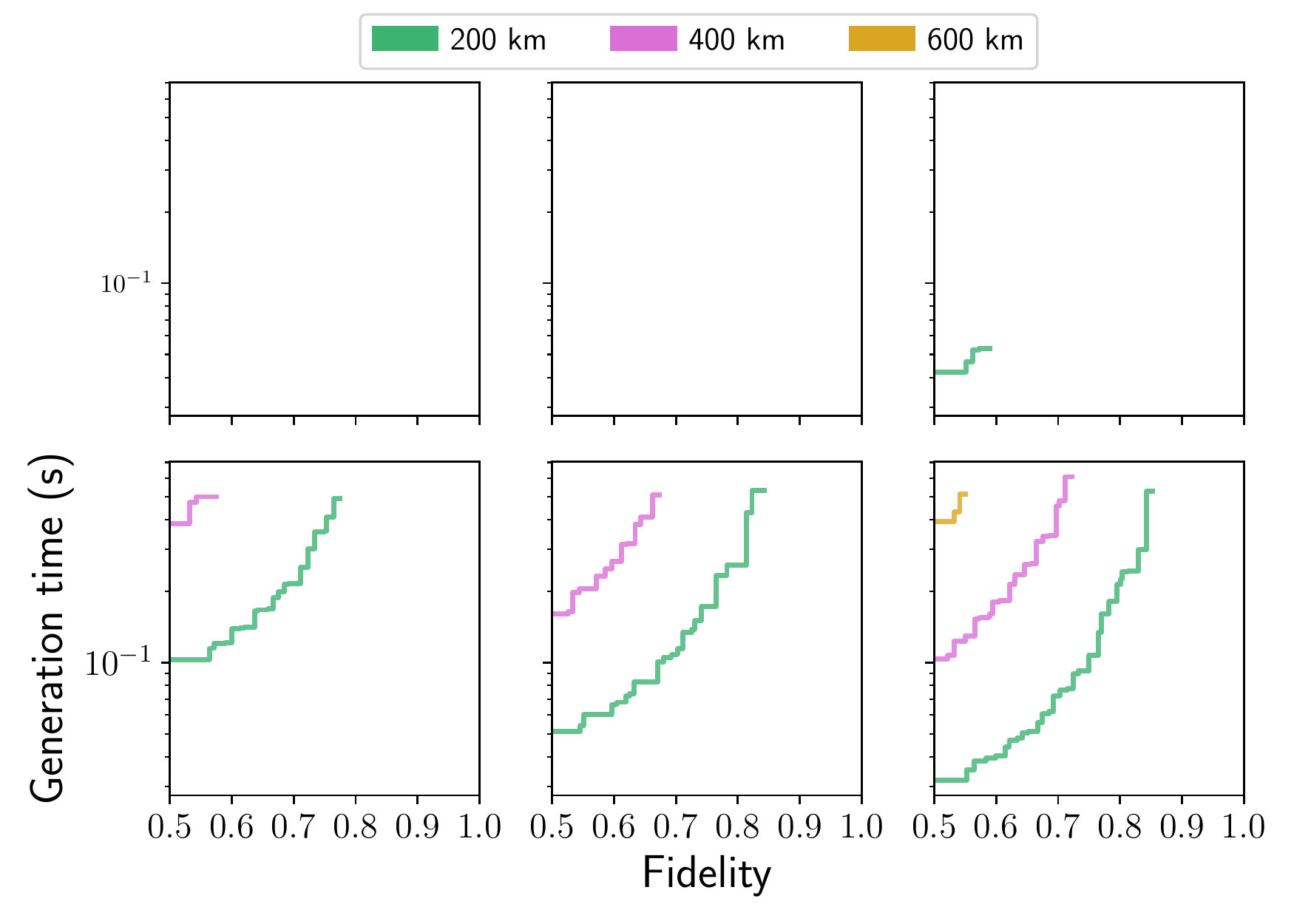}};
	\begin{scope}[x={(image.south east)},y={(image.north west)}]
	\node[fill=white!10, label={[font=\footnotesize, fill=white!10, rotate=0]below:\hspace*{1.8mm}\vspace*{0.9mm}}]
	at (0.001, 0.725) {\hspace*{1.85mm}\vspace*{1.0mm}};
	\node[label={[font=\scriptsize, rotate=0, scale = 1.15]below:$10^{-1}$}]
	at (-0.0091, 0.800) {$$};
	\node[fill=white!10, scale=1.15, label={[font=\scriptsize, scale=1.15, fill=white!10, rotate=0]below:\hspace*{2.55mm}\vspace*{1.25mm}}]
	at (-0.075, 0.815-0.5853) {\hspace*{2.55mm}\vspace*{1.25mm}};
	\node[label={[font=\scriptsize, rotate=0, scale = 1.15]below:$10^{-1}$}]
	at (-0.0751, 0.835-0.54) {$$};
	\end{scope}
	\end{tikzpicture}
	\vspace{5mm}
	\caption{Results of the heuristic optimisation for total distances of 200, 400 and 600 kilometres, for multiplexed platforms and ten intermediate nodes. We use parameter set 2 for multiplexed platforms (see Table~\ref{tab:ParametersMP}) as a baseline, where we set the success probability of the Bell state measurements to $\frac{3}{4}, \frac{7}{8}, \frac{15}{16}$  in the first, second, and third column, respectively. We set the efficiency coherence time $T_{\textrm{coh}}$ to 1 and 10 in the first and second row, respectively.}
	\label{fig:MPparameterexploration-10-repeaters}
\end{figure}

\begin{figure}
	\centerfloat
	\includegraphics[clip,  width = 0.49\textwidth, trim = 2mm 3mm 2.5mm 3mm]{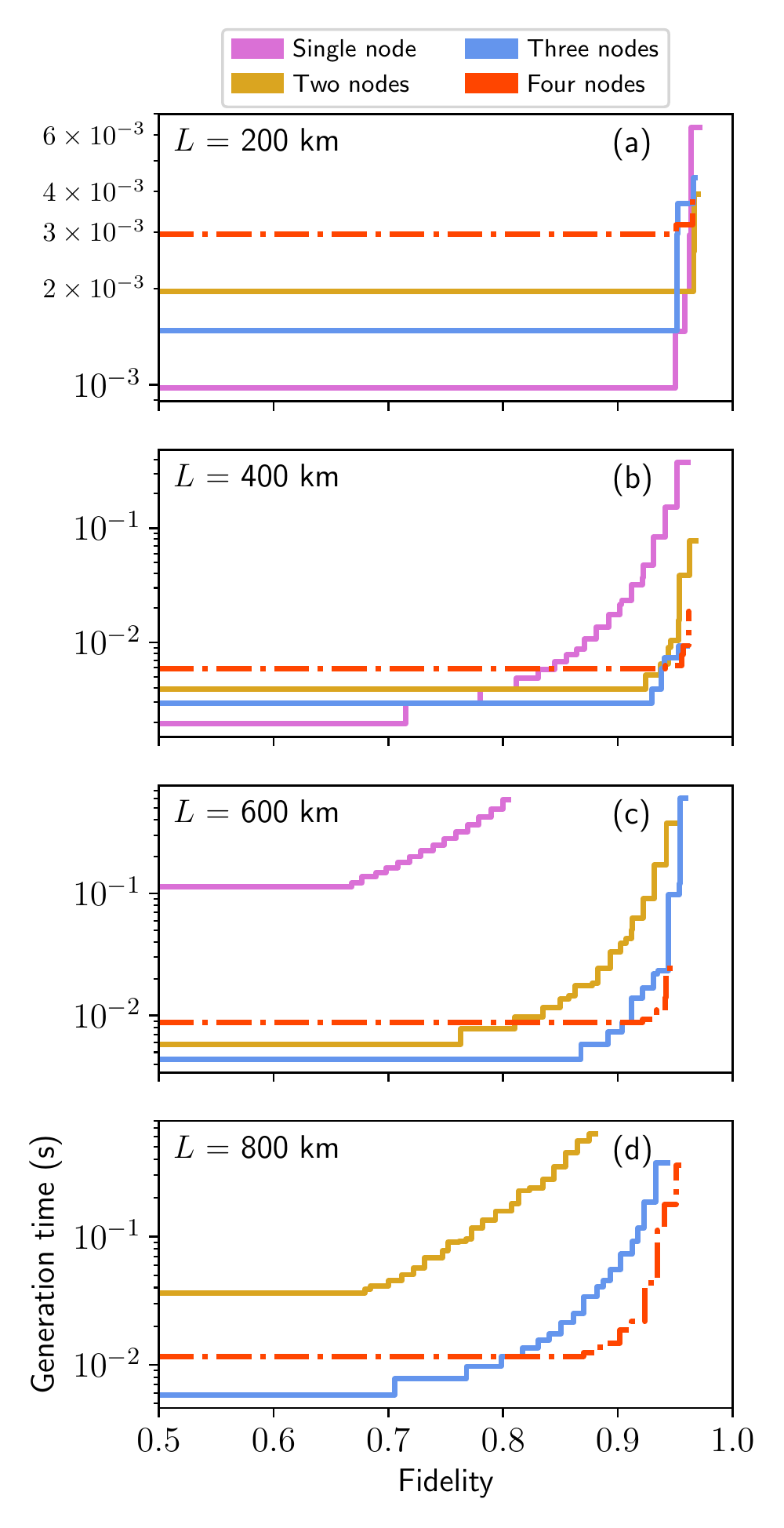}
	\caption{Results of the achieved fidelity and generation time for total distances of 200 (a), 400 (b), 600 (c) and 800 (c) kilometre using parameter set 4 (see Table \ref{tab:parameter_sets_MP}) for multiplexed platforms, where we consider having 1 (purple), 2 (yellow), 3 (blue) or 4 (orange) of such intermediate nodes.}
	\label{fig:long_distances_MP_nodes}
\end{figure}

\begin{figure}
	\centering
	\vspace*{2mm}
	\begin{subfigure}{0.35\textwidth}
		\begin{tikzpicture}
		\centering
		\node[anchor=south west,inner sep=-0mm] (image) at (-1,0) {\includegraphics[clip,  scale = 0.9, trim = 2.8mm 1.5mm 2.5mm 7mm]{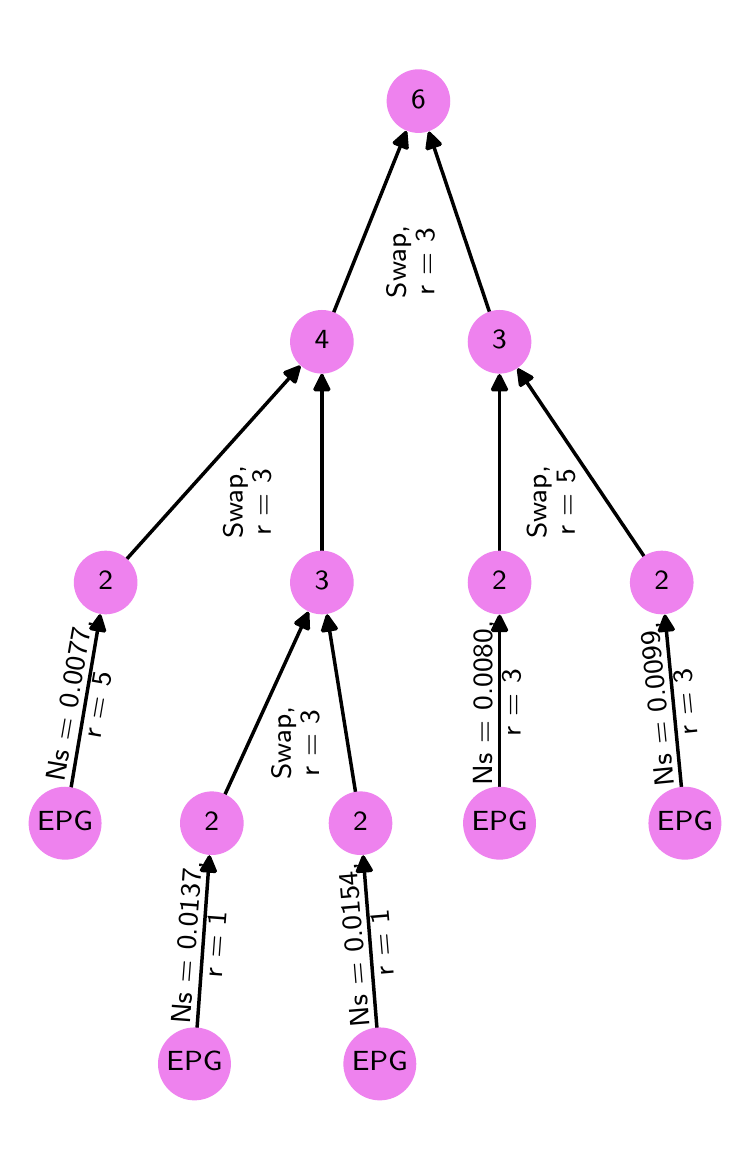}};
		\node[scale = 1.15] at (5.5101, 8.415) {(a)};
		\end{tikzpicture}
	\end{subfigure}
	\vspace{-3.6mm}
	\hrule
	\vspace*{0.5mm}
	\begin{subfigure}{0.35\textwidth}
		\begin{tikzpicture}
		\centering
		\node[anchor=south west,inner sep=-0mm] (image) at (-1,0) {\includegraphics[clip,  scale = 0.9, trim = 2.8mm 6.5mm 2.9mm 6mm]{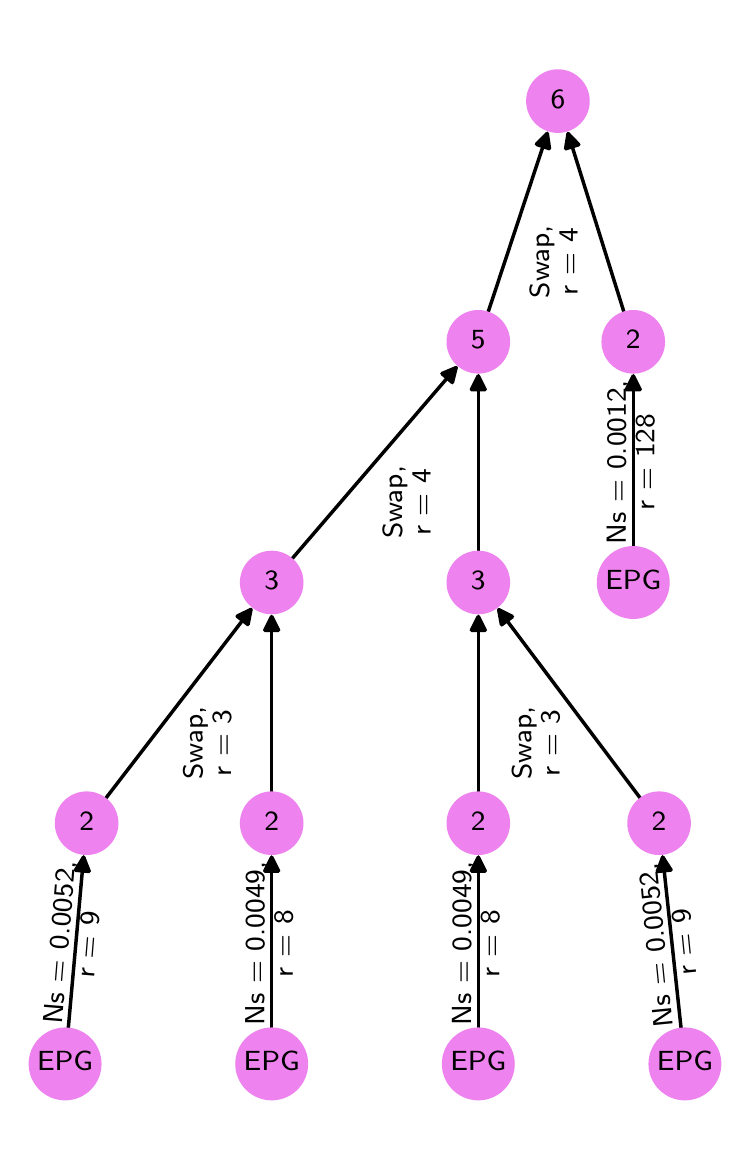}};
		\node[scale = 1.15] at (5.5101, 7.815) {(b)};
		\end{tikzpicture}
	\end{subfigure}
	\caption{Visual representation of the schemes with the lowest non-trivial fidelity (a) and highest fidelity (b) respectively, for a distance of 800 kilometres with multiplexed platforms using parameter set 4 (see Table \ref{tab:parameter_sets_MP}) and four intermediate nodes. The `N$_\textrm{s}$= N$_\textrm{s}^*$' indicates the elementary pair generation (EPG) protocol with mean photon number N$_\textrm{s}^*$ used for multiplexed platforms discussed in the main text. The `$r$' here indicates the number of rounds the corresponding subtree is attempted. Note that the second scheme requires a swap between a link of length five and two at the end.}
	\label{fig:schemerepr321}
\end{figure}

\subsubsection{Long-distance entanglement generation with MP platforms}
We observe by comparing Figs.~\ref{fig:intermediate_distances_IP_nodes} and~\ref{fig:intermediate_distances_MP_nodes} that multiplexed platforms start to outperform information processing platforms for distances of around $\sim$200 km. Here we are interested in whether multiplexed platforms still perform well for even greater distances, which is the relevant scenario for large-scale quantum networks. 

Let us first focus on the effect of the efficiency coherence times and Bell state measurement efficiency on long distance entanglement generation. In Fig.~\ref{fig:MPparameterexploration-10-repeaters} we investigate a repeater chain with 10 nodes with the parameters from set 2, the success probabilities of the Bell state measurements given by $\frac{3}{4}, \frac{7}{8}$ or $\frac{15}{16}$ (corresponding to a number of ancillary photons $2, 6$ and $14$, respectively), and the efficiency coherence time $T_{\textrm{coh}}$ set to 1 or 10. We find that, even with the most optimistic parameters it is not possible to generate entanglement for distances of 800 kilometre with ten nodes. 

This leads to our results shown Fig.~\ref{fig:long_distances_MP_nodes}, where we plot the heuristic optimisation results using parameter set 4, for distances of 200, 400, 600 and 800 kilometre, and the number of nodes running from one to four. We find indeed that, even for a distance of 800 kilometres, entanglement can still be generated at a high fidelity ($\sim0.95$). This, combined with the fact that entanglement generation for the same distance is not possible in Fig.~\ref{fig:MPparameterexploration-10-repeaters}, suggests that it is essential to also increase the number of modes and the success probabilities to generate entanglement over large distances.

We give more detail of two schemes found from the optimisation of Fig.~\ref{fig:long_distances_MP_nodes}. In particular, in Fig.~\ref{fig:schemerepr321} we give the schemes that achieve the lowest non-trivial fidelity and highest fidelity, respectively. As expected, the second scheme uses smaller values of the mean-photon number $N_s$ for the elementary pair generation. This increases the fidelity of the elementary pairs, at the cost of a lower success probability. Indeed, the number of attempts for the elementary pair generation range from $1$ to $5$ and from $8$ to as high as $128$, for the schemes in (a) and (b), respectively.

Here, we also note that there is a non-trivial interplay between the exponential decrease in output efficiency and performing more rounds (i.e.~attempting more times to generate the elementary pairs) to increase the success probability. As we show in Appendix~\ref{sec:effdecoherence}, the requirement that each step succeeds with probability at least $p_\textrm{min}$ can lead to a scenario where under a slight change of the network/parameters, entanglement suddenly cannot be generated anymore.

Interestingly, we observe that the second scheme in~\ref{fig:schemerepr321} requires a swap between links of lengths as five and two at the end. This shows that, as with information processing platforms, exploring more complex asymmetric schemes provides a benefit over more simplistic schemes.

\subsection{Long-distance entanglement generation using a combination of IP and MP platforms}
\label{sec:resultsIPMP}
Here we investigate combining the strengths of information processing platforms with those of multiplexed platforms. For this, we generate the elementary pairs with multiplexed platforms, after which all the operations are performed with information processing platforms. We optimise then over the same protocols as was done for information processing and multiplexed platforms, see Sections~\ref{sec:IPresults} and~\ref{sec:MPresults}.
We expect that, with sufficiently good parameters, the combination of the two outperforms the individual platforms, and that we can distribute entanglement over significantly larger distances.

\begin{figure}
	\centerfloat
	\includegraphics[clip, trim =  3mm 3mm 4mm 2mm, width = 0.5\textwidth]{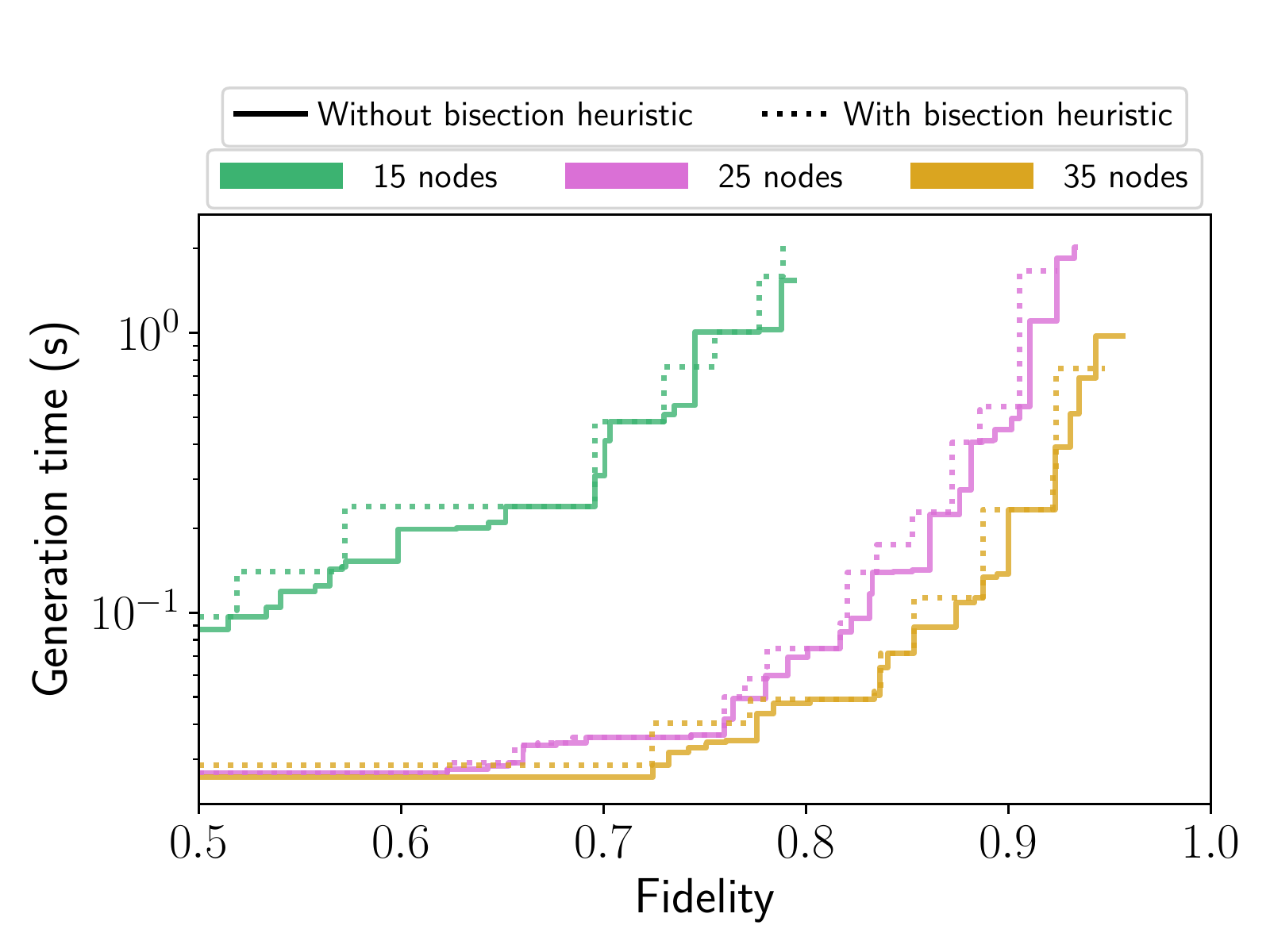}
	\caption{Optimisation results for a total distance of 4000 kilometres, using a combination of multiplexed and information processing platforms. We use parameter sets 4 from both the multiplexed and information processing part of the platform, see Tables \ref{tab:parameter_sets_IP} and \ref{tab:parameter_sets_MP}. The solid lines are the optimisation without the bisection heuristic discussed in~\ref{sec:heuristics}, while the dotted lines are with the bisection heuristic.}
	\label{fig:MPlongdistance}
\end{figure}

Using the parameter set 4 of both platforms, we plot the results for 15, 25 and 35 nodes in Fig.~\ref{fig:MPlongdistance}, for a total distance of 4000 kilometre. Furthermore, we also plot a comparison here when the optimisation includes the bisection heuristic, see Section~\ref{sec:heuristics}.

From Fig.~\ref{fig:MPlongdistance} we observe that, by combining both the strengths from multiplexing and information processing platforms, it is possible to generate entanglement with a high fidelity near-deterministically over large distances by using a large number of nodes. We find that the optimisation results with the bisection heuristic are similar to the results without, while being significantly faster to perform. We find for the cases of 15, 25, and 35 intermediate nodes that the algorithm runtime drops from an order of magnitude of $\sim$100 minutes to $\lesssim$10 minutes. We thus find that the bisection heuristic allows for a faster heuristic optimisation, without the resultant schemes becoming significantly worse than without the bisection heuristic.

\begin{figure}
	\centerfloat
	\begin{tikzpicture}
	\centering
	\node[anchor=south west,inner sep=-7mm] (image) at (-1,0) {\includegraphics[width=0.5\textwidth]{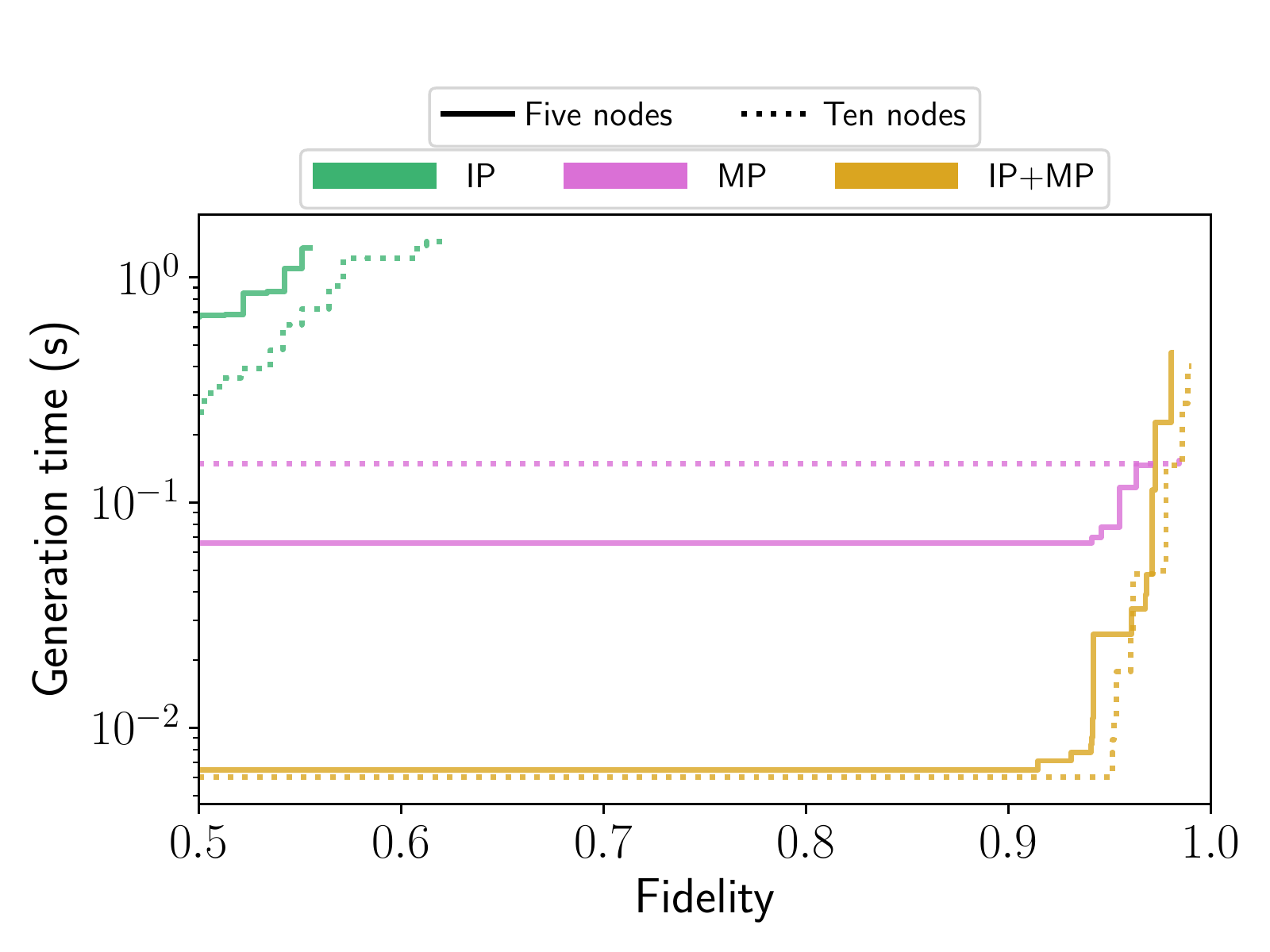}};
	\begin{scope}[x={(image.south east)},y={(image.north west)}]
	\draw (0.9501, 0.099) node[cross=3pt, color=blue] {};
	\draw (0.9508, 0.099) node[cross=3pt, color=blue] {};
	\draw (1.0711, 0.515) node[cross=3pt, color=red] {};
	\draw (1.0703, 0.515) node[cross=3pt, color=red] {};
	\end{scope}
	\end{tikzpicture}
	\vspace{3mm}
	\caption{Results of the heuristic optimisation for a total distance of 800 kilometres, where we compare the three implementations considered in this paper, using five (solid) or ten (dashed) intermediate nodes. We use parameter sets 4 from information processing (IP) platforms, multiplexed (MP) platforms and the combination of the two (IP+MP). The two crosses in the plot indicate the schemes depicted in Figs.~\ref{fig:MP_800_lowestfidelity_scheme_MP_IP} and~\ref{fig:MP_800_almosthighestfidelity_scheme_MP_IP}, respectively.}
	\label{fig:800comp}
\end{figure}
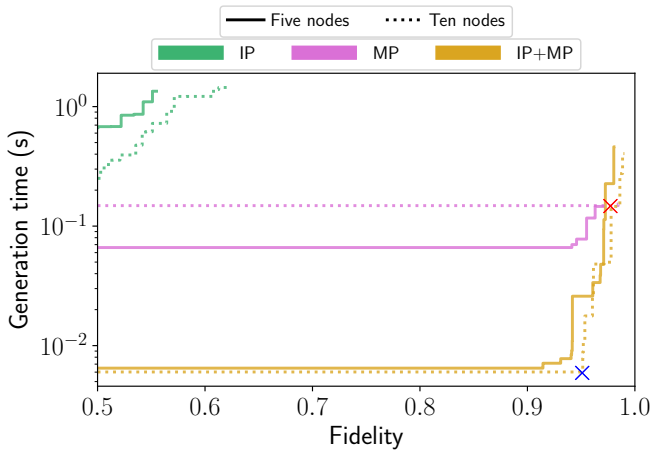

We conclude our results with a plot comparing entanglement generation with the three implementations considered in this paper for a distance of 800 kilometres and five or ten intermediate nodes. We find in Fig.~\ref{fig:800comp} that, for large distances, the combination of information processing and multiplexed platforms outperforms the individual platforms. In fact, it can generate target fidelities below $\sim0.9$ an order of magnitude faster than the multiplexed platform. We see that, as expected, information processing platforms perform significantly worse, where the maximum fidelity is limited to around $\sim0.6$. This is due to the effects of losses during elementary pair generation becoming too strong. This can of course be counteracted by using more nodes, but this results in too much decoherence. This suggests that, for large distances, multiplexed platforms outperform information processing platforms for near-deterministic entanglement generation.

We depict the two schemes corresponding to the two crosses found in Fig.~\ref{fig:800comp} in Figs.~\ref{fig:MP_800_lowestfidelity_scheme_MP_IP} and~\ref{fig:MP_800_almosthighestfidelity_scheme_MP_IP} in Appendix~\ref{sec:additional_results}, respectively. The first of these (blue cross) corresponds to the lowest non-trivial fidelity achieved, while the second one (red cross) corresponds to a state with fidelity of $F=0.9605$, generated in time $T=17.7$ milliseconds. A higher fidelity was not chosen, due to those schemes becoming too big to fit on a page, demonstrating the non-trivial nature of the optimisation performed here.

\section{Conclusions}
\label{sec:conclusions}

The future quantum internet has the potential to change our information society by enabling the implementation of quantum communication tasks. For many of these tasks the key resource is the availability of high fidelity entanglement at the necessary rates. However, given the complex relation between experimental parameters, entanglement distribution protocols and quantum network design, it is unclear what are the necessary parameters to distribute entanglement except for the most basic scenarios. Here, we develop an algorithm to partially answer this question. In particular, our algorithm optimises the near-deterministic distribution of entanglement over chains of quantum repeaters which are abstractly characterised by a small set of relevant parameters.

Even in this abstract setting, the number of possible protocols for a given quantum repeater chain is too large to attempt brute-force optimisation. To make optimisation feasible, we introduce a number of heuristics that render optimisation feasible by dramatically reducing the runtime of the algorithm. Moreover, the heuristics can also be interpreted as approximate rules for protocol design as numerical results show that optimal protocols follow the heuristics. We could expect these heuristics to apply to more dynamic schemes, where the information of the current present entanglement in the network is used to make decisions on the fly by the network.

Any realistic quantum repeater network will be asymmetric in the distances between the nodes and the experimental parameters. We have applied our algorithm to an asymmetric repeater chain, and have found that our optimisation results strongly outperform the results from a simplified optimisation over symmetric/hierarchical schemes, such as those presented in~\cite{briegel1998quantum,Duer1998}.

We have used the algorithm not only for optimising entanglement distribution, but also for parameter exploration. In particular, we have optimised entanglement distribution for several parameter regimes investigating the most relevant parameters for both information processing and multiplexed platforms. For both, we find that success probabilities (e.g.~the emission probabilities, detector efficiencies, etc.) have a strong impact on performance.

In contrast with previous work, our focus on near-deterministic schemes allowed us to make exact statements about the generation time and fidelities of the distributed states. The ability to deliver states with high probability at specific times could be of benefit for routing entanglement in a network.

In conclusion, here we have developed an algorithm that allows to efficiently optimise and explore the parameter space for near-deterministic entanglement distribution over repeater chains. We have investigated a number of representative platforms but the algorithm is not particular to these choices. We make the source code publicly available~\cite{repo} to facilitate the investigation of other implementations, parameters and/or error models.

\section{Acknowledgements}
The authors would like to thank Filip Rozp\k{e}dek, Tim Coopmans, Matthew Skrzypczyk, Joshua Slater, Guus Avis, Mohsen Falamarzi, Dani\"el Bouman and Gl\'aucia Murta Guimar\~aes for helpful discussions for this project. This work was supported by the Dutch Technology Foundation (STW), the Netherlands Organization for Scientific Research (NWO) through a VIDI grant (S.W.), the European Research Council through a Starting Grant (S.W.), the QIA project (funded by European Union's Horizon 2020, Grant Agreement No. 820445) and the Netherlands Organization for Scientific Research (NWO/OCW), as part of the Quantum Software Consortium program (project number 024.003.037/3368).

\FloatBarrier
\phantom{1mm}
  \bibliographystyle{unsrt}
\bibliography{library}
\onecolumngrid

\appendix
\newpage

\section{Complexity of the algorithm}
\label{sec:complexity}
Here we discuss the complexity of the algorithm. For this, first we bound from below the number of schemes that a brute-force approach without heuristics would need to explore. We then incorporate the heuristics and derive an upper bound on the number of schemes of the algorithm as described in Section~\ref{sec:heuristics}. We finalise by deriving an upper bound on the number of schemes in the particular case of `symmetric' repeater chains. That is, chains where each node has the same parameters and adjacent nodes are connected by identical links. 

\subsection{A lower bound on the complexity of the brute-force algorithm}\label{sec:complexlb}

Here we derive two lower bounds on the number of schemes considered by a brute-force algorithm. The two lower bounds are given by $\mathcal{O}\left(\left(r_{\textrm{discr}}\cdot \left|\mathcal{E}\right|\cdot \left|\mathcal{S}\right|\cdot \left|\mathcal{D}\right|\right)^{2^{m\cdot n}}\right)$ and $\mathcal{O}\left(\left(\left(r_\textrm{discr}\right)^2\cdot \left|\mathcal{E}\right|\cdot \left|\mathcal{S}\right|\right)^n\right)$. These bounds correspond to the case with and without distillation protocols considered, respectively. Here, $n$ denotes the number of elementary links in the repeater chain (i.e.~one less than the number of nodes), $m$ denotes the maximum number of distillation rounds,  $r_\textrm{discr}$ the maximum different values of the number of attempts, and $\left|\mathcal{E}\right|$, $\left|\mathcal{S}\right|$ and $\left|\mathcal{D}\right|$ denote the number of elementary pair generation, swapping and distillation protocols, respectively.

To make the analysis tractable, while still obtaining a strict lower bound on the number of schemes, we analyse a simpler algorithm that explores a reduced set of swapping schemes.
At level $i$, instead of exploring all combinations of adjacent links with a combined length $i$, this algorithm only considers swapping between the leftmost link of length $i-1$ and one of length 1, i.e.~the entanglement is propagated by one elementary link at each level.
Furthermore, we will assume the worst case scenario, where all generated schemes have success probability greater than $p_\textrm{min}$, meaning that all of them will be stored.

At the elementary pair level, the algorithm considers $r_\textrm{discr}$ different values of attempts per elementary pair generation protocol. In other words, for each of the $\left|\mathcal{E}\right|$ elementary pair generation protocols that can be performed, there are $r_\textrm{discr}$ different choices of $r$, leading to a total of $\left|\mathcal{E}\right| \cdot r_\textrm{discr}$ schemes. Performing distillation for one round for each of the distillation protocols in $\mathcal{D}$ results then in a total of $r_{\textrm{discr}} \cdot \left|\mathcal{D}\right| \cdot \left(\left|\mathcal{E}\right| \cdot r_\textrm{discr}\right)^{2} + \left|\mathcal{E}\right| \cdot r_\textrm{discr} \geq r_{\textrm{discr}}^3 \cdot \left|\mathcal{D}\right|\cdot \left|\mathcal{E}\right|^2$ schemes for each of the $n$ elementary links. In general, if there are $\zeta$ different schemes over a link, the algorithm will explore (to highest order) $r_\textrm{discr}\cdot \left|\mathcal{D}\right| \cdot \zeta^2$ schemes for one distillation round, such that $m$ repeated rounds of distillation creates at least $\left(r_\textrm{discr} \cdot \left|\mathcal{D}\right|\right)^{2^m-1}\cdot \zeta^{2^m}$ schemes. In other words, we have the following map,

\begin{equation}
	\zeta \rightarrow \left(r_\textrm{discr} \cdot \left|\mathcal{D}\right|\right)^{2^m-1}\cdot \zeta^{2^m}\label{eq:distmap}\ .
\end{equation}

In particular, at the elementary link level, the algorithm considers at least $\zeta_{\textrm{init}} \equiv \left(r_\textrm{discr} \cdot \left|\mathcal{D}\right|\right)^{2^m-1}\cdot \left(\left|\mathcal{E}\right|\cdot r_{\textrm{discr}}\right)^{2^m}$ schemes after $m$ distillation rounds.

Let us now continue to swapping. As mentioned above, for each link of length $i$, the simplified algorithm combines the $\zeta$ schemes of the link of length $i-1$ with the $\zeta_{\textrm{init}}$ schemes stored for an elementary link. We obtain the following map on the number of schemes, 
\begin{align}
	\zeta \rightarrow r_\textrm{discr} \cdot \left|\mathcal{S}\right| \cdot \zeta_{\textrm{init}} \cdot \zeta \label{eq:swapmap}\ .
\end{align}

Thus, starting from a number $\zeta$ of schemes, the composition of swapping with a scheme on an elementary link (equation \eqref{eq:swapmap}) and then distilling (equation \eqref{eq:distmap}) gives us the following map for the lower bound on the number of schemes

\begin{align}
	\zeta &\rightarrow 
	\left(r_\textrm{discr} \cdot \left|\mathcal{D}\right|\right)^{2^m-1}\cdot \left(r_\textrm{discr} \cdot \left|\mathcal{S}\right| \cdot \zeta_{\textrm{init}}\cdot \zeta\right)^{2^m}\nonumber\\
	&= \Omega \cdot \zeta^{2^m} \label{eq:totalmap}\ ,
\end{align}
where we define $\Omega \equiv \left(r_\textrm{discr} \cdot \left|\mathcal{D}\right|\right)^{2^m-1}\cdot \left(r_\textrm{discr} \cdot \left|\mathcal{S}\right| \cdot \zeta_{\textrm{init}}\right)^{2^m}$, which is independent from $\zeta$. Repeating the above map in equation \eqref{eq:totalmap} $n-1$ times on $\zeta_{\textrm{init}}$ (the lower bound of schemes stored over an elementary link) yields the following lower bound,

\begin{align}
	\Omega^{1+2^m+2^{2\cdot m} +  \ldots + 2^{m \cdot \left(n-2\right)}} \cdot \zeta^{2^{m\left(n-1\right)}} = \Omega^{\frac{2^{m\left(n-1\right)}-1}{2^m-1}}\cdot \zeta^{2^{m\left(n-1\right)}}\label{eq:algcomplexalmostfinal}\ .
\end{align}
Expanding \eqref{eq:algcomplexalmostfinal} gives
\begin{align}
	\Omega^{\frac{2^{m\left(n-1\right)}-1}{2^m-1}}\cdot \zeta^{2^{m\left(n-1\right)}}=~& \left(\left(r_\textrm{discr} \cdot \left|\mathcal{D}\right|\right)^{2^m-1}\cdot \left(r_\textrm{discr} \cdot \left|\mathcal{S}\right| \cdot \left(\left(r_\textrm{discr} \cdot \left|\mathcal{D}\right|\right)^{2^m-1}\cdot (\left| \mathcal{E}\right| \cdot r_\textrm{discr})^{2^m}\right)\right)^{2^m}\right)^{\frac{2^{m\left(n-1\right)}-1}{2^m-1}} \nonumber \\
	\cdot~& \left(\left(r_\textrm{discr} \cdot \left|\mathcal{D}\right|\right)^{2^m-1}\cdot \left(r_\textrm{discr}\cdot \left|\mathcal{E}\right|\right)^{2^m}\right)^{2^{m\left(n-1\right)}}\nonumber \\
	=~& \left(r_\textrm{discr}\right)^{\left(2^{m\left(n-2\right)}\right)\cdot \left(4^{m+1}-1\right)} \cdot \left|\mathcal{E}\right|^{\left(2^{m\cdot n+1}\right)\cdot \left(2^m+1\right)}\cdot \left|\mathcal{S}\right|^{2^{m\left(n-1\right)}} \cdot \left|\mathcal{D}\right|^{\left(2^{m\left(n-2\right)}\right)\cdot \left(2^{2m+1}-2^m-1\right)}\nonumber \\
	=~& \mathcal{O}\left(\left(r_{\textrm{discr}}\cdot \left|\mathcal{E}\right|\cdot \left|\mathcal{S}\right|\cdot \left|\mathcal{D}\right|\right)^{2^{m\cdot n}}\right)\label{eq:algcomplexfinal}\ .
\end{align}
We note that this bound becomes trivial when no distillation is performed, i.e.~$m=0$. This is due to the fact that lower order terms were ignored in the number of schemes after distilling. We treat the $m=0$ case separately here. For the case of no distillation, we perform $r_\textrm{discr} \cdot \left|\mathcal{S}\right|$ different swap protocols for $n-1$ times. Since we start with a total number of $r_\textrm{discr}\cdot \left|\mathcal{E}\right|$ schemes on the elementary links, the total number of schemes is then given by

\begin{align}
	\left(r_\textrm{discr} \cdot \left|\mathcal{S}\right|\right)^{n-1}\cdot  \left(r_\textrm{discr}\cdot \left|\mathcal{E}\right| \right)^n =~&\left(r_\textrm{discr}\right)^{2n-1}\left(\left|\mathcal{E}\right|\cdot \left|\mathcal{S}\right|\right)^{n-1}\nonumber\\
	=~&
	\mathcal{O}\left(\left(\left(r_\textrm{discr}\right)^2\cdot \left|\mathcal{E}\right|\cdot \left|\mathcal{S}\right|\right)^n\right)\label{eq:algcomplexfinalnodistill}\ .
\end{align}

We see that with distillation (i.e.~$m\geq 1$) the number of schemes to consider grows super-exponentially in the number of elementary links $n$, while without distillation $m=0$ the number of schemes grows exponentially. It is clear that a brute-force optimisation becomes infeasible for any reasonable number of protocols (i.e.~$\left|\mathcal{E}\right|$, $\left|\mathcal{S}\right|$, $\left|\mathcal{D}\right|$), number of distillation rounds $m$ and elementary links $n$.

\subsection{An upper bound on the complexity of the heuristic algorithm}\label{sec:complexub}
In this section we consider the complexity with the heuristics implemented. The upper bounds we find scales as $\mathcal{O}\left(n^2\log(n)\right)$ for an arbitrary repeater chain, where $n$ is the number of elementary links in the repeater chain. As discussed in the main text, the optimisation can be simplified for the scenario of a repeater chain where every node has exactly the same parameters and the distance between each of the repeaters is equal. For such a symmetric repeater chain, we find a scaling of $\mathcal{O}\left(n\log(n)\right)$.

Let us first briefly discuss the effects the heuristics have on the complexity, before upper bounding the number of schemes. First off, we note here that in the worst-case scenario, all the schemes are incomparable, leading to no pruning. Secondly, the coarse-graining of the fidelity and probability imposes an upper limit on the considered schemes. The coarse-graining fixes the maximum stored schemes to be $\lceil\frac{\left(1-F_{\textrm{threshold}}\right)}{\varepsilon_F}\rceil\lceil\frac{\left(1-p_{\textrm{min}}\right)}{\varepsilon_p}\rceil$ per link. For instance, the number of schemes for elementary pair generation does not change with the heuristic, namely it remains $n\left|\mathcal{E}\right|\cdot r_\textrm{discr}$ in total. However, at most $\lceil\frac{\left(1-F_{\textrm{threshold}}\right)}{\varepsilon_F}\rceil\lceil\frac{\left(1-p_{\textrm{min}}\right)}{\varepsilon_p}\rceil$ of these are stored per elementary link. 

Let us now consider swapping. The algorithm restricts the creation of a link of length $i$ to swapping two links of length $\frac{i}{2} \pm \log(i-1)$, leading to at most $2\lfloor \log(i-1)\rfloor+1$ different options, see equation \ref{eq:nodisplength}. The banded swapping heuristic further reduces swapping to schemes that verify equation \ref{eq:banded_swap} from the main text,
\begin{equation}
	\left|\frac{\log(F_1)}{i_1} - \frac{\log(F_2)}{i_2} \right| \leq \varepsilon_{\textrm{swap}}\nonumber \ .
\end{equation}
This equation becomes in the asymptotic limit
\begin{equation}
	\frac{F_1}{F_2} \leq \exp(\frac{i\cdot \varepsilon_{\textrm{swap}}}{2})\ ,\nonumber 
\end{equation}
since $i_1 \sim i_2 \sim \frac{i}{2}$, and where we have assumed without loss of generality that $F_1\geq F_2$. Now note that $F_\textrm{threshold}\geq \frac{1}{2}$, which implies that $\frac{1}{2} \leq \frac{F_1}{F_2} \leq 2$. We thus have that, in the asymptotic limit, the banded swapping heuristic becomes void if $\varepsilon_{\textrm{swap}}$ is fixed. This means that asymptotically the algorithm considers the full $r_\textrm{discr} \cdot  \frac{\left(1-F_{\textrm{threshold}}\right)^2\left(1-p_{\textrm{min}}\right)^2}{\left(\varepsilon_F\varepsilon_{p}\right)^2}$ schemes for swapping.

The last heuristic is banded distillation. For a fixed distillation protocol, it reduces the number of schemes for performing distillation to at most $2\cdot \lceil\frac{\varepsilon_{\textrm{distill}}}{\varepsilon_F}\rceil \lceil \frac{1-p_{\textrm{min}}}{\varepsilon_{p}}\rceil$. Since there are $\lceil\frac{\left(1-F_{\textrm{threshold}}\right)}{\varepsilon_F}\rceil\lceil\frac{\left(1-p_{\textrm{min}}\right)}{\varepsilon_p}\rceil$ stored schemes, we find the following upper bound on the considered schemes for distillation for a single link, $2\cdot r_\textrm{discr}\cdot \left|\mathcal{D}\right| \cdot m \cdot  \frac{\varepsilon_{\textrm{distill}}\left(1-F_{\textrm{threshold}}\right)\left(1-p_{\textrm{min}}\right)^2}{\left(\varepsilon_F\varepsilon_{p}\right)^2}$. We have removed here and in what follows the ceiling functions, since we are interested in the asymptotic complexity and increasing readability. \\

Combining the previous arguments, we find the following upper bound:
\begin{align}
	&n \left|\mathcal{E}\right| \cdot r_{\textrm{discr}}+ r_\textrm{discr}\sum_{i=2}^{n}\left(n-i+1\right)\nonumber\\
	~&\cdot \left(\left(2\cdot \lfloor\log\left(i-1\right)\rfloor+1\right) \cdot \left|\mathcal{S}\right| \cdot \frac{\left(1-F_{\textrm{threshold}}\right)^2\left(1-p_{\textrm{min}}\right)^2}{\left(\varepsilon_F\varepsilon_{p}\right)^2} + 2\cdot m \cdot  \left|\mathcal{D}\right| \cdot \frac{\varepsilon_{\textrm{distill}}\left(1-F_{\textrm{threshold}}\right)\left(1-p_{\textrm{min}}\right)^2}{\left(\varepsilon_F\varepsilon_{p}\right)^2}\right)\nonumber \\
	= ~& r_{\textrm{discr}}\left(n \left|\mathcal{E}\right| + \frac{\left(1-F_{\textrm{threshold}}\right)\left(1-p_{\textrm{min}}\right)^2}{\left(\varepsilon_F\varepsilon_{p}\right)^2}\sum_{i=2}^{n}\left(n-i+1\right)\cdot  \left|\mathcal{S}\right|\cdot \left(\left(2\cdot \lfloor \log\left(i\right)\rfloor+1\right)\left(1-F_{\textrm{threshold}}\right) + 2\cdot m \cdot \left|\mathcal{D}\right|  \cdot \varepsilon_{\textrm{distill}}\right)\right)\nonumber \\
	\sim ~& r_{\textrm{discr}}\left(n \left|\mathcal{E}\right| + \frac{\left(1-F_{\textrm{threshold}}\right)\left(1-p_{\textrm{min}}\right)^2}{\left(\varepsilon_F\varepsilon_{p}\right)^2}\left(2\left(1-F_{\textrm{threshold}}\right)\cdot  \left|\mathcal{S}\right| \cdot n^2\log\left(n\right)+ 2\cdot m\cdot \left|\mathcal{D}\right| \cdot  \varepsilon_{\textrm{distill}}\cdot \frac{n^2}{2}\right)\right)\nonumber \\
	=~& r_{\textrm{discr}}\left(n \left|\mathcal{E}\right| + n^2\frac{\left(1-F_{\textrm{threshold}}\right)\left(1-p_{\textrm{min}}\right)^2}{\left(\varepsilon_F\varepsilon_{p}\right)^2}\left(2\left(1-F_{\textrm{threshold}}\right) \cdot \left|\mathcal{S}\right| \cdot \log\left(n\right)+ m\cdot \left|\mathcal{D}\right| \cdot \varepsilon_{\textrm{distill}}\right)\right)\label{eq:comp2}
\end{align}
We observe that the algorithm is $\mathcal{O}\left(n^2\log\left(n\right)\right)$, where the pre-factor is given by $2 \cdot \left|\mathcal{S}\right| \cdot r_{\textrm{discr}}\left(\frac{\left(1-F_{\textrm{threshold}}\right)\left(1-p_{\textrm{min}}\right)}{\varepsilon_F\varepsilon_{p}}\right)^2$.\\

In the case of a symmetric repeater chain (i.e.~every node has exactly the same parameters and the distance between each of the repeaters is equal) we can simplify the optimisation by exploiting the symmetry. That is, the optimisation done over a link of length $i$ only needs to be done once, as opposed to $n-i+1$ times in the general setting. Furthermore, there are only $\left \lfloor \log\left(i-1\right) \right \rfloor + 1$ unique ways to perform swapping. The number of schemes is then upper bounded by

\begin{align}
	&\left|\mathcal{E}\right| \cdot r_{\textrm{discr}}\nonumber\\
	+~& r_\textrm{discr}\sum_{i=2}^{n}\left(\lfloor \left(\log\left(i-1\right)\rfloor + 1\right) \cdot \left|\mathcal{S}\right| \cdot \frac{\left(1-F_{\textrm{threshold}}\right)^2\left(1-p_{\textrm{min}}\right)^2}{\left(\varepsilon_F\varepsilon_{p}\right)^2} + 2\cdot m \cdot \left|\mathcal{D}\right| \cdot  \frac{\varepsilon_{\textrm{distill}}\left(1-F_{\textrm{threshold}}\right)\left(1-p_{\textrm{min}}\right)^2}{\left(\varepsilon_F\varepsilon_{p}\right)^2}\right)\nonumber \\
	\sim~& r_{\textrm{discr}}\cdot \left|\mathcal{E}\right| + r_{\textrm{discr}}\cdot \frac{\left(1-F_{\textrm{threshold}}\right)\left(1-p_{\textrm{min}}\right)^2}{\left(\varepsilon_F\varepsilon_{p}\right)^2}\left(\left(1-F_{\textrm{threshold}}\right)\cdot \left|\mathcal{S}\right| \cdot n\log\left(n\right)+ n\cdot m\cdot \left|\mathcal{D}\right| \cdot \varepsilon_{\textrm{distill}}\right)\label{eq:comp3}
\end{align}
We observe that for the symmetric scenario, the algorithm is $\mathcal{O}\left(n\log\left(n\right)\right)$, where the pre-factor is given by $r_{\textrm{discr}}\cdot \left|\mathcal{S}\right| \cdot \left(\frac{\left(1-F_{\textrm{threshold}}\right)\left(1-p_{\textrm{min}}\right)}{\varepsilon_F\varepsilon_{p}}\right)^2$. As mentioned before, the algorithm developed supports both the general and symmetric case.

\section{Analysis of the heuristics}
\label{sec:heuristics_results}
The algorithm detailed in this paper uses four different parameters to reduce the search space of the optimisation, namely $\varepsilon_{F}$, $\varepsilon_{p}$, $\varepsilon_{\textrm{swap}}$, and $\varepsilon_{\textrm{distill}}$, see \ref{sec:heuristics}. The parameters $\varepsilon_F$ and $\varepsilon_p$ are responsible for the coarse-graining in the algorithm, while the parameters $\varepsilon_{\textrm{swap}}$ and $\varepsilon_{\textrm{distill}}$ govern the restrictions on the states used for swapping an distillation, respectively. In this section we investigate the heuristics and how they affect the algorithm runtime and accuracy of the optimisation, which we use to settle on values for $\varepsilon_{F}$, $\varepsilon_{p}$, $\varepsilon_{\textrm{swap}}$, and $\varepsilon_{\textrm{distill}}$. The objective here is to find a good trade-off between the algorithm runtime and the accuracy of the algorithm. We first investigate the coarse-graining - i.e. we vary $\varepsilon_{F}$ and $\varepsilon_{p}$. Afterwards, we investigate the effects of $\varepsilon_{\textrm{swap}}$ and $\varepsilon_{\textrm{distill}}$ on the optimisation results. Finally, we compare the banded swapping heuristic with the naive heuristic, i.e.~where we require the two states to be close in fidelity.

\begin{figure}[h!]
	\centering
	\begin{minipage}{.47\textwidth}
		\centering
		\includegraphics[clip, trim = 4mm 0mm 0mm 0mm, width = 0.95\textwidth]{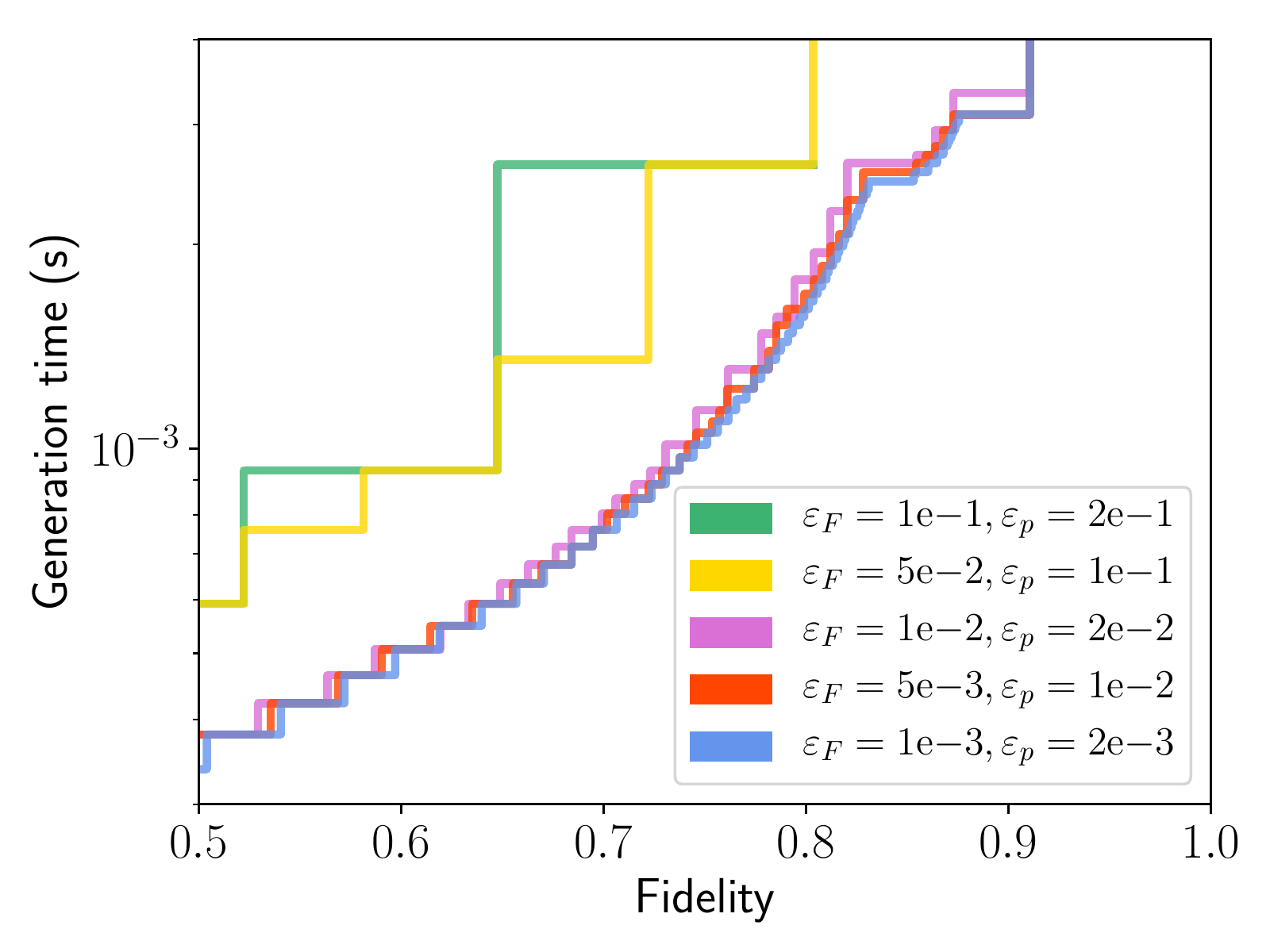}
		\captionof{figure}{Optimised schemes for a distance of 15 kilometres using a single node with the IP parameter set 2 (see Table~\ref{tab:parameter_sets_IP}) for several different pairs of $\varepsilon_F$ and $\varepsilon_p$. Note that as $\varepsilon_F$ and $\varepsilon_p$ approach zero, the curves converge.}
		\label{fig:varepsf}	
	\end{minipage}%
	\qquad
	\begin{minipage}{.47\textwidth}
		\centering
		\includegraphics[clip, trim = 4mm 0mm 0mm 0, width = 0.95\textwidth]{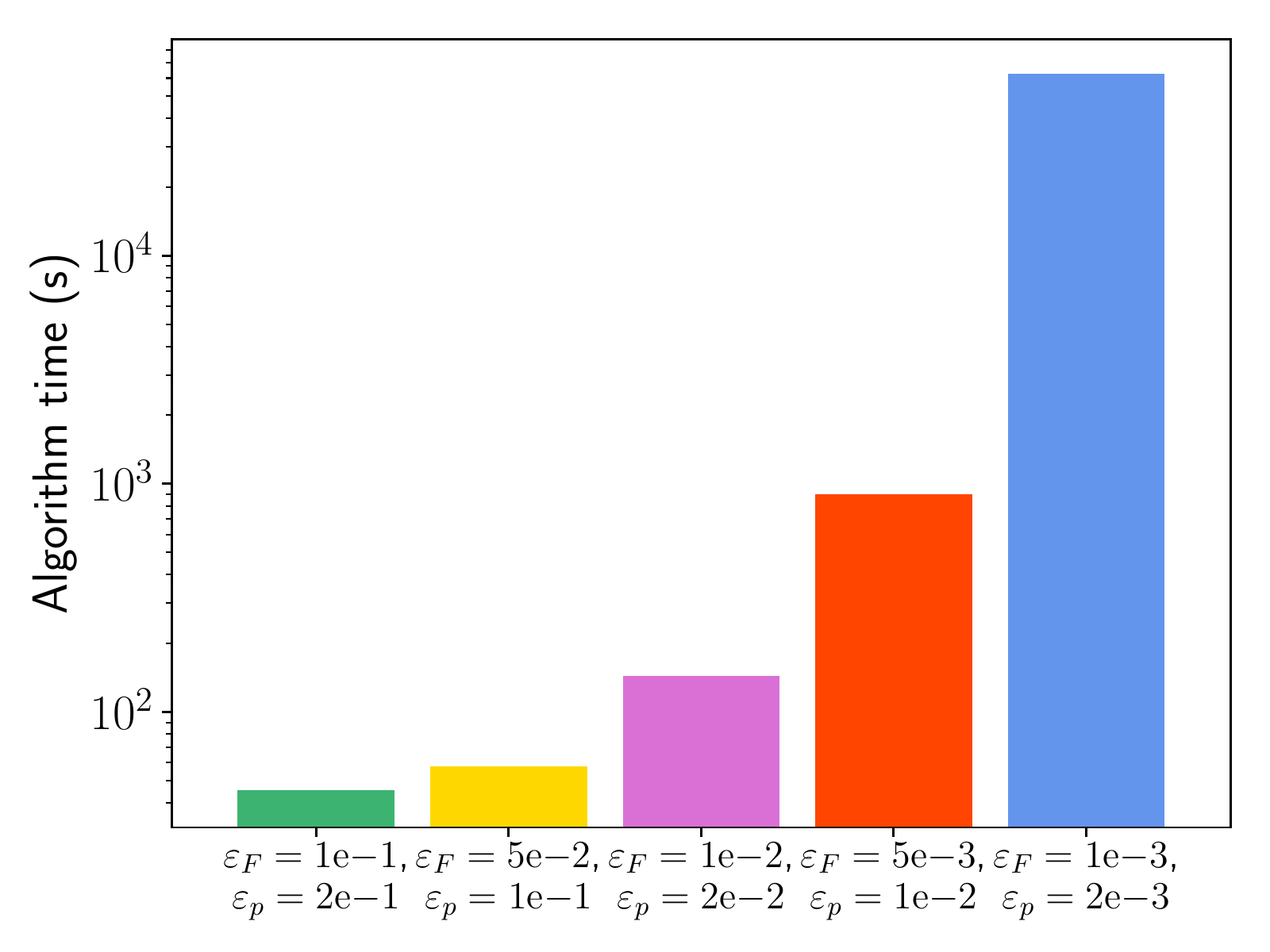}
		\captionof{figure}{The runtime of the algorithm for the optimisations performed in Fig.~\ref{fig:varepsf}. Notice the logarithmic scale, indicating the strong growth rate as $\varepsilon_F$ and $\varepsilon_p$ become smaller.}
		\label{fig:varepsf_time}
	\end{minipage}
\end{figure}

We first vary $\varepsilon_{F}$ and $\varepsilon_{p}$ simultaneously when optimising over schemes for a distance of 6 kilometres and a single repeater with the IP parameter set 2, of which the results can be seen in Fig.~\ref{fig:varepsf} and \ref{fig:varepsf_time}. As expected, there is a trade-off between the accuracy of the algorithm and its running time as $\varepsilon_F$ and $\varepsilon_p$ are varied. While a good trade-off between the accuracy and the runtime depends on each specific case, we use these results to settle in this paper for $\varepsilon_F = 0.01$ and $\varepsilon_p = 0.02$. We settle for these parameters since the important characteristics of the generation time as a function of the fidelity appear to be similar when a more fine-grained optimisation is implemented, without the runtime becoming infeasible.

\begin{figure}
	\centering
	\begin{subfigure}[b]{0.49\textwidth}
		\includegraphics[clip, trim = 4mm 0mm 0mm 0mm, width = 0.98\textwidth]{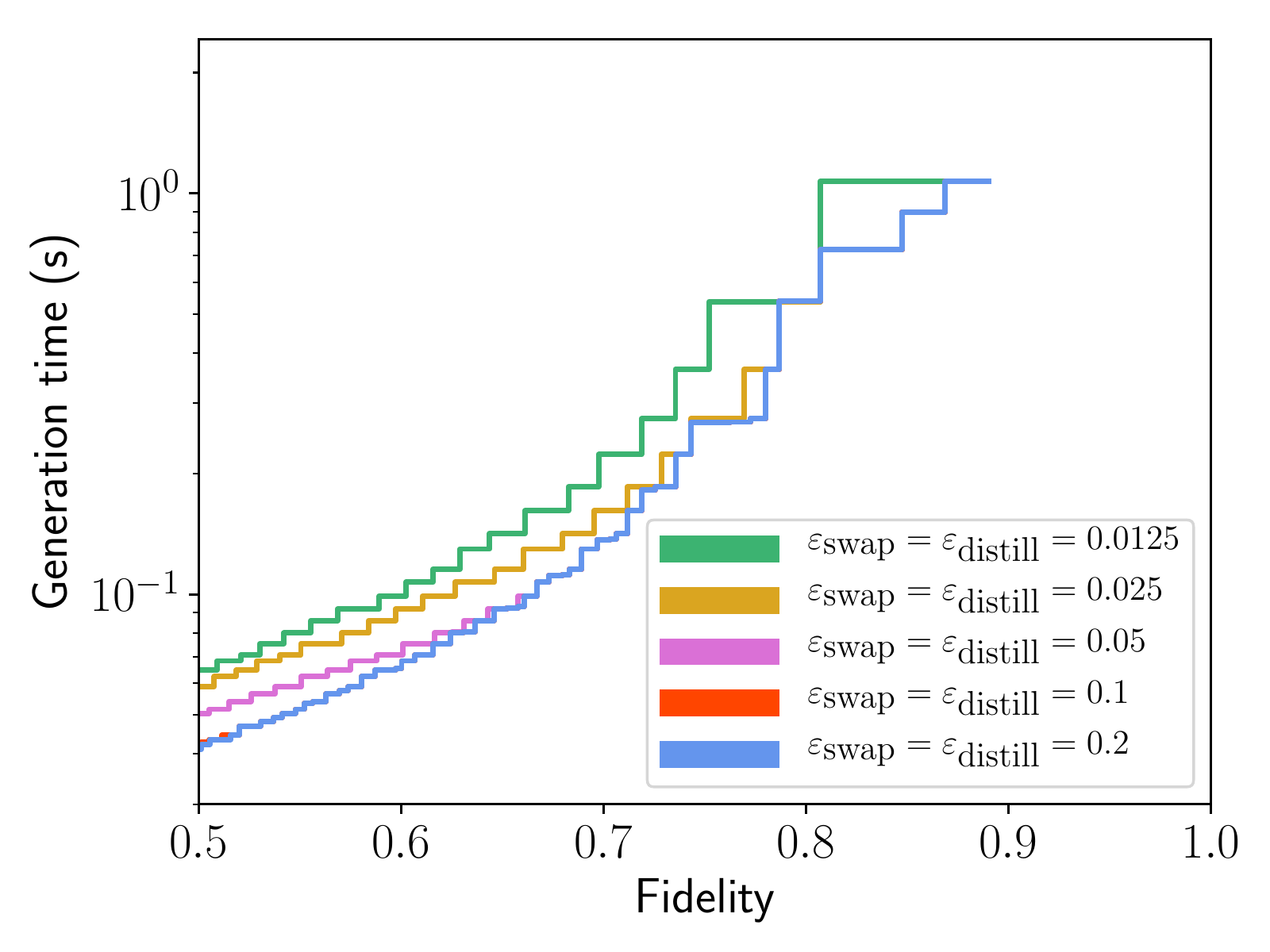}
		\caption{Optimisation results with banded swapping heuristic. The difference between $\varepsilon_{\textrm{swap}}$ = $\varepsilon_{\textrm{distill}} = 0.1$ and $0.2$ is minimal, differing only slightly for very low fidelities.}
		\label{fig:varepsds}
	\end{subfigure}
	~ 
	\begin{subfigure}[b]{0.49\textwidth}
		\includegraphics[clip, trim = 4mm 0mm 0mm 0mm, width = 0.98\textwidth]{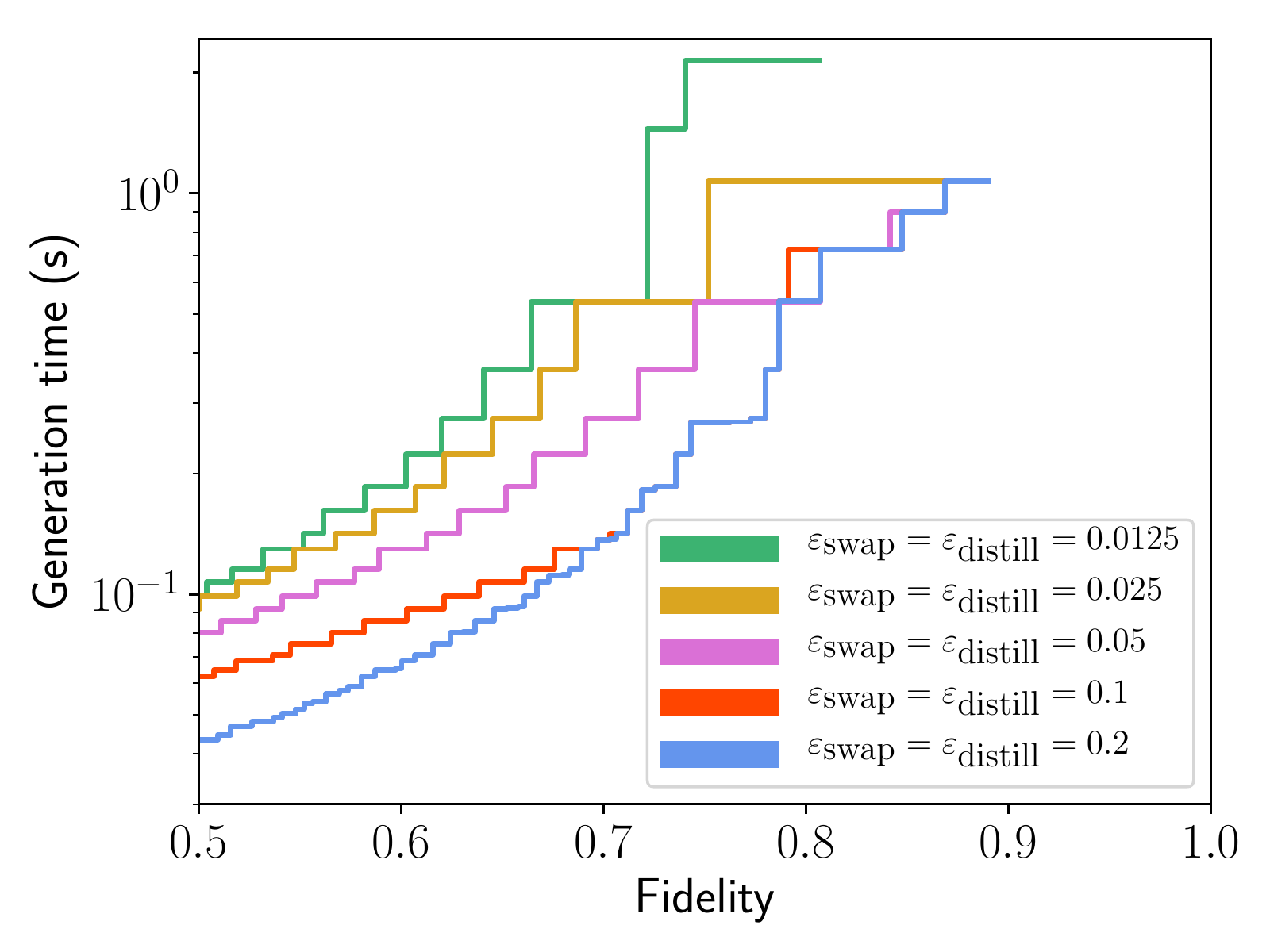}
		\caption{Optimisation results with naive swapping heuristic. Note that the curves converge in a significantly poorer fashion than in Fig.~\ref{fig:varepsds}.}
		\label{fig:varepsds2}
	\end{subfigure}
	\caption{Optimised schemes for a distance of 300 kilometres using four intermediate nodes with the IP parameter set 4 (see Table~\ref{tab:parameter_sets_IP}) for several different pairs of $\varepsilon_{\textrm{swap}}$ and $\varepsilon_{\textrm{distill}}$. Note that the curves converge in a significantly poorer fashion than in Fig.~\ref{fig:varepsds}.}
\end{figure}

\begin{figure}
	\centering
	\includegraphics[clip, trim = 0mm 0mm 0mm 0mm, width = 0.5\textwidth]{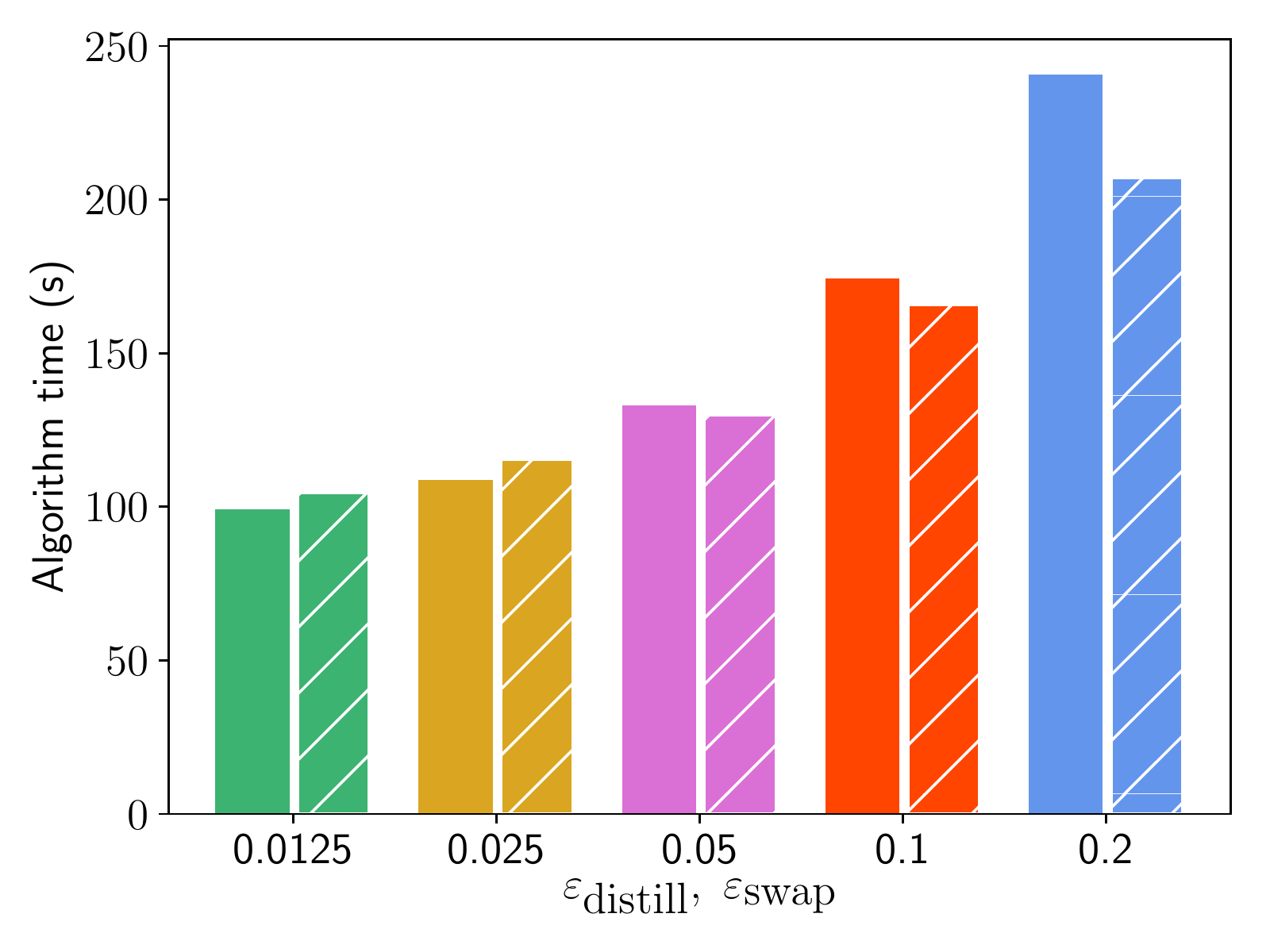}
	\caption{Algorithm runtimes for several different values of $\varepsilon_{\textrm{swap}}$ and $\varepsilon_{\textrm{distill}}$. The right, solid bars are for the optimisation with the heuristic for swapping found in Eq.~\ref{eq:banded_swap}, while the left, hatched bars are for the optimisation where we only swap between states that are $\varepsilon_{\textrm{swap}}$-close in fidelity. We observe that both heuristics lead to approximately the same runtime behaviour, while the results with the banded swapping heuristic are closer to optimal.}
	\label{fig:varepstime}	
\end{figure}

In Figs.~\ref{fig:varepsds} and \ref{fig:varepsds2} we perform an optimisation for several different values of $\varepsilon_\textrm{swap}$ and $\varepsilon_\textrm{distill}$, using parameter set 4 with four intermediate nodes for a distance of 300 kilometre. In Fig.~\ref{fig:varepsds} we use the banded swapping heuristic (see equation \ref{eq:banded_swap}), while in Fig.~\ref{fig:varepsds2} we only swap between states that are $\varepsilon_{\textrm{swap}}$-close in fidelity. We observe that the optimisation results in Fig.~\ref{fig:varepsds2} are significantly worse than those in Fig.~\ref{fig:varepsds}, while Fig.~\ref{fig:varepstime} indicates that the runtimes are comparable for both heuristics. We use these results to settle on $\varepsilon_\textrm{swap} = \varepsilon_\textrm{distill}  = 0.05$. Furthermore, we find that the heuristic plays primarily a role for smaller fidelities. This implies that only for small fidelities there is a benefit in swapping between states with disparate fidelities.

\section{Average noise due to storage}
\label{sec:avg_noise}
Here we discuss the average noise induced when repeating a protocol with success probability $p$ until success or until a maximum number of $r$ attempts. 
Denote the quantum channel corresponding to storing for a single round by $\Lambda$. 
The average noise channel $\mathbb{E}\left[\Lambda\right]$ corresponds to having the channel $\underbrace{\Lambda \circ \Lambda \ldots \circ \Lambda}_{r-j}$ with probability $\frac{p\left(1-p\right)^{j-1}}{1-(1-p)^r}$ for $1\leq j \leq r$. Note that we can calculate the average channel instead of the average density matrix at the output, due to the linearity of quantum channels.

We consider two types of noise in this paper, depolarising and dephasing. These types of noise occur naturally in quantum information processing systems, and have the following exponential behaviour,

\begin{gather}
	\mathcal{N}_{d}(\rho) = e^{-\frac{t}{\lambda_d}}\rho + \left(1-e^{-\frac{t}{\lambda_d}}\right)\frac{\mathds{I}}{2}\ ,~\textrm{(depolarising)}\nonumber\\
	\mathcal{N}_{Z}(\rho) = \frac{1+e^{-\frac{t}{\lambda_Z}}}{2}\rho + \frac{1-e^{-\frac{t}{\lambda_Z}}}{2}\frac{\mathds{I}}{2}\ ,~\textrm{(dephasing)}\nonumber
\end{gather}

Thus, if we want to calculate the average amount of noise for depolarising and dephasing, it suffices to calculate the average of $e^{-c\cdot \left(k-j\right)}$ with probability distribution $\frac{p\left(1-p\right)^{j-1}}{1-(1-p)^r}$, $j = \lbrace 1, \ldots, r\rbrace$, where $c\equiv \frac{T_\textrm{attempt}}{T_{\textrm{depol}/\textrm{deph}}}$ quantifies the noise experienced in a single attempt for depolarising and dephasing, respectively.  We find thus that the average channels correspond to having the exponential terms in the above channels set to

\begin{align}
	\mathds{E}\left[e^{-c\cdot (r-j)}\right] =&\sum_{j = 1}^{r}\frac{p\left(1-p\right)^{j-1}}{1-(1-p)^r} \cdot  e^{-c\cdot (r-j)} \nonumber\\
	=&\frac{pe^{c}\left(\left(1-p\right)^r-e^{-cr}\right)}{\left(1-\left(1-p\right)^r\right)\left(e^c\left(1-p\right)-1\right)}\label{eq:avgexp}\ .
\end{align}

Finally, the decay in the success probability for retrieving a state from a memory for multiplexed platforms is given by $\mathds{E}\left[e^{-c\cdot (r-j)}\right]$, where $c = \frac{T_{\textrm{attempt}}}{T_\textrm{coh}}$.

\section{Modelling of elementary pair generation for multiplexed platforms}
\label{sec:ellinkgenMP}
Here we detail the calculations performed to derive the analytical form of the resultant state during elementary pair generation for multiplexed platforms, and the success probability (see equation \ref{eq:tittelprob}). We first discuss the effects of the losses on the state emitted by the PDC sources. Secondly, the Bell state measurement and the resulting post-measurement state are discussed. We will close with a brief discussion on the post-selection of having zero photons. 
We model all losses in the setup as a pure-loss channel. Since we restrict ourselves to at most two-photon excitations in each mode, we truncate the Kraus operators from~\cite{ivan2011operator} to the $\lbrace \ket{0}, \ket{1}, \ket{2}\rbrace$ subspace and find the explicit matrix form of the truncated Kraus operators. They are
\begin{align}
	A_0 = \begin{bmatrix}
		1 & 0 & 0\\
		0 & \sqrt{1-\gamma} & 0\\
		0 & 0 & 1-\gamma
	\end{bmatrix},~A_1 = \begin{bmatrix}
		0 & \sqrt{\gamma} & 0\\
		0 & 0 & \sqrt{2\left(1-\gamma\right)\gamma}\\
		0 & 0 & 0
	\end{bmatrix},~A_2 = \begin{bmatrix}
		0 & 0 & \gamma \\
		0 & 0 & 0\\
		0 & 0 & 0
	\end{bmatrix}\ \nonumber,
\end{align}
where $\gamma = 1-\eta$ is the loss parameter. Note that, even after truncation, these Kraus operators still form a channel since $\sum_{i=0}^{2}A_i^{\dagger}A_i = \mathbb{I}$. 
We now let four such channels act on the state
\begin{align}
	\ket{\psi_{N_s}} &= \sqrt{p_0} \ket{00, 00} + \sqrt{\frac{p_1}{2}}\left(\ket{10, 01} + \ket{01, 10}\right) + \sqrt{\frac{p_2}{3}}\left(\ket{20, 02} - \ket{11, 11} + \ket{02, 20}\right)\nonumber\ ,
\end{align}
where the early and late photonic modes in the direction towards the memory each evolve under a truncated pure-loss channel with parameter $\gamma_1$, and similarly for the two modes going towards the beamsplitter station with parameter $\gamma_2$. This results in a state $\rho(N_s, \gamma_1, \gamma_2)_{a_0a_1b_0b_1}$ between the memory and the photon just before the beamsplitter. The same situation holds for the other PDC source, such that the total state just before the beamsplitter is $\rho(N_s, \gamma_1, \gamma_2)_{a_0a_1b_0b_1}\otimes \rho(N_s, \gamma_1, \gamma_2)_{c_0c_1d_0d_1}$, where we have assumed the prepared states have equal mean photon number and experience equal losses.
Instead of applying the unitary corresponding to the beamsplitter and then applying the POVMs for the detectors, we can apply the inverse of the beamsplitter unitary on the POVMs corresponding to success. Since we assume photon number resolving detectors, we find our POVM elements corresponding to success to be 
\begin{align}
	\mathbb{I}_{a_0a_1}	 \otimes \ket{\Psi^+}\hspace{-1mm}\bra{\Psi^+}_{b_0c_0}\otimes \ket{\Psi^+}\hspace{-1mm}\bra{\Psi^+}_{b_1c_1} \otimes \mathbb{I}_{d_0d_1}\nonumber\ ,\\
	\mathbb{I}_{a_0a_1}	 \otimes \ket{\Psi^-}\hspace{-1mm}\bra{\Psi^-}_{b_0c_0}\otimes \ket{\Psi^+}\hspace{-1mm}\bra{\Psi^+}_{b_1c_1} \otimes \mathbb{I}_{d_0d_1}\nonumber\ ,\\
	\mathbb{I}_{a_0a_1}	 \otimes \ket{\Psi^+}\hspace{-1mm}\bra{\Psi^+}_{b_0c_0}\otimes \ket{\Psi^-}\hspace{-1mm}\bra{\Psi^-}_{b_1c_1} \otimes \mathbb{I}_{d_0d_1}\nonumber\ ,\\
	\mathbb{I}_{a_0a_1}	 \otimes \ket{\Psi^-}\hspace{-1mm}\bra{\Psi^-}_{b_0c_0}\otimes \ket{\Psi^-}\hspace{-1mm}\bra{\Psi^-}_{b_1c_1} \otimes \mathbb{I}_{d_0d_1}\nonumber\ .
\end{align}
We find that the post-measurement states for each of these POVM element is equivalent up to local unitaries, such that we only have to consider the first one. 
While one could call the whole process described so far elementary pair generation, the state will have a fidelity equal to half or less for any $N_s > 0$. The reason for this is that there has not been any post-selection on detecting a valid click pattern on the detectors when performing, say, Bell state measurements. For this reason, we apply the following POVM to post-select on having non-zero photons at each side of the memory,
\begin{align}
	\mathbb{I}_{a_0a_1b_0b_1} - \left(\ket{00}\hspace{-1mm}\bra{00}_{a_0a_1}\otimes \mathbb{I} + \mathbb{I} \otimes \ket{00}\hspace{-1mm}\bra{00}_{d_0d_1} - \ket{0000}\hspace{-1mm}\bra{0000}_{a_0a_1d_0d_1} \right)\nonumber \ .
\end{align}
The resultant state is too cumbersome to report here, but can be found in the accompanying Mathematica and Python scripts. We find the success probability to be given by
\begin{align}
	p_{\textrm{succ}} &= 4 \cdot p_{\textrm{bsm}} \cdot p_{\textrm{non-zero photons}} \nonumber\\
	& =4 \cdot \eta^2\frac{3p_1^2-4\left(4\eta -3\right)p_1p_2 +4p_2 \left(1+\left(3-8\eta + 4\eta^2\right)\right)}{24} \nonumber \\
	&\cdot p_\textrm{app}^2\frac{\left(p_1+4\left(\eta-1\right)p_2\left(p_{\textrm{app}}-2\right)\right)\left(3p_1 + 4\left(\eta-1\right)p_2\left(p_\textrm{app}-2\right)\right)}{4p_2 + \left(p_1 +\left(2-4\eta \right)p_2\right)\left(3p_1 + \left(6-4\eta\right)p_2\right)}\label{eq:tittelprob}\ .
\end{align}

\section{The interplay between the number of modes and the fidelity for multiplexed platforms}
\label{sec:interplay}
Here we investigate the interplay between the number of modes, the fidelity, and the losses in the fibre for MP platforms. We assume here that the only source of noise is from the PDC source, and there are no or negligible losses locally. We take the state derived in the previous section (but which is too cumbersome to report here), and set $p_\textrm{app} = 1$. The fidelity of the resultant state is then calculated to be

\begin{align}
	F=\frac{3}{4 (\eta -1)^2 {N_s}^4+24 (\eta -1)^2 {N_s}^3+4 \left(9 \eta^2-20 \eta +11\right) {N_s}^2-24 (\eta -1) {N_s}+3}\nonumber \ .
\end{align}
Solving for $N_s$, we find that
\begin{align}
	N_s = \frac{1}{2} \left(\sqrt{\frac{-9 \eta  F+5 F+2 \sqrt{F (F+3)}}{F-\eta  F}}-3\right)\nonumber \ .
\end{align}

Let us now input the above relation into equation~\ref{eq:tittelprob} where we set $p_1$ and $p_2$ according to equations~\ref{eq:photondistribution}. We find a success probability of

\begin{align}
	p = \frac{32 \eta ^2 \left(\sqrt{\frac{-9 F \eta +5 F+2 \sqrt{F (F+3)}}{F-F \eta
		}}-3\right)^2}{F \left(\sqrt{\frac{-9 F \eta +5 F+2 \sqrt{F (F+3)}}{F-F \eta
		}}-1\right)^6}\ \label{eq:nsF}\ .
\end{align}

Since we need on the order of $\frac{1}{p}$ modes, we find that we need on the order of

\begin{align}
	\frac{1}{p} = \frac{F \left(\sqrt{\frac{-9 F \eta +5 F+2 \sqrt{F (F+3)}}{F-F \eta
		}}-1\right)^6}{32 \eta ^2 \left(\sqrt{\frac{-9 F \eta +5 F+2 \sqrt{F (F+3)}}{F-F \eta
		}}-3\right)^2} = 
	\frac{F \left(\sqrt{\frac{2 \sqrt{F (F+3)}}{F}+5}-1\right)^6}{32\eta ^2 \left(\sqrt{\frac{2
				\sqrt{F (F+3)}}{F}+5}-3\right)^2} + \mathcal{O}\left(\eta^{-1}\right)\nonumber
\end{align}

modes to achieve a fidelity of $F$ for $\eta \approx 0$. The $\eta^2= \exp(-\frac{L}{L_0})$ term in the denominator is given by the total losses of the fibre. The contribution due to the fidelity is then given by

\begin{align}
	\frac{F \left(\sqrt{\frac{2 \sqrt{F (F+3)}}{F}+5}-1\right)^6}{32 \left(\sqrt{\frac{2
				\sqrt{F (F+3)}}{F}+5}-3\right)^2} =\frac{32}{\left(1-F\right)^2} + \mathcal{O}\left(\left(1-F\right)^{-1}\right)\nonumber \ .
\end{align}

We thus find that the number of minimum required modes scales as $\frac{e^{\frac{L}{L_0}}}{\left(1-F\right)^2}$, where $L$ is the internode distance.

\section{The effect of efficiency decoherence for multiplexed platforms}
\label{sec:effdecoherence}
In this section we explore the effects the exponential decrease of the output efficiency has on the ability of performing schemes with probability greater than $p_\textrm{min}$. While for information processing platforms it is always possible to achieve any success probability by performing as many attempts $r$ as required, this is not the case for MP platforms due to the decrease in output efficiency over time. Here we derive conditions on the efficiency coherence times of the memories for a given $p_\textrm{min}$, generation time $T$ and success probability $p$ of the underlying schemes, such that $p_\textrm{min}$ can be achieved.

Since there are two memories used for state storage, the success probability of emitting both states again is modelled as given by 

\begin{align}
	p_{\textrm{single success}}&= \left(1-\left(1-p\right)^r\right) \cdot \mathds{E}\left[e^{-\left(c_1+c_2\right)\cdot (r-j)}\right]\nonumber \\
	&= \frac{pe^{\left(c_1+c_2\right)}\left(\left(1-p\right)^r-e^{-{\left(c_1+c_2\right)}r}\right)}{e^{\left(c_1+c_2\right)}\left(1-p\right)-1}\label{eq:psuc1}\ .
\end{align}

We are interested in when the above quantity cannot be larger than $p_\textrm{min}$. To this end, we take the derivative of equation \ref{eq:psuc1} with respect to $r$ and set it to zero to find the maximum value of success probability. Setting $c = c_1 +c_2$, we find

\begin{gather}
	\frac{e^c p \left(c e^{-c r}+(1-p)^r \log (1-p)\right)}{e^c (1-p)-1} = 0\ \nonumber,\\
	\rightarrow r = \frac{c-\log \left(-\frac{e^c \log (1-p)}{c}\right)}{c+\log (1-p)}\label{eq:eqforr}
	\ .
\end{gather}
However, since $r$ needs to be an integer equal to or greater than one, we choose the ceiling or floor of equation \ref{eq:eqforr}, whichever maximises the resultant $p_\textrm{succ}$.
Furthermore, since we cannot perform distillation, our main concern is the drop in success probability after performing a Bell state measurement. This motivates us to set $p = 1-\frac{1}{2^{N+1}}$~\cite{lutkenhaus1999bell, vaidman1999methods, ewert20143, grice2011arbitrarily, lee2015nearly1}. Setting equation~\ref{eq:psuc1} equal to $p_\textrm{min}=0.9$, we numerically find that $N = 0$ gives $c\approx 0.023$, $N=1$ gives $c \approx 0.053$, $N=2$ gives $c \approx 0.101$ and $N=3$ gives $c = \infty$. Obviously, the assumption here is that the initial success probability is given by $1-\frac{1}{2^{N+1}}$, which is not true due to other losses in the system. However, it is clear that increasing $N$ can increase the total time significantly during which entanglement can be generated in a near-deterministic fashion. In particular, we find that the sum of the reciprocals of the efficiency coherence times of the memories should be \emph{at least} $\frac{1}{c}$ times the generation time of a scheme for multiplexed platforms to successfully generate entanglement near-deterministically. This results in factors of approximately $43$, $19$ and $10$ times the generation time for Bell state measurement success probabilities of $\frac{1}{2}$, $\frac{3}{4}$ and $\frac{7}{8}$, respectively.

\newpage

\section{Additional optimisation results}
\label{sec:additional_results}

This section contains the additional figures mentioned in the main text. 

\begin{figure}[h!]
	\centerfloat
	\vspace*{2mm}
	\includegraphics[clip,  width = 0.8\textwidth, trim = 3.9mm 3.5mm 3.5mm 3.6mm]{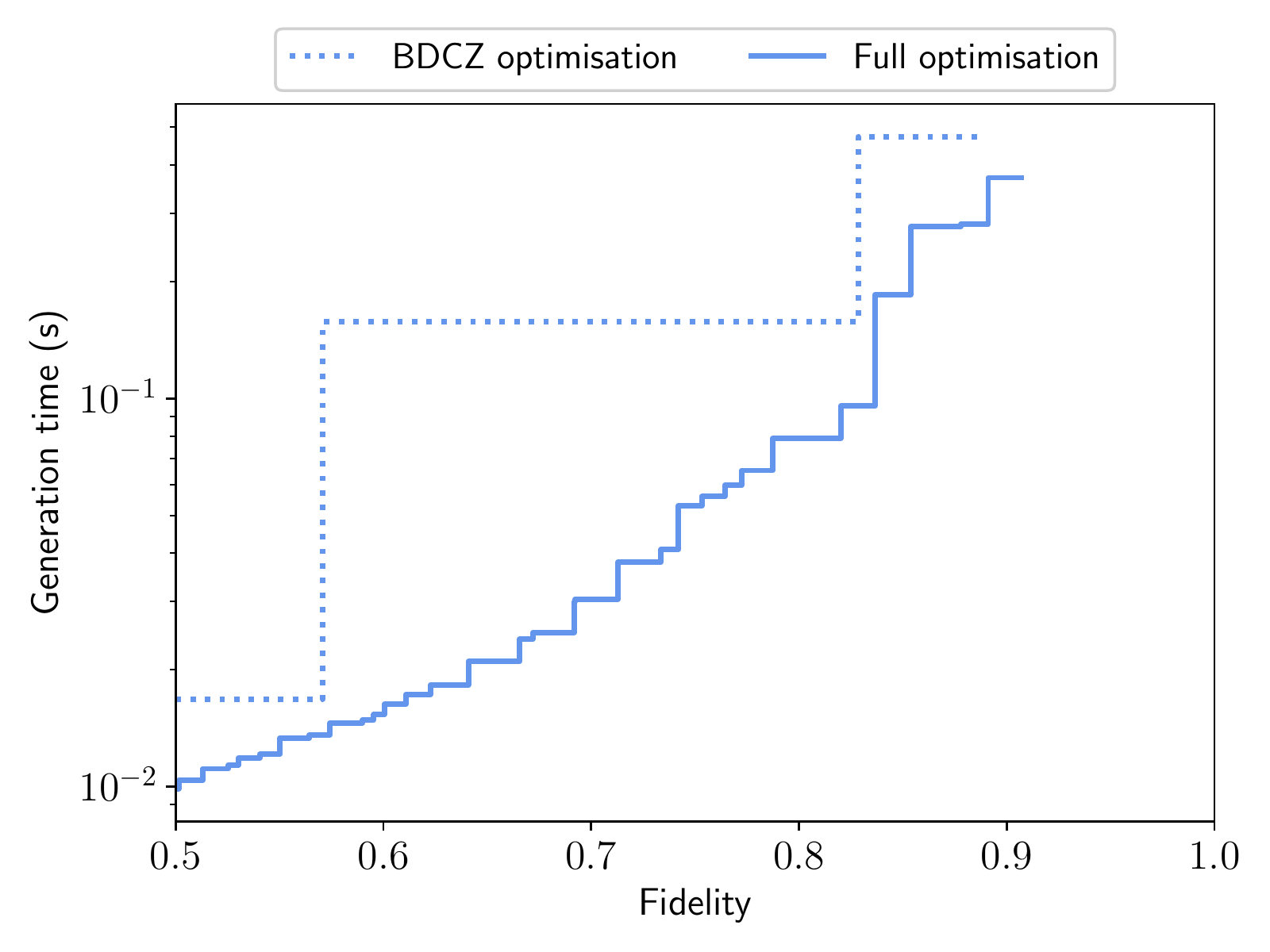}
	\vspace*{-2mm}
	\caption{Comparison between the optimisation results of the full optimisation with the optimisation over BDCZ schemes. We consider a repeater chain over 200 kilometres with three intermediate nodes. The parameters used are parameter set 4 for the intermediate nodes and parameter set 2 for Alice and Bob. BDCZ schemes are those schemes that only perform swapping and distillation between two schemes that have used the same sequences of protocols. Contrary to the comparison with BDCZ schemes in~\cite{jiang2007optimal} we allow for an optimisation over the different ways of generating elementary pairs, i.e.~we vary the number of attempts $r$ and the $\theta$ parameter. We observe that the schemes found with the full optimisation outperform BDCZ schemes, achieving a faster generation time by a factor of $\sim$10, and extending the maximal achievable fidelity by a small margin. }
	\label{fig:bdcz_comp}
\end{figure}

First, we compare the optimisation results of the full optimisation with an optimisation over BDCZ schemes only in Fig.~\ref{fig:bdcz_comp}. BDCZ schemes are those schemes that for each connection and distillation step only combines two schemes that have used the same sequence of protocols, as in~\cite{briegel1998quantum,Duer1998}. We consider an asymmetric repeater chain with three intermediate nodes over a distance of 200 kilometre. We model the behavior of the intermediate nodes with parameter set 4 and Alice and Bob with parameter set 2 (see Table \ref{tab:parameter_sets_IP}). The full optimisation yields schemes that can achieve faster generation rates (by approximately a factor of 10) than achievable with the BDCZ schemes.

\begin{figure}[h!]
	\centerfloat
	\vspace*{2mm}
	\includegraphics[clip,  width = 1.02\textwidth, trim = 16.2mm 16mm 6.5mm 16.8mm, angle=270]{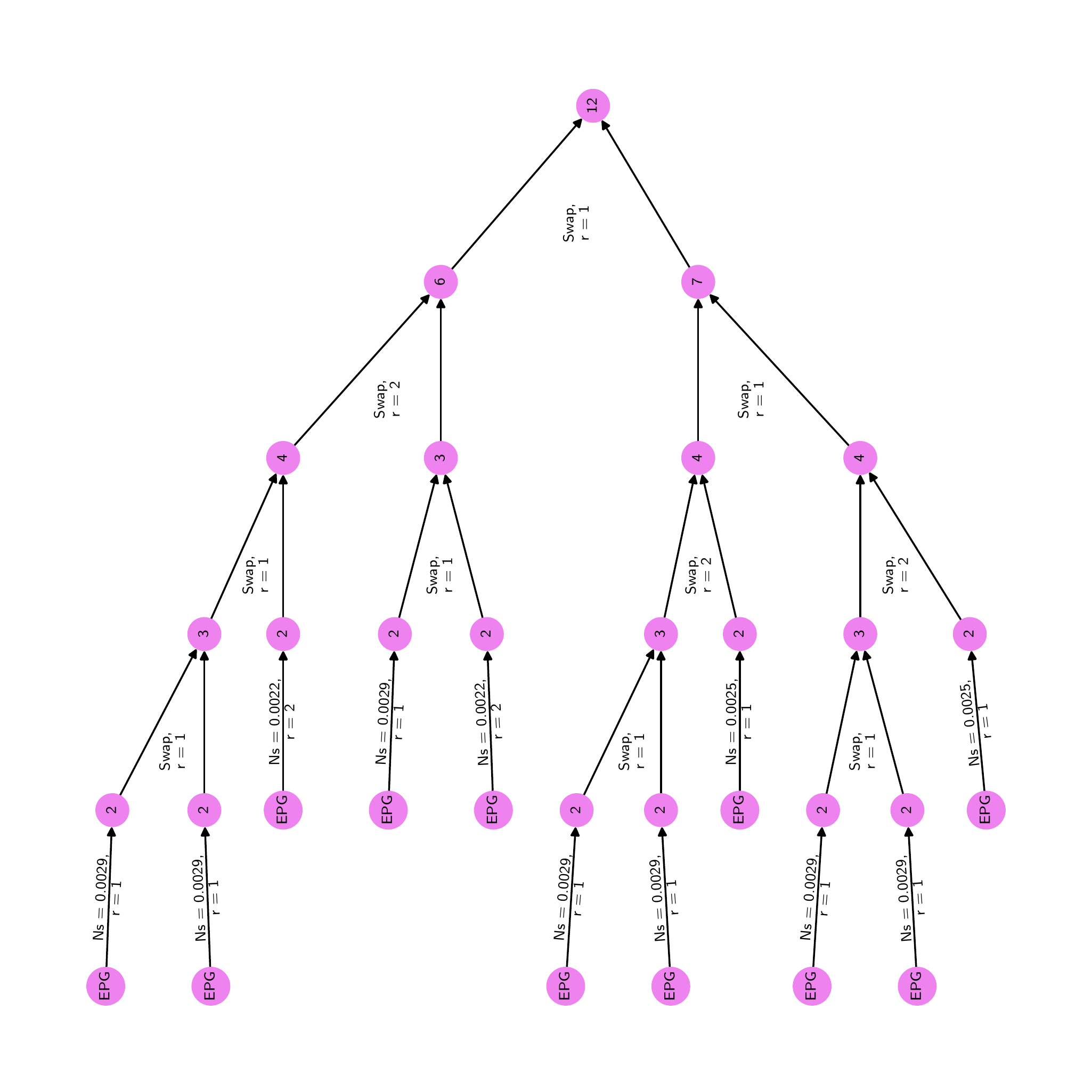}
	\vspace*{-9mm}
	\caption{Visual representation of the scheme with the lowest non-trivial achieved fidelity for a distance of 800 kilometres with a combination of information processing and multiplexed platforms using parameter sets 4 (see Tables \ref{tab:parameter_sets_IP} and \ref{tab:parameter_sets_MP}) and ten intermediate nodes. Elementary pair generation is indicated by EPG, and the mean photon number used is indicated by the $N_s$.}
	\label{fig:MP_800_lowestfidelity_scheme_MP_IP}
\end{figure}

\begin{figure}
	\centerfloat
	\includegraphics[clip,  width = 1.19\textwidth, trim = 18mm 18mm 18.5mm 18mm, angle=270]{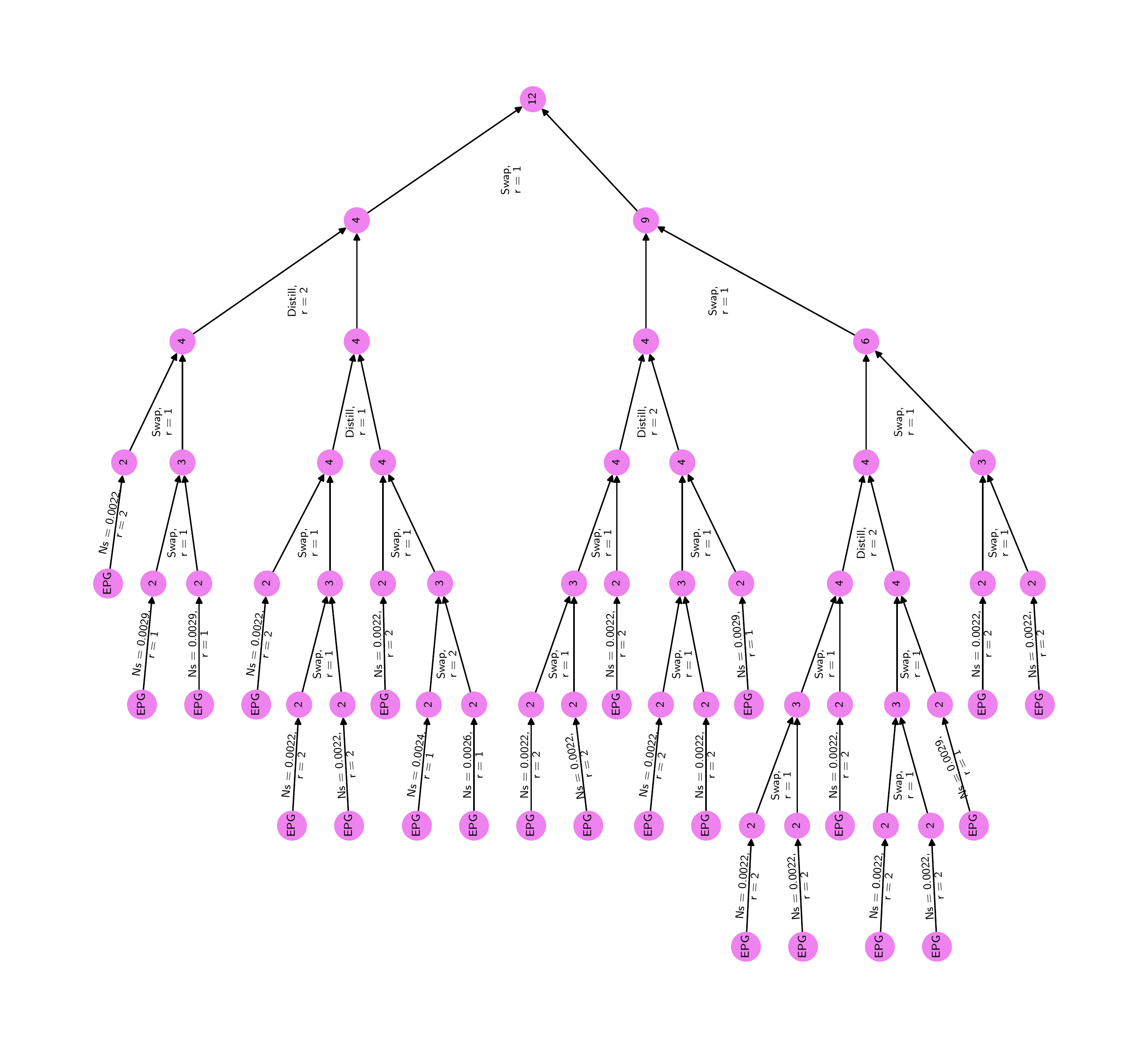}
	\caption{Visual representation of the scheme that achieves a fidelity of $F=0.9605$ in time $T=17.7$ milliseconds (indicated by the cross in Fig.~\ref{fig:800comp}), for a distance of 800 kilometres with a combination of information processing and multiplexed platforms using parameter sets 4 (see Tables \ref{tab:parameter_sets_IP} and \ref{tab:parameter_sets_MP}) and ten intermediate nodes. Elementary pair generation is indicated by EPG, and the mean photon number used is indicated by the $N_s$. The structure of the scheme is non-hierarchical, which can most clearly be seen in the final swap operation, which happens between two links of lengths four and nine.}
	\label{fig:MP_800_almosthighestfidelity_scheme_MP_IP}
\end{figure}

\begin{center}
	\begin{figure}
		\vspace*{-0mm}
		\includegraphics[clip, trim =  3.6mm 3mm 4mm 4mm, width = 0.8\textwidth]{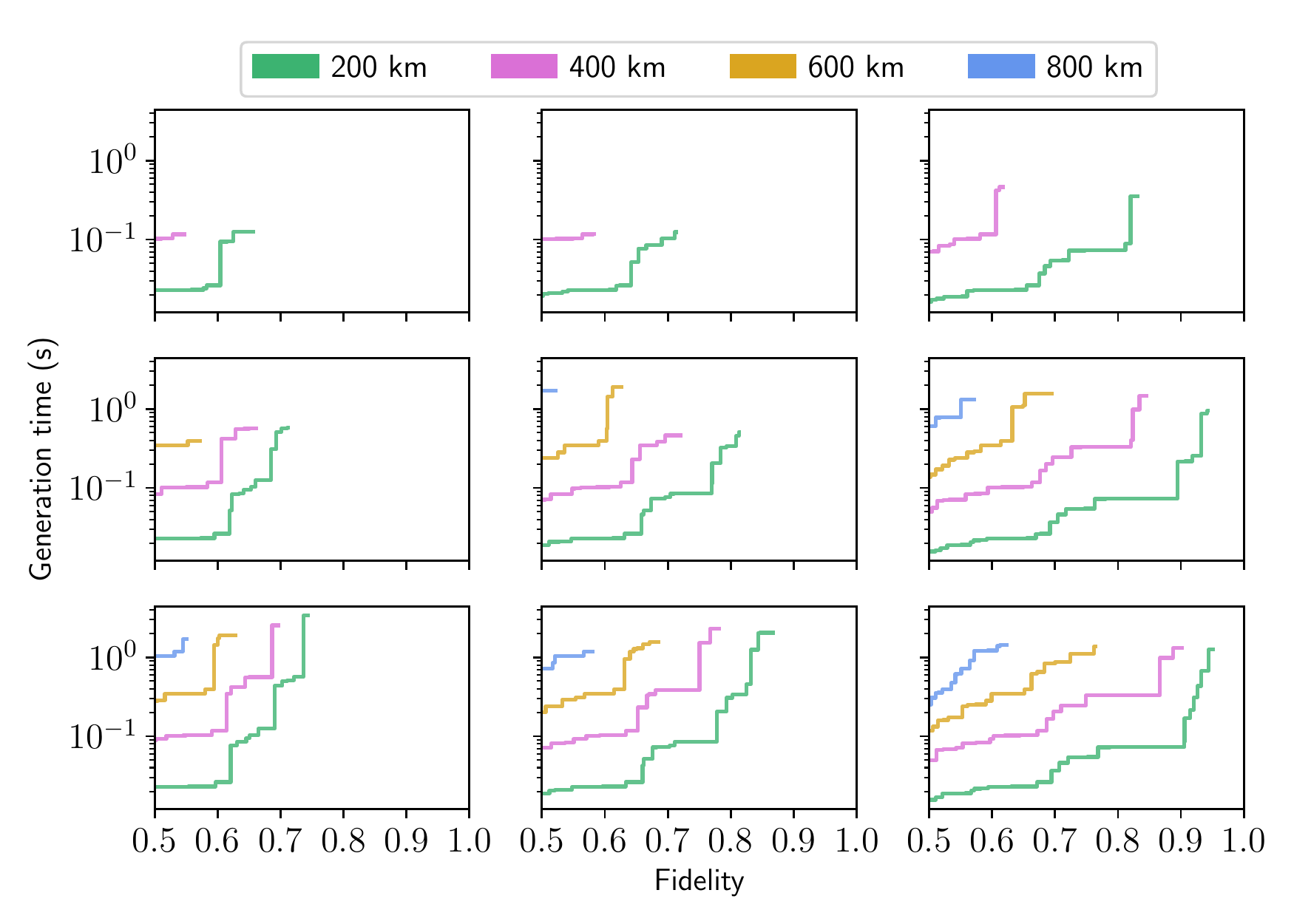}
		\captionof{figure}{Optimisation results for total distances of 200, 400, 600 and 800 kilometres, using ten intermediate nodes. We use IP parameter set 2 as a baseline, where we set the gate fidelities to be 0.99, 0.995 and 0.999 in the first, second and third column respectively. We set the coherence times $T_{\textrm{deph}},~T_{\textrm{depol}}$ to 10, 50 and 100 seconds in the first, second and third row, respectively.}
		\label{fig:IPparameterexploration-10-repeaters}
	\end{figure}
\end{center}

Second, we consider two visualisations (Fig.~\ref{fig:MP_800_lowestfidelity_scheme_MP_IP} and \ref{fig:MP_800_almosthighestfidelity_scheme_MP_IP}) of the schemes found for a distance of 800 kilometres with a combination of information processing and multiplexed platforms using parameter sets 4 (see Tables \ref{tab:parameter_sets_IP} and \ref{tab:parameter_sets_MP}) and ten intermediate nodes. Finally, Fig.~\ref{fig:IPparameterexploration-10-repeaters} contains the results found while performing a parameter exploration for total distances of 200, 400, 600 and 800 kilometres, using ten intermediate nodes for information processing platforms.

\end{document}